\documentclass[preprint,prd,tightenlines,nofootinbib,eqsecnum,superscriptaddress,showpacs,preprintnumbers]{revtex4-1}

\usepackage{amsmath}
\usepackage{amsfonts}
\usepackage{amssymb}
\usepackage{bm}
\usepackage{color}
\usepackage[colorlinks]{hyperref}

\definecolor{CiteColor}{rgb}{0,0.6,0}
\hypersetup{citecolor=CiteColor}

\newcommand{\Hzero}{H^{(0)}}
\newcommand{\Hint}{H^{(1)}}
\newcommand{\HI}{H_{\rm int}}
\newcommand{\Heff}{{{\cal H}}}
\newcommand{\cG}{{\cal G}}
\newcommand{\bn}{{\bf n}}
\newcommand{\bw}{{\bf w}}
\newcommand{\A}{{\sf A}}
\newcommand{\B}{{\sf B}}
\newcommand{\atISSO}{_{\rm ISSO}}
\newcommand{\ud}{\mathrm{d}}
\newcommand{\ui}{\mathrm{i}}
\newcommand{\subgamma}{{(\gamma)}} 
\newcommand{\rmd}{{\rm d}}

\begin{document}

\title{Hamiltonian Formulation of the Conservative Self-Force Dynamics in the Kerr Geometry}

\def\addYITP{Yukawa Institute for Theoretical Physics, Kyoto University, Kyoto 606-8502, Japan}
\def\addCENTRA{CENTRA, Departamento de F\'{\i}sica, 
Instituto Superior T\'ecnico,\\
Universidade de Lisboa -- UL, Avenida Rovisco Pais 1, Portugal}
\def\addGuelph{Department of Physics, University of Guelph, 
Guelph, Ontario, N1G 2W1, Canada}
\def\addNeoNeet{1500-5, Nishikiwa, Ube, Yamaguchi 755-0151, Japan}
\def\addKyoto{Department of Physics, Kyoto University, Kyoto 606-8502, Japan}
\def\addLUTh{LUTH, Observatoire de Paris, PSL Research University, CNRS,
Universit\'e Paris Diderot, Sorbonne Paris Cit\'e, 92190 Meudon, France}
\def\addRoch{Center for Computational Relativity and Gravitation, Rochester Institute of Technology, \\
85 Lomb Memorial Drive, Rochester, New York 14623, USA}
\def\addRyukoku{Faculty of Law, Ryukoku University, Kyoto 612-8577, Japan
}
\def\addKyushu{Faculty of Arts and Science, Kyushu University, Fukuoka 819-0395, Japan}

\author{Ryuichi~Fujita}
\affiliation{\addCENTRA}

\author{Soichiro~Isoyama}
\affiliation{\addGuelph}

\author{Alexandre~Le~Tiec}
\affiliation{\addLUTh}

\author{Hiroyuki~Nakano}
\affiliation{\addRyukoku}
\affiliation{\addKyoto}
\affiliation{\addRoch}

\author{Norichika~Sago}
\affiliation{\addKyushu}

\author{Takahiro~Tanaka}
\affiliation{\addKyoto}
\affiliation{\addYITP}

\date{\today}

\preprint{KUNS-2648,YITP-16-122}

\begin{abstract}

We formulate a Hamiltonian description of the orbital motion of a point particle in Kerr spacetime for generic (eccentric, inclined) orbits, which accounts for the effects of the conservative part of the gravitational self-force. This formulation relies on a description of the particle's motion as geodesic in a certain smooth effective spacetime, in terms of (generalized) action-angle variables. Clarifying the role played by the gauge freedom in the Hamiltonian dynamics, we extract the gauge-invariant information contained in the conservative self-force. We also propose a possible gauge choice for which the orbital dynamics can be described by an effective Hamiltonian, written solely in terms of the action variables. As an application of our Hamiltonian formulation in this gauge, we derive the conservative self-force correction to the orbital frequencies of Kerr innermost stable spherical (inclined or circular) orbits. This gauge choice also allows us to establish a ``first law of mechanics'' for black-hole-particle binary systems, at leading order beyond the test-mass approximation.


\end{abstract}
\maketitle

\section{Introduction}
\label{sec:Intro}

\subsection{Motivation}

Following the detections by the LIGO
observatories \cite{Abbott:2016blz,Abbott:2016bis,Abbott:2016ter}
of gravitational waves from the coalescence of binary black holes, the era of 
gravitational-wave astronomy has finally begun. While a worldwide network of
ground-based detectors is under development, future missions 
such as LISA \cite{Seoane:2013qna,Seoane:2017drz} and 
(B-)DECIGO/BBO \cite{Seto:2001qf, BBO, BDECIGO} 
\footnote{B-DECIGO was previously known as Pre-DECIGO~\cite{Nakamura:2016hna}.}
will bring gravitational-wave physics into space.
The performances of the LISA Pathfinder spacecraft,
a proof-of-concept technological mission for a future gravitational-wave antenna
in space, exceeded all expectations \cite{Armano:2016}.
An important class of sources for those space-based observatories is inspiralling
compact-object binaries with a large hierarchy of masses,
dubbed extreme mass-ratio inspirals (EMRIs) \cite{AmaroSeoane:2007aw}.

Owing to the large mass ratios of these sources, the effects of radiation reaction
are weak, such that many orbital (and gravitational wave) cycles 
will occur in the observable frequency band of detectors 
before the inspiral ends.
This will provide unparalleled precision to 
probe the near-horizon spacetime region \cite{Barausse:2014tra}  
and to test general relativity in the strong-field regime \cite{Berti:2015}, 
and allow measuring the astrophysical parameters of the
sources (including the masses and spins) with exquisite accuracy \cite{Barack:2004}.
However, in order to extract such information about the black hole geometry, 
it is necessary to develop highly accurate theoretical templates of
the waveforms \cite{Cutler:1992tc,Lindblom:2008}. 
(Much progress has already been achieved towards this objective;
see for instance Ref.~\cite{Buonanno:2014aza} and references therein.)
This, in turn, requires a detailed understanding of the long-term
radiative evolution of the orbital phase, for generic orbits around
astrophysical Kerr black holes.

The program to develop template waveforms for EMRIs
based on black hole perturbation theory is motivated
by the initial success in studying linear perturbations
of the Kerr geometry \cite{Teukolsky:1973ha}.
That theory was first used to compute the fluxes of energy and angular momentum
due to the emission of gravitational waves \cite{Kojima:1984cj}
(see also Refs.~\cite{Mino:1997bx,Sasaki:2003xr} and references therein), from
which the average rates of change of the energy and angular momentum of the
orbital motion of the smaller body can be inferred. 
More recently, the formalism required to calculate 
the rate of change of the particle's
Carter constant \cite{Carter:1968rr}, the third constant of motion 
in Kerr spacetime, was developed \cite{Mino:2003yg,Sago:2005gd} 
and implemented \cite{Sago:2005fn,Ganz:2007rf,Fujita:2009us,Sago:2015rpa};
see also Refs.~\cite{Hughes:2005qb,Drasco:2005kz}.

However, the knowledge of the rates of change of the constants of motion, which are no
longer constant once the effects of radiation reaction are taken into
account \cite{Cutler:1994pb,Finn:2000sy,Hughes:2001jr,Fujita:2009us}, is
not enough to devise sufficiently accurate predictions for the waveforms.
The knowledge of the local \textit{gravitational self-force} affecting the motion of the small
compact body is necessary. The original expression for the gravitational self-force, 
as established in Refs.~\cite{Mino:1996nk,Quinn:1996am},
was rather formal and difficult to evaluate explicitly.
Later, several practical methods for computing the self-force
have been proposed \cite{Barack:1999wf,Lousto:1999za,
Burko:2000xx,Barack:2001gx,Mino:2001mq,Anderson:2005gb,
Vega:2011wf,Pound:2013faa}
and successfully implemented in the Schwarzschild case
\cite{Barack:2007tm,Detweiler:2008ft,Sago:2008id,
Barack:2010tm,Wardell:2014kea,Osburn:2014hoa,Merlin:2014qda} 
and the Kerr case \cite{Shah:2012gu,vandeMeent:2015lxa,vandeMeent:2016pee}.
See Refs.~\cite{Barack:2009ux,Poisson:2011nh,Harte:2014wya,
Pound:2015tma,Wardell:2015kea} for recent reviews.

In the context of self-forced orbital evolution, it is convenient to split the self-force
into a \textit{dissipative} component and a \textit{conservative} component.
At linear order in the mass ratio, the former is obtained
from the half-retarded minus half-advanced metric perturbation,
which is free from the issue of the divergence of the self-field at the particle's location. 
The latter is the self-force mediated by the time-symmetric metric perturbation,
i.e., the half-retarded plus half-advanced one, regularized
by subtracting a properly chosen singular piece that diverges along the orbit \cite{Detweiler:2002mi}.
Indeed, detailed analyses reveal that the dissipative part of the first-order self-force is responsible for the average rates of change of the constants of motion, and that accuracy requirements for EMRI waveforms demand the knowledge of the conservative part of the self-force to first order in the mass ratio, as well as the dissipative part up to second order \cite{Mino:2005an,Tanaka:2005ue,Hinderer:2008dm,Isoyama:2012bx}. The impact of a resonance, when the frequencies of the radial and polar motions are in a small integer ratio \cite{Flanagan:2010cd,Flanagan:2012kg,Isoyama:2013yor}, must also be taken into account. Although every EMRI with a large eccentricity is expected to pass through at least one low-order resonance as it sweeps through the frequency band of LISA \cite{Ruangsri:2013hra,Berry:2016bit}, a resonance is not expected to be sustained for a long period of
time \cite{vandeMeent:2013sza}, so the first-order conservative self-force and
the second-order dissipative self-force will be highly relevant
to construct sufficiently accurate templates. The formalism required to compute the second-order self-force has
recently been developed \cite{Pound:2012nt,Pound:2012dk,Gralla:2012},
and its implementation is currently underway \cite{Pound:2014xva,Pound:2015wva,Miller:2016hjv}.

In this paper, we focus on the effects of the \textit{first-order conservative self-force}
on a particle's motion in a Kerr background spacetime, i.e., we consider 
the self-field mediated by the time-symmetric part of the Green function.
Beyond the long-term objective of devising accurate template waveforms
for EMRIs, the conservative self-force dynamics provides
``exact'' results in a particular limit of the general relativistic two-body problem. 
Indeed, over recent years, self-force theory has been used to compute various
quantities that characterize post-geodesic, conservative effects 
on the strong-field dynamics of compact binary systems, 
at linear order in the mass ratio, such as the ``redshift'' variable \cite{Detweiler:2008ft,Sago:2008id,Shah:2010bi,Shah:2012gu}, the frequencies of innermost stable circular orbits \cite{Barack:2009ey,Isoyama:2014mja}, the circular-orbit periastron advance \cite{Barack:2010,Barack:2011,vandeMeent:2016hel}, the geodetic spin precession frequency \cite{Dolan:2013roa,Akcay:2016dku}, as well as tidal invariants \cite{Dolan:2014pja,Nolan:2015vpa}. Such results provide accurate strong-field benchmarks for comparisons to the predictions from post-Newtonian calculations \cite{Blanchet:2013haa} and full numerical-relativity simulations \cite{Choptuik:2015mma}, and help refine semi-analytical models of inspiralling compact-object binaries over the full range of parameters, e.g. effective-one body models \cite{Buonanno:1998gg,Buonanno:2000ef}, thus stimulating synergy between gravitational self-force theory and other approaches to the dynamics of compact binaries in general relativity \cite{Blanchet:2010,Blanchet:2010b,Damour:2010,LeTiec:2011,Akcay:2012,LeTiec:2012b,Barausse:2012,LeTiec:2013b,Akcay:2015,Akcay:2016,LeTiec:2014lba}.

Since those approximation methods and numerical techniques
employ different coordinate systems to perform explicit
calculations, such comparisons crucially rely on the use 
of gauge-invariant (physical) quantities.
However, although both analytical and numerical 
self-force calculations are now performed with extremely high accuracy
\cite{vandeMeent:2015lxa,Bini:2016qtx,Kavanagh:2016idg,Bini:2016dvs,vandeMeent:2016pee},
the self-force at a given moment of time does not (in general) have any
gauge-invariant meaning \cite{Barack:2001ph,Gralla:2011zr,Pound:2015fma}. 
Therefore, one of the main challenges of this synergy is 
to devise a systematic framework allowing one
to identify gauge-invariant quantities and relationships 
that characterize the conservative self-force effects on
the generic orbital motion of a particle in Kerr spacetime.

\subsection{Summary}

In this paper, we address this challenge by formulating a Hamiltonian 
description of the conservative self-force dynamics of a particle in Kerr spacetime,
at linear order in the small mass ratio.
Our approach is based on the self-consistent formulation 
of self-forced motion, as reviewed e.g. in Ref.~\cite{Pound:2015tma}, 
together with a description of the particle's motion as geodesic
in a certain smooth effective spacetime, in terms of (generalized) 
action-angle variables \cite{Schmidt:2002qk,Hinderer:2008dm}.

One of the main goals of this paper is to identify
the gauge-invariant pieces of information that are contained
in the conservative self-force. Having clarified the role
played by the gauge freedom in the Hamiltonian dynamics, 
we identify the quantities that characterize the gauge-invariant
effects of the conservative self-force on the orbital motion.
As expected, we find that the three aforementioned constants
of motion lose their gauge-invariant meaning once the self-force
effect is taken into account. On the contrary, the long-time averaged
frequencies of the radial, azimuthal angle, 
and zenithal angle oscillations, as well as the so-called ``redshift'' 
variable, are all proven to be gauge invariant
for the conservative dynamics.

Another objective of this paper is to use the gauge freedom
to simplify the description of the conservative self-force dynamics.
Indeed, in order to account for the long-time evolution of an orbit subject to
the conservative self-force, an appropriate definition of 
the constants of motion is needed, by specifying the gauge. 
In this paper, we prove the existence of a class of ``canonical'' gauges 
in which the entire effect of the conservative self-force on the particle's 
motion is described by means of an (integrable) effective
Hamiltonian, written solely in terms of action variables 
that are conserved along the orbit.

As a first application of this effective Hamiltonian,
we derive a simple formula yielding the conservative self-force correction 
to the orbital frequencies of Kerr innermost stable spherical orbits
(ISSOs),\footnote{Spherical orbits are also known as circular (inclined) orbits 
of constant Boyer-Lindquist coordinate radius~\cite{Wilkins:1972rs,Hughes:1999bq}.}
in terms of a gauge-invariant ``redshift'' variable. 
Because the concept of an ISSO is gauge invariant, 
its frequency shift likewise has a gauge-invariant meaning. Moreover
this effective Hamiltonian allows us to establish a ``first law of mechanics''
for black-hole-particle binary systems, which is valid at linear order 
beyond the test-mass approximation, without assuming any expansion 
with respect to the spin of the background black hole 
or the velocity of the satellite particle. Interestingly,
this Hamiltonian first law is analogous to the various first laws of binary mechanics 
that were previously established in the context of arbitrary mass-ratio 
compact binaries \cite{LeTiec:2012,Blanchet:2013,LeTiec:2015,Blanchet:2017}.
This part of our analysis also provides a detailed account of 
our short report \cite{Isoyama:2014mja}, where the formula for ISSOs 
and the first law were established in the circular equatorial orbit case, 
without giving the details of the derivations.

Independently, Vines and Flanagan \cite{Vines:2015} have recently
proven that, for generic stable bound orbits in a Schwarzschild background,
the dynamics of a point mass subject to the conservative piece
of the osculating-geodesic-sourced self-force \cite{Pound:2007th}
is Hamiltonian and integrable.
The question of whether this result extends to the Kerr case was left open,
primarily due to an incomplete understanding of the gauge freedom in the
dynamical system and complications arising in the case of resonant orbits.
Earlier work on a Hamiltonian formulation of the geodesic motion of a test
particle in quasi-Kerr spacetimes and the self-forced motion of a particle 
in a Schwarzschild background can also be found 
in Refs.~\cite{Glampedakis:2005cf,Yang:2014}.

The remainder of this paper is organized as follows. 
In Sec.~\ref{sec:Hamilton},
we specify the Hamiltonian, which is the sum of a background
and an interaction Hamiltonian, in terms of generalized action-angle variables.
In Sec.~\ref{sec:NoSC}, we show that for generic orbits there is no secular growth
of the action variables under the effect of the conservative piece of the self-force.
In Sec.~\ref{sec:GT}, we discuss the gauge transformations
of various variables, and identify gauge-invariant quantities and relations.
In Sec.~\ref{sec:special}, the effective Hamiltonian is obtained
in a class of gauges that can be chosen consistently. 
In Sec.~\ref{sec:gauge-reg}, we discuss in details the consistency
and the regularity of gauge transformations 
in the equatorial and spherical orbit limits.
We then identify the ISSO condition for general inclined orbits
in Sec.~\ref{sec:ISSO-shift}, and derive a ``first law'' of
binary mechanics in Sec.~\ref{sec:1stlaw}.
Section \ref{sec:conclusion} is devoted to a summary of this work 
and a discussion of future prospects.

A number of technical topics are relegated to appendices:
the particular case of resonant orbits is discussed in App.~\ref{app:resonant},
a scaling transformation used throughout this work is related to the reparameterization
invariance of the particle's action in App.~\ref{app:scaling},
further details on our special gauge choice are given in App.~\ref{app:gauge},
and the Fourier decompositions of the interaction Hamiltonian 
and the canonical variables are presented in App.~\ref{app:Fourier}.

Throughout this paper we use geometrized units $G = c = 1$,
as well as a metric signature $(-,+,+,+)$.
Greek indices $\alpha,\beta,\mu,\nu,\dots$ denote coordinate components
in Boyer-Lindquist coordinates $(t,r,\theta,\phi)$ and
parenthesis around indices are used for the totally symmetric part of 
a given tensor. The mass of the background Kerr black hole is $M$,
that of the particle is $\mu$, and the mass ratio is
$\eta \equiv \mu/M$. The symbol $\simeq$ is used to denote an equality
that is satisfied exactly in the test-mass limit,
but only approximately so in the perturbative case.

\section{Hamiltonian mechanics of a point particle's motion in the perturbed Kerr geometry}
\label{sec:Hamilton}

We consider the motion of a point particle with mass $\mu$ 
in a Kerr background spacetime,
while taking into account the effects of the conservative part
of the gravitational self-force. 
Using standard Boyer-Lindquist coordinates $(t,r,\theta,\phi)$,
the Kerr metric of mass $M \gg \mu$ and spin $S \equiv aM$ reads 
\begin{align}\label{Kerr} 
 g_{\mu \nu}^{(0)} \, \ud x^{\mu} \ud x^{\nu} = 
 &- \left( 1 - \frac{2 M r}{\Sigma} \right) \ud t^2 
 - \frac{4 M a r \, {\sin^2 \theta} }{\Sigma} \, \ud t \, \ud \phi 
 + \frac{\Sigma}{\Delta} \, \ud r^2 \nonumber \\ &+ \Sigma \, \ud \theta^2 
 + \left( r^2 + a^2 + \frac{2 M a^2 r}{\Sigma} \sin^2 \theta \right)
 \sin^2 \theta \, \ud \phi^2 \,,
\end{align}
where
\begin{subequations}\label{Kerr-hojo}
 \begin{align}
  \Sigma &\equiv r^2 + a^2 \cos^2 \theta \,, \\
  \Delta &\equiv r^2 - 2 M r + a^2 \,. 
 \end{align}
\end{subequations}
The Kerr metric possesses two Killing vectors $\chi_{(t)}^{\mu}$
and $\chi_{(\phi)}^{\mu}$,
as well as one irreducible Killing tensor $K^{\mu \nu}$, which are given by
\begin{subequations}\label{killing-v}
 \begin{align}
  \chi_{(t)}^{\mu} &\equiv (\partial_t)^{\mu} = (1, 0, 0, 0) \,, \label{killing-t}\\
  \chi_{(\phi)}^{\mu} &\equiv (\partial_{\phi})^{\mu} = (0, 0, 0, 1) \,, \label{killing-phi} \\
  K^{\mu \nu} &\equiv 2 \Sigma \, \ell^{( \mu} n^{\nu)} + r^2 g^{\mu \nu}_{(0)} \,, \label{CKY}
 \end{align}
\end{subequations}
where $\ell^{\mu} \equiv \left( r^2 + a^2, \Delta, 0, a \right)/ \Delta$
and $n^{\mu} \equiv  \left(r^2 + a^2, -\Delta, 0, a \right)/ (2 \Sigma)$ 
are two radial null components of the Kinnersley tetrad.
In this paper, we discuss only \textit{bound} orbits, whose 
radial and polar motions are restricted in the regions 
$r_{\rm min} \leqslant r \leqslant r_{\rm max}$ and 
$\theta_{\rm min} \leqslant \theta \leqslant \pi - \theta_{\rm min}$, 
respectively. 

At first order in the mass ratio $\eta$, the motion of
the particle is accelerated with respect
to the background Kerr metric, under the effect of the gravitational
self-force. Equivalently, this motion can be described as being \textit{geodesic} in a properly regularized
effective metric given by $g_{\mu \nu}^{(0)}(x)+ h_{\mu \nu}^{(\text{R})}(x;\,\gamma)$. 
Here $\gamma$ denotes the source orbit
and $h_{\mu \nu}^{(\text{R})} = O(\eta)$ is
the (regular) R-part of the metric perturbation,
a specific vacuum solution of the linearized Einstein equation,
obtained by subtracting from the retarded metric perturbation
a singular piece that can be specified from local information about
the orbit \cite{Mino:1996nk,Detweiler:2002mi,Pound:2009sm,Harte:2011ku}.
See Refs.~\cite{Pound:2012nt,Pound:2012dk} for an extension of
the definition of $h_{\mu \nu}^{(\text{R})}$ to second order in $\eta$.

In this paper, we restrict our attention to the conservative part of the first-order self-force.
In this context, the conservative self-force refers to the force mediated by the time-symmetric part
of the metric perturbation ($=$ ``half retarded'' $+$ ``half advanced'' metric perturbation).
Given a \textit{fixed} source orbit $\gamma$, we denote the properly regularized time-symmetric
part of the metric perturbation simply by $h_{\mu \nu}(x;\gamma)$, for brevity. Thus, we consider
the geodesic motion of a particle in the effective metric
\begin{equation}
\label{h-R0}
 {g}_{\mu \nu}(x; \gamma) \equiv g_{\mu \nu}^{(0)} (x)
 + h_{\mu \nu}(x; \gamma) \,.  
\end{equation}
We note that the source orbit $\gamma$ can be
expanded in powers of the small mass ratio $\eta$ around the neighboring
``osculating'' geodesic \cite{Pound:2007th,Pound:2015tma}. 
Although this osculating geodesic deviates secularly from the true
(physical) orbit---because, for example, of the loss of energy through gravitational
radiation---, we show in Sec.~\ref{sec:NoSC} that no such secular deviation occurs in the conservative setup.
This osculating geodesic is different from the true orbit $\gamma$, but the deviation remains small everywhere, and contributes only at higher orders in the perturbation, which is beyond the scope of this paper.

To discuss the geodesic motion of the particle in this effective metric, we adopt the framework of four-dimensional Hamiltonian mechanics \cite{Carter:1968rr,Schmidt:2002qk,Hinderer:2008dm}.
The canonical Hamiltonian of a particle of mass $\mu$ following geodesic motion
in the metric \eqref{h-R0} is
\begin{equation}
\label{Hamiltonian-G0}
H(x,u;\gamma) \equiv 
\frac{1}{2} \, g^{\mu \nu}(x;\gamma) \,u_{\mu} u_{\nu} \,,
\end{equation}
where we indicated explicitly the functional dependence
on the source trajectory $\gamma$. 
Hamilton's equations for the canonical position $x^\mu$ and momentum 
$u_\mu$ then read
\begin{equation}
\label{H-eq0}
 \dot x^{\nu} = 
 \left(\frac{\partial H}{\partial u_{\nu}}\right)_{\!x} \,,
 \quad
 \dot u_{\nu} = - \left(\frac{\partial H}{\partial x^{\nu}}\right)_{\!u} \,,
\end{equation}
where the overdot stands for the derivative with respect to the proper time $\tau$,
as measured in the effective metric \eqref{h-R0}. An action principle associated to
the Hamiltonian \eqref{Hamiltonian-G0} is discussed in App.~\ref{app:scaling}.
Substituting for the Hamiltonian \eqref{Hamiltonian-G0}
into Hamilton's equations of motion \eqref{H-eq0}, the canonical momentum $u_{\mu}$
is found to be related to the four-velocity
$\dot x^\nu = \ud x^\nu / \ud \tau$ of the particle
by $g^{\mu \nu} u_{\mu} = \dot x^\nu$.
Thus, for physical orbits (on-shell),
the canonical momentum has to be normalized according to
\begin{equation}
\label{norm-G}
 g^{\mu \nu} u_{\mu} u_{\nu} = - 1 \,. 
\end{equation}
Whenever we need to specify this geodesic or $\gamma$ explicitly in terms of 
the canonical variables, we will denote it as $x^{\mu}(\tau)$ and $u_{\mu}(\tau)$, 
with the argument $\tau$ associated explicitly. 

We now expand the Hamiltonian~\eqref{Hamiltonian-G0} as 
\begin{equation}
\label{Hamiltonian-eff}
 H = H^{(0)} (x, u) + \Hint (x, u; \gamma) \,,
\end{equation}
which is the sum of the background Hamiltonian $H^{(0)}(x,u)$,
simply given by the expression \eqref{Hamiltonian-G0}
with the substitution $g^{\mu\nu} \to g_{(0)}^{\mu\nu}$,
and the interaction Hamiltonian ($\propto\eta$)
\begin{equation}
\label{Hamiltonian-int}
 \Hint (x, u; \gamma)
 \equiv 
 - \frac{1}{2} {h}^{\mu \nu} (x; \gamma) \, u_{\mu} u_{\nu} \,. 
\end{equation}
The canonical variables evaluated along the orbit are expanded 
in a similar manner,
namely $x^{\mu}(\tau) =x^{\mu}_{(0)}(\tau) + x^{\mu}_{(1)}(\tau)$ 
and $u_\mu(\tau) = u_\mu^{(0)}(\tau) + u_\mu^{(1)}(\tau)$.
Moreover, since we study first-order perturbations,
thereafter we implicitly assume that all the variables are substituted 
by their background values when the accuracy is sufficient.

To exploit the symmetries of the background Kerr spacetime, 
it is convenient to perform a canonical transformation from the canonical variables $(x^\mu,u_\mu)$ to the (generalized) action-angle variables $(w^{\alpha},J_{\alpha})$ \cite{Schmidt:2002qk,Hinderer:2008dm}.
We perform this canonical transformation in two steps. 
The first step is the transformation from $(x^{\mu}, u_{\mu})$ 
to a new set of canonical coordinates $(X^{\mu}, P_{\mu})$, such that 
the momenta $P_{\mu}$ become constants of motion 
for the geodesics in the background Kerr geometry. 
Timelike geodesics in Kerr spacetime possess 
three independent constants of motion,
besides the normalization \eqref{norm-G} of the four-velocity.
Indeed, the Killing vectors and the Killing tensor
in \eqref{killing-v} allow us to define 
the specific energy $\hat E$,  
the specific azimuthal angular momentum $\hat L_z$, and 
the specific Carter constant $\hat Q$ \cite{Carter:1968rr} as
\begin{equation}
 \hat E \equiv  -\chi^{\mu}_{(t)} u_{\mu} \,,
 \quad
 \hat L_z \equiv \chi^{\mu}_{(\phi)}  u_{\mu} \,, 
 \quad
 \hat Q\equiv  K^{\mu \nu} u_{\mu} u_{\nu} \,.
\label{COM}
\end{equation}
These are all constant for the unperturbed Kerr geodesics. 
As a simple choice of $P_\mu$, we set\footnote{Beware that
$\hat{\mu}$ is a dimensionless quantity,
different from the rest mass $\mu$ of the particle.
$\hat \mu$ is one of the canonical variables, 
while $\mu$ is just an external parameter.}
\begin{equation}\label{def-P}
 P_0 = -\frac{\hat\mu^2}{2} \equiv \frac{1}{2} g^{\mu\nu}_{(0)}u_{\mu} u_{\nu} \,,
 \quad
 P_1 = \hat E \,, 
 \quad
 P_2 = \hat L_z \,, 
 \quad
 P_3 = \hat Q \,. 
\end{equation}
The generating function $W(x, P)$
for the canonical transformation from $(x^{\mu}, u_{\mu})$
to $(X^{\mu}, P_{\mu})$ is the one obtained by 
solving the Hamilton-Jacobi equation for the timelike geodesics in 
the background Kerr spacetime,  
\begin{equation}
\label{HJ-eq}
 g^{\mu \nu}_{(0)} \, \frac{\partial W}{\partial x^{\mu}}
 \frac{\partial W}{\partial x^{\nu}} + \hat\mu^2 = 0 \,.
\end{equation}
The solution is given by~\cite{Carter:1968rr}
\begin{equation}
\label{def-W}
 W(x, P) = -P_1 \, t + P_2 \, \phi
 + \int^r \frac{\sqrt{R (r', P)}}{\Delta(r')} \, \ud r' 
 + \int^{\theta} \sqrt{\Theta (\theta', P)} \, \ud \theta' \,,
\end{equation}
with
\begin{subequations}\label{def-R-Theta}
 \begin{align}
  R(r,P) &\equiv \left\{ (r^2 + a^2) P_1 - a P_2 \right\}^2
  - \Delta \left\{  P_3 - 2P_0 r^2  \right\} , \\
  \Theta (\theta, P) &\equiv P_3 - \left( a P_1 -P_2 \right)^2 +
  \left\{ \left(2P_0 + P_1^2\right) a^2 - \frac{ P_2^2}{\sin^2 \theta}
  \right\} \cos^2 \theta \,.
 \end{align}
\end{subequations}
It should be understood that $\sqrt{R (r, P)}$ and $\sqrt{\Theta (\theta, P)}$ 
in \eqref{def-W} are positive (resp. negative) when $\dot r$ and $\dot \theta$ are 
positive (resp. negative). 
More precisely, one should think of two Riemann surfaces for $\sqrt{R(r, P)}$ 
as a function of $r$, with the branch cuts
extending from $r_{\rm min}$ to $-\infty$ and from $r_{\rm max}$ to $+\infty$. 
In its increasing phase, $r$ is on one Riemann surface,
while in its decreasing phase it evolves on the other surface. 
The same is true for $\sqrt{\Theta (\theta, P)}$. 
Then, the desired canonical transformation is specified as 
\begin{equation}
\label{canonical-trans}
 u_{\mu} = \left(\frac{\partial W} {\partial x^{\mu}}\right)_{\!P} \,,
 \quad
 X^{\mu} = \left(\frac{\partial W}{\partial P_{\mu}}\right)_{\!x} \,.
\end{equation}
More explicitly, up to sign, we have 
\begin{equation}\label{def-p0}
 u_t  = - P_1 \,, 
 \quad
 u_{\phi} = P_2 \,, 
 \quad
 u_r = \frac{\sqrt{R(r, P)}}{\Delta(r)} \,, 
 \quad
 u_{\theta} = \sqrt{{\Theta}(\theta, P)} \,,
\end{equation} 
and the new coordinates $X^{\mu}$ are expressed as
\begin{subequations}\label{def-X0}
 \begin{align}
 X^{0} &= \int^r \!\! \frac{r'^2}{\sqrt{R(r', P)}} \, \ud r'
 + \int^{\theta} \!\! \frac{a^2 \cos^2 \theta'}{\sqrt{{\Theta} (\theta', P)}}
 \, \ud \theta' \,, \\
 X^{1} &= -t + \frac{1}{2} \int^r\!\! \frac{\ud r'}{\Delta \sqrt{R(r', P)}}
 \frac{\partial R (r', P)}{\partial P_1}\,
 + \frac{1}{2}\int^{\theta}\!\! \frac{\ud \theta'}{\sqrt{{\Theta} (\theta', P)}}
 \frac{\partial \Theta (\theta', P)}{\partial P_1} \,, \\
 X^{2} &= \phi  + \frac{1}{2}\int^{r}\!\! \frac{\ud r'}{\Delta \sqrt{R(r', P)}}
 \frac{\partial R (r', P)} {\partial P_2}\, 
 + \frac{1}{2}\int^{\theta}\!\! \frac{\ud \theta'}{\sqrt{{\Theta} (\theta', P)}}
 \frac{\partial {\Theta} (\theta', P)}{\partial P_2} \,, \\ 
 X^{3} &=  - \frac{1}{2} \int^{r}\!\! \frac{\ud r'}{\sqrt{R (r', P)}} 
 + \frac{1}{2} \int^{\theta}\!\! \frac{\ud \theta'}{\sqrt{\Theta (\theta', P)}} \,,
 \end{align}
\end{subequations}
where the lower limits of integration for $r$ and $\theta$
are $r_{\rm min}$ and $\theta_{\rm min}$, respectively.
Here, $r_{\rm min}$ and $\theta_{\rm min}$ are functions of 
$P_\mu$ that give the second largest real zeros of ${R (r, P)}$
and ${\Theta (\theta, P)}$, respectively.
Since $H \simeq P_0$, Hamilton's equations of motion \eqref{H-eq0} imply
$\dot X^0\simeq 1$ and $\dot X^i\simeq 0$ for $i=1,2,3$. 
Thus, except for $X^0$, the coordinates $X^\mu$ are also constants 
of motion for background Kerr geodesics.

The next step is to transform from the phase-space coordinates
$(X^{\mu}, P_{\mu})$ to the generalized action-angle variables
$(w^{\alpha}, J_{\alpha})$, i.e., canonical coordinates
that are well adapted to integrable dynamical systems.
For Kerr geodesics, the actions
$J_\alpha = \frac{1}{2\pi} \oint u_\alpha \, \ud x^\alpha$ are
defined by~\cite{Schmidt:2002qk}
\begin{equation}\label{def-J0}
 J_t \equiv - P_1 \,, 
 \quad
 J_{\phi} \equiv P_2 \,, 
 \quad
 J_{r} \equiv 
 \frac{1}{2 \pi} \oint \frac{\sqrt{R(r, P)}}{\Delta(r)} \, \ud r \,, 
 \quad
 J_{\theta} \equiv 
 \frac{1}{2 \pi}  \oint \sqrt{\Theta (\theta, P)} \, \ud \theta \,,
\end{equation}
where $\oint$ denotes the integral over one cycle, namely
twice the integral over the allowed region of motion
where both $R(r, P)$ and $\Theta(\theta, P)$ are positive. 
By definition, the action variables are 
functions of $P_\mu$ only, i.e., 
\begin{equation}
\label{JvsP}
 J_{\alpha} = J_{\alpha}(P)\,. 
\end{equation} 
Moreover, as shown in Ref.~\cite{Schmidt:2002qk},
the relation \eqref{JvsP} can be inverted, yielding $P_{\mu} = P_{\mu}(J)$.
Therefore, we are allowed to rewrite the generating function in
Eq.~\eqref{def-W} as 
\begin{equation}
\label{def-WJ}
 {\cal W}(x, J) \equiv W(x, P(J)) \,,
\end{equation}
which generates the desired canonical transformation from 
$(x^{\mu}, u_{\mu})$ to $(w^{\alpha}, J_{\alpha})$.
The angle variables, the canonical variables conjugate to the actions, 
are defined by\footnote{As pointed out by Hinderer and Flanagan \cite{Hinderer:2008dm},
there exists a freedom to redefine the origin of the angle variables, as well as an ambiguity related 
to the choice of rotational frame. We define the angle variables $w^\alpha$ unambiguously by explicitly specifying 
the form of the generating function \eqref{def-WJ}.}
\begin{equation}
\label{def-q0}
w^{\alpha} 
= 
\left(\frac{\partial {\cal W}}{ \partial J_{\alpha}}\right)_{\!x} \,. 
\end{equation}

It is important to recognize that the generating function ${\cal W}$
can be decomposed as 
\begin{equation}
 {\cal W}(x, J)
 = t J_t+\phi J_\phi+2\pi N^r J_r+ 2\pi N^\theta J_\theta 
 +\tilde{\cal W}(r,\theta,J) \,,
\end{equation}
\newpage\noindent
where $N^r$ and $N^\theta$ are integer parts of 
the cycles of radial and azimuthal oscillations, respectively,
and the function
\begin{equation}\label{def-tildeW}
 \tilde {\cal W}(r, \theta, J)
 \equiv  \int_{r_{\rm min}}^r \!\! \frac{\sqrt{R (r', P(J))}}{\Delta(r')} \, \ud r' 
 + \int_{\theta_{\rm min}}^{\theta} \!\! \sqrt{\Theta (\theta', P(J))} \, \ud \theta'
\end{equation}
takes care only of the last incomplete cycle.
It is a quadratic-valued function of $r$ and $\theta$,
as we distinguish the increasing and decreasing phases of $r$ and $\theta$.
Using the above expression for the generating function ${\cal W}$, we obtain
\begin{equation}\label{wtransform}
 w^{\alpha} = \breve w^{\alpha} + \biggl(\frac{\partial\tilde{\cal W}}{\partial J_{\alpha}}\biggr)_{\!r,\theta} \,, 
\end{equation}
with 
\begin{equation}\label{wtransform2}
 \breve w^{t} = t \,, 
 \quad
 \breve w^\phi = \phi \,, 
 \quad
 \breve w^r = 2\pi N^r \,, 
 \quad
 \breve w^\theta = 2\pi N^\theta \,.
\end{equation}
From this, we find that $r$ and $\theta$ are both functions of 
$w^r$, $w^\theta$ and $J_{\alpha}$, and are both periodic
with respect to $w^r$ and $w^\theta$ with period $2\pi$:
\begin{subequations}\label{rth-w}
 \begin{align}
  r(w^r, w^\theta, J) &= r(w^r+2N^r\pi, w^\theta+2N^\theta\pi, J) \,, \\
  \theta(w^r, w^\theta, J) &= \theta(w^r+2N^r\pi, w^\theta+2N^\theta\pi, J) \,.
 \end{align}
\end{subequations}

In terms of generalized action-angle variables,
Hamilton's canonical equations read
\begin{equation}
 \label{H-eq-qJ0}
 \omega^{\alpha} \equiv \dot w^{\alpha}
 = \left(\frac{\partial H}{\partial J_{\alpha}}\right)_{\!w} \,, 
 \quad
 \dot J_{\alpha} = - \left(\frac{\partial H}{\partial w^{\alpha}}\right)_{\!J} \,. 
\end{equation}
For the background Kerr spacetime, the Hamiltonian is
a function of the action variables only,
$H^{(0)}=H^{(0)}(J)$, such that both $J^{(0)}_{\alpha}(\tau)$ and
$\omega_{(0)}^{\alpha}(\tau)
\equiv(\partial H^{(0)}/\partial J_\alpha)|_{J=J^{(0)}(\tau)}$
are constant.
However, we should stress that the arguments of $H^{(0)}(J)$ and
$\omega_{(0)}^{\alpha}(J)$ in Eq.~\eqref{H-eq-qJ0} with 
\eqref{Hamiltonian-eff} are phase-space coordinates; 
they are not the on-shell solution $J_{\alpha}(\tau)$ and are 
not truncated.

Before closing this section, we discuss the meaning of a shift
of the initial values of the angles $w^{\alpha}(\tau)$,
especially for background Kerr geodesics. 
First, the change of the orbit induced by a shift of the initial value
$w_{\rm I}^{t}\equiv w^{t}(\tau_{\rm I})$
or $w_{\rm I}^\phi\equiv w^\phi(\tau_{\rm I})$ of 
$w^t(\tau)$ or $w^\phi(\tau)$ can be absorbed by 
a shift of the origin of the coordinate $t$ or $\phi$, 
because of Eqs.~\eqref{wtransform2}.
Therefore, such a shift is physically irrelevant.
By contrast, a shift of $w_{\rm I}^r\equiv w^r(\tau_{\rm I})$ 
or $w_{\rm I}^\theta \equiv w^\theta(\tau_{\rm I})$
changes the phase of the radial or the zenithal angle oscillation, 
as can be seen from Eqs.~\eqref{rth-w}.
Thus, the change of the orbit induced by an infinitesimal shift 
of $w_{\rm I}^r$ or $w_{\rm I}^\theta$ 
cannot be absorbed by a corresponding infinitesimal shift of coordinates.
However, for a generic orbit, we can always find a total reflection point 
where both $u^{(0)}_r(\tau)$ and $u^{(0)}_\theta(\tau)$
vanish simultaneously to an arbitrarily high accuracy \cite{Mino:2003yg}.
Hence, when we shift $w_{\rm I}^r$ or $w_{\rm I}^\theta$,
the orbit will not be different from the original orbit globally,
if we also allow a shift of the origin of the proper time $\tau$.
The above argument implies that a shift of the initial values of
the angles $w^\alpha$ is irrelevant
when a long-time average over the orbit is performed. 
This argument does not, however, apply to 
\textit{resonant orbits}, for which the ratio of the frequencies 
$\omega^r(\tau)$ and $\omega^\theta(\tau)$ is a rational number
\cite{Mino:2005an,Tanaka:2005ue,Flanagan:2010cd,Flanagan:2012kg,
Isoyama:2013yor,Ruangsri:2013hra,vandeMeent:2013sza,Brink:2013nna,Brink:2015roa,Berry:2016bit}.
The special case of resonant orbits is discussed in App.~\ref{app:resonant}, 
but we shall not discuss such orbits in the bulk of this paper, 
since generic orbits are non-resonant.

\section{No secular change in the action variables}
\label{sec:NoSC}

In this section we show that, for generic (non-resonant) orbits, 
the action variables do not evolve secularly
when the motion is geodesic in the time-symmetric effective metric \eqref{h-R0}. 
First, we note that the proper time derivative of the actions $J_{\alpha}$
can be split into two parts,
\begin{equation}
 \dot J_{\alpha} = \langle \dot J_{\alpha} \rangle 
 + \delta \dot J_{\alpha} \,,
\end{equation}
where $\langle \dot J_{\alpha} \rangle$ and $\delta \dot J_{\alpha}$
are the time averaged and the oscillating components, respectively, 
and the long-time average of the latter, 
$\langle \delta \dot J_{\alpha}\rangle$, vanishes by definition. 
More precisely, we define the long-time average 
of a function $f(x, u)$ along the orbit parameterized by $\tau$ as 
\begin{equation}
 \langle f \rangle \equiv \lim_{T\to\infty}\frac{1}{2T}\int_{-T}^{T}
 \ud\tau\, {f}(x^{(0)}(\tau),u^{(0)}(\tau)) \,,
\end{equation}
where the quantity $f$ is assumed to be manifestly of $O(\eta)$.
Thus, the orbit in the arguments of $f$ is replaced
by the ``neighboring'' Kerr geodesic,
neglecting the higher-order corrections in the small mass ratio. 
Making use of  Eqs.~\eqref{Hamiltonian-int} and~\eqref{H-eq-qJ0}, 
the average rate of change of the actions is then given by 
\begin{align}\label{avJdot}
 \langle \dot J_{\alpha} \rangle 
 &= -\left\langle \left(
 \frac{\partial \Hint}{\partial w^{\alpha}} \right)_{\!J}
 \right\rangle \nonumber \\
 &= \frac{1}{2}\,\lim_{T\to\infty}\frac{1}{2T}\int_{-T}^{T} \ud \tau \left[ 
 \left(
 \frac{\partial}{\partial w^{\alpha}}
 \int \ud \tau' \,
 G \bigl(x, u; x^{(0)}(\tau'), u^{(0)}(\tau')\bigr)
 \right)_{\!J} \right]_{\substack{x=x^{(0)}(\tau) \\ u=u^{(0)}(\tau)}} \,,
\end{align}
where we defined
\begin{equation}\label{symmetricG}
G(x,u;x',u')
 \equiv{\mu} \,
 u_{\mu} u_{\nu}\, 
 G^{\,\mu\nu\,\rho\sigma}_{\rm (sym-S)}\left(x; x'\right)
 \,u'_\rho u'_\sigma \,.
\end{equation}
Here, $G^{\mu\nu\,\rho\sigma}_{\rm (sym-S)}(x; x')$ 
is the (regularized) symmetric part of the gravitational Green's function 
defined in terms of 
the retarded Green's function $G^{\mu\nu\,\rho\sigma}_{\rm (ret)}(x; x')$, 
the advanced Green's function $G^{\mu\nu\,\rho\sigma}_{\rm (adv)}(x; x')$, 
and a certain (self/singular) S-part of the Green's function
$G^{\mu\nu\,\rho\sigma}_{\rm (S)}(x; x')$~\cite{Detweiler:2002mi} by
\begin{equation}\label{def-G-symS}
G^{\mu\nu\,\rho\sigma}_{\rm (sym-S)}(x; x')
\equiv
\frac{1}{2}
\left\{
G^{\mu\nu\,\rho\sigma}_{\rm (ret)}(x; x') 
+ 
G^{\mu\nu\,\rho\sigma}_{\rm (adv)}(x; x')
\right\}
-
G^{\mu\nu\,\rho\sigma}_{\rm (S)}(x; x')\,.
\end{equation}
It is symmetric in its arguments and pairs of indices, 
namely, 
$G^{\mu\nu\,\rho\sigma}_{\rm (sym-S)}(x; x')
=
G^{\rho\sigma\,\mu\nu}_{\rm (sym-S)}(x'; x)$.
Throughout this paper, we assume the existence of 
$G^{\mu\nu\,\rho\sigma}_{\rm (sym-S)}(x; x')$ 
in any gauge where the corresponding time-symmetric
part of the metric perturbation $h_{\mu \nu}(x;\gamma)$ 
is well defined~\cite{Pound:2015fma}, 
and we will only use this property of 
$G^{\mu\nu\,\rho\sigma}_{\rm (sym-S)}(x; x')$ 
without requiring its explicit expression.
\footnote{The general explicit form of the Green's function~\eqref{def-G-symS} 
requires to make a gauge choice for $h_{\mu \nu}(x;\gamma)$. 
For instance, assuming that $x$ is in the normal convex neighborhood of $x'$, 
the method to explicitly construct 
$G^{\mu\nu\,\rho\sigma}_{\rm (sym-S)}(x; x')$ in the Lorenz gauge 
is described in details in Sec. 16.2 of
Ref.~\cite{Poisson:2011nh}.} 
As a result, since the source of the metric perturbation--the 
energy-momentum tensor of the point mass--is proportional 
to $u_{\mu} u_{\nu}$, the expression \eqref{symmetricG} 
before differentiation with respect to
$w^\alpha$ is symmetric under the exchange of $\tau$ and $\tau'$.  

One can easily prove that the averaged quantity \eqref{avJdot} vanishes 
as follows. 
Let us consider the double integral over the trajectory of
the time-symmetric Green's function \eqref{symmetricG}, namely
\begin{equation}
\lim_{T \to \infty}
 \int^{T}_{-T} \ud\tau 
 \int^{T}_{-T} \ud\tau' \, 
 G \bigl(x^{(0)}(\tau), u^{(0)}(\tau); x^{(0)}(\tau'), u^{(0)}(\tau') \bigr) 
 \,. 
\label{intG}
\end{equation}
In this expression, it is enough to substitute the background trajectory
$\gamma^{(0)}$ because Eq.~\eqref{intG} is already $O(\eta)$. 
As mentioned in the previous section,
a shift of the initial phases $w_{\rm I}^{\alpha}$
does not cause any physically relevant change.
Thus, the total derivative of Eq.~\eqref{intG}
with respect to the initial values $w_{\rm I}^{\alpha}$ must vanish. 
This means that 
\begin{align}\label{zero-Jdot}
 0 &=
\lim_{T \to \infty}
 \int^{T}_{-T} \ud\tau 
 \int^{T}_{-T} \ud\tau'
 \left.
 \left( \frac{\partial}{\partial w^{\alpha}} + \frac{\partial}{\partial w'^{\alpha}} \right)  G(x, u; x', u')
 \right|_{\substack{x=x^{(0)}(\tau), \, x'=x^{(0)}(\tau') \\ u=u^{(0)}(\tau), \, u'=u^{(0)}(\tau')}}
 \cr
 &=
 2 
\lim_{T \to \infty}
 \int^{T}_{-T} \ud\tau 
 \int^{T}_{-T} \ud\tau' \, 
 \left.
 \frac{\partial}{\partial w^{\alpha}} G(x, u; x', u')
 \right|_{\substack{x=x^{(0)}(\tau), \, x'=x^{(0)}(\tau') \\ u=u^{(0)}(\tau), \, u'=u^{(0)}(\tau')}} \,, 
\end{align}
where we used the symmetry property $G(x,u;x',u')=G(x',u';x,u)$
of the Green's function \eqref{symmetricG} in the second equality.
Thus, we find that $\langle \dot J_{\alpha} \rangle = 0$, i.e.,
there is no secular change in the actions 
under the effect of the conservative self-force.
We note, however, that the above argument does not apply for resonant orbits, 
which we discuss separately in App.~\ref{app:resonant}. 

While the conservative self-force setup effectively ``turns off'' the gravitational radiation, our result \eqref{zero-Jdot} is not trivial and should not be confused with that based on the usual balance argument in the context of gravitational radiation reaction \cite{Gal'tsov:1982zz}. Going back to Eqs.~\eqref{def-J0}, 
an absence of secular change in $J_t$ and $J_{\phi}$ is naturally expected 
because these variables are directly related to the Killing vectors \eqref{killing-t} and \eqref{killing-phi} 
of the background Kerr spacetime.
However, this is not the case for $J_{r}$ and $J_{\theta}$,
which involve the Killing tensor \eqref{CKY} 
through Carter's constant $P_3 = {\hat Q}$ [recall Eq.~\eqref{def-P} above]. 
Indeed, there is no known conservation law associated with 
this Killing tensor, and a ``Carter-constant balance argument''
does not work to obtain Eq.~\eqref{zero-Jdot}.

\section{Gauge transformations in the Hamiltonian formulation}
\label{sec:GT}

As is now well-established, any description of the self-forced motion can be altered 
by a gauge transformation~\cite{Barack:2001ph,Gralla:2011zr,Pound:2015fma}.
In this section, we clarify how our Hamiltonian formulation of self-forced motion 
depends inherently on one's choice of gauge, 
discussing the action of gauge transformations
of the form $\bar x^{\mu} = x^{\mu} + \xi^{\mu}(x)$
on several variables.

Under such a gauge transformation, the metric transforms as
\begin{equation}\label{dg}
 \delta_\xi g_{\mu\nu}(x)
 \equiv \bar g_{\mu\nu}(x) -g_{\mu\nu}(x)
 = -2\xi_{(\mu; \nu)} \,,
\end{equation}
where the semicolon $(;)$ denotes the covariant differentiation
associated with the background metric $g^{(0)}_{\mu\nu}$. 
Importantly, to avoid any spurious secular deviation from 
the original orbit, we restrict the generator $\xi^\mu$
to remain small everywhere in spacetime.

We also introduce a different symbol $\hat\delta_\xi$ to denote 
the shifts of quantities that are evaluated along an arbitrary trajectory 
parameterized by the proper time $\tau$.
The action of this gauge transformation on the orbit $x^{\mu}(\tau)$
is obviously given by 
\begin{equation}
 \hat\delta_\xi x^{\mu}(\tau)
 \equiv \bar x^{\mu}(\tau) - x^{\mu}(\tau)
 = \xi^{\mu} \,. 
\label{eq:hdx_x}
\end{equation}
It should be noted that we take the difference
for the same value of $\tau$ here. 
Along the orbit, the metric and the inverse metric transform as 
\begin{subequations}
 \begin{align}
  \hat\delta_\xi g_{\mu\nu}(\tau) &\equiv \bar g_{\mu\nu}(\bar x(\tau))
  - g_{\mu\nu}(x(\tau))
  = - 2 g^{(0)}_{\rho(\mu} \,\xi^{\rho}{}_{,\nu)} \,, \label{htransform} \\
  \hat\delta_\xi g^{\mu\nu}(\tau) &\equiv \bar g^{\mu\nu}(\bar x(\tau))
  - g^{\mu\nu}(x(\tau))
  = 2 g_{(0)}^{\rho(\mu} \,\xi^{\nu)}{}_{,\rho} \,.
 \end{align}
\end{subequations}
In the above, the colon $(,)$ denotes the ordinary partial derivative,
and not the covariant one, because the location at
which we evaluate the metric also shifts under this transformation.

To find the transformation of the canonical momentum
$u_{\mu}=g_{\mu\nu} \dot x^\nu$, we combine Eq.~\eqref{htransform}
with $\hat\delta_\xi \dot x^{\mu}(\tau)
= \dot {\bar x}^{\mu}(\tau) - \dot x^{\mu}(\tau) 
= \dot\xi^{\mu}(x(\tau)) = \dot{x}^{\rho} \xi^{\mu}{}_{,\rho}$ to obtain
\begin{equation}
 \hat \delta_\xi u_{\mu}(\tau) = - u_{\nu} \xi^\nu{}_{,\mu} \,.
 \label{eq:hdx_u}
\end{equation}
Since the trajectory that we consider is arbitrary, 
the above transformation for $(x^{\mu}, u_{\mu})$
can be viewed as the coordinate transformation in phase space. 
In this sense, the transformation of $u_\mu$ can be understood 
as the ordinary gauge transformation of a constant one-form field. 
Moreover, by introducing the function
\begin{equation}\label{Xi}
 \Xi(x, u) \equiv u_{\mu} \xi^{\mu}(x)
\end{equation}
of the phase-space coordinates $(x^\mu,u_\mu)$,
the above gauge transformation can also be understood
as the infinitesimal canonical transformation induced
by the generator \eqref{Xi}.
Indeed, Eqs.~\eqref{eq:hdx_x} and \eqref{eq:hdx_u} can be rewritten as
\begin{equation}\label{dx-du}
 \hat\delta_\xi x^{\mu} = \left( \frac{\partial \Xi}{\partial u_\mu} \right)_{\!x} \,,
 \quad
 \hat \delta_\xi u_{\mu} = - \left( \frac{\partial \Xi}{\partial x^\mu} \right)_{\!u} \,.
\end{equation}
Therefore, a gauge transformation in four-dimensional spacetime is equivalently realized as 
an infinitesimal canonical transformation in eight-dimensional phase space.

\subsection{Transformation of the actions $J_\alpha$}
\label{subsec:dJ}

We now examine how the action-angle variables $(w^\alpha,J_\alpha)$
are affected by the gauge transformation \eqref{dg}.
First, we focus on the action variables.
We will show that one can eliminate the oscillating component of $J_\alpha$ 
by a gauge transformation, and that the averaged part of $J_\alpha$
can also be changed freely,
except for one constraint corresponding to the value of 
$\langle P_0 \rangle$ along the orbit. 

The transformation of the action variables $J_\alpha$
is straightforwardly obtained as 
\begin{align}
 \hat\delta_\xi J_\alpha &=
 \left(\frac{\partial J_\alpha}{\partial x^{\mu}}\right)_{\!u} \hat\delta_\xi x^{\mu}
 + \left(\frac{\partial J_\alpha}{\partial u_{\mu}}\right)_{\!x} \hat\delta_\xi u_{\mu}
 \cr
 &= -\left(\frac{\partial u_\mu}{\partial w^{\alpha}}\right)_{\!J}
 \left( \frac{\partial \Xi}{\partial u_\mu} \right)_{\!x}
 - \left(\frac{\partial x^{\mu}}{\partial w^{\alpha}}\right)_{\!J}
 \left( \frac{\partial \Xi}{\partial x^\mu} \right)_{\!u}
 = - \left(\frac{\partial \Xi}{\partial w^\alpha}\right)_{\!J} \,, 
\label{eq:dJ}
\end{align}
where we used Eq.~\eqref{dx-du}. To understand how much freedom we
have in changing the values of the actions $J_\alpha$ by the gauge transformation,
it is more convenient to work with the equivalent set of variables
$P_\mu$ [recall Eq.~\eqref{JvsP}].
By making use of the identities
\begin{subequations}
 \begin{align}
  \left(\frac{\partial P_\nu}{\partial x^{\mu}}\right)_{\!P}
  &= \left(\frac{\partial P_\nu}{\partial u_{\rho}}\right)_{\!x}
  \left(\frac{\partial u_{\rho}}{\partial x^{\mu}}\right)_{\!P}
  + \left(\frac{\partial P_\nu}{\partial x^{\mu}}\right)_{\!u} = 0 \,, \\
  \left(\frac{\partial \Xi}{\partial x^{\mu}}\right)_{\!P}
  &= \left(\frac{\partial \Xi}{\partial u_{\rho}}\right)_{\!x}
  \left(\frac{\partial u_{\rho}}{\partial x^{\mu}}\right)_{\!P}
  + \left(\frac{\partial \Xi}{\partial x^{\mu}}\right)_{\!u} \,, \\
  \left(\frac{\partial u_{\rho}}{\partial x^{\mu}}\right)_{\!P}
  &= \left(\frac{\partial^2 W}{\partial x^{\mu} \partial x^{\rho}}\right)_{\!P}
  = \left(\frac{\partial u_{\mu}}{\partial x^{\rho}}\right)_{\!P} \,,
 \end{align}
\end{subequations}
the gauge transformation of the constants of motion $P_\nu$ can be written as 
\begin{equation}\label{gauge-P}
 \hat\delta_\xi P_\nu = - {\cal Q}^{\mu}_{(\nu)}
 \left(\frac{\partial \Xi}{\partial x^{\mu}}\right)_{\!P} \,,
 ~~\text{with}~~
 {\cal Q}^{\mu}_{(\nu)} \equiv
 \left(\frac{\partial P_\nu}{\partial u_{\mu}}\right)_{\!x}
 = \left( u^{\mu}, -\chi_{(t)}^{\mu}, \chi_{(\phi)}^{\mu},
 2u_{\rho} K^{\rho\mu} \right) . 
\end{equation}

First of all, by construction the matrix ${\cal Q}^{\mu}_{(\nu)}$
is non-degenerate. Indeed, at a generic point along the orbit,
one can shift the phase-space coordinates $P_\nu(x,u)$ 
by changing $u_{\mu}$, except at reflection points 
where $u_r$ or $u_\theta$ vanishes. 
Thus, as long as one can vary the four components of
$\left({\partial \Xi/\partial x^{\mu}}\right)_P$ freely, 
$P_\nu$ can also be shifted as desired. 
While ${\cal Q}^{\mu}_{(\nu)}$ becomes degenerate 
at the reflection points, the $r$ and $\theta$ components of
$\left({\partial \Xi/\partial x^{\,\mu}}\right)_P$ diverge 
at these points. Because of this, 
the degrees of freedom to shift $P_\nu$ are not reduced there.
The only constraint for $\hat{\delta}_\xi P_\nu$ 
comes from the averaged part of $P_0$ along the orbit, 
which is gauge invariant. Indeed, from Eq.~\eqref{gauge-P} the transformation of $P_0$ 
is given by $\hat\delta_\xi P_0 |_\gamma = -\dot \Xi$, which implies 
\begin{equation}\label{gauge-P0}
\hat{\delta}_\xi \langle P_0 \rangle = - \bigl\langle { \dot \Xi  } \bigr\rangle 
= - \lim_{T\to\infty}\frac{1}{2T} \int_{-T}^{T} \ud\tau\, \dot \Xi = \lim_{T\rightarrow\infty} \frac{1}{2T} \left[ {\Xi (-T) - \Xi (T)} \right] .
\end{equation}
Since the factor in square brackets on the right-hand side must remain
small---recall that we require the generator $\xi^{\mu}$ of 
the gauge transformation to remain small everywhere---, 
while $1/(2T)$ vanishes in the limit where $T\to \infty$, 
we find that $\hat\delta_\xi \langle P_0 \rangle = 0$.

In summary, except for the value of $\langle P_0 \rangle$ along the orbit, 
one can change $P_\mu,$ and thus $J_\alpha$, freely at the level of
first-order perturbations, by making use of the gauge transformations.

\subsection{Transformation of the angles $w^{\alpha}$}
\label{transformationofw}

We move on to the gauge transformation of the angle variables $w^{\alpha}$. 
Here, we will show that the oscillating part of the frequencies
$\omega^\alpha \equiv \dot w^\alpha$ 
can be eliminated by a gauge transformation.

The gauge transformation of the angle variables $w^{\alpha}$ is given by \begin{align}
 \hat\delta_\xi w^\alpha 
 &=
 \left(\frac{\partial w^\alpha}{\partial x^{\mu}}\right)_{\!u} \hat\delta_\xi x^{\mu}
 + \left(\frac{\partial w^\alpha}{\partial u_{\mu}}\right)_{\!x} \hat\delta_\xi u_{\mu} 
 \cr
 &= \left(\frac{\partial u_\mu}{\partial J_{\alpha}}\right)_{\!w}
 \left( \frac{\partial \Xi}{\partial u_\mu} \right)_{\!x}
 + \left(\frac{\partial x^{\mu}}{\partial J_{\alpha}}\right)_{\!w}
 \left( \frac{\partial \Xi}{\partial x^\mu} \right)_{\!u}
 =\left(\frac{\partial \Xi}{\partial J_\alpha}\right)_{\!w} \,, 
\label{eq:hdxwt1}
\end{align}
where we used Eqs.~\eqref{def-q0} and \eqref{dx-du}.
Alternatively, it is also instructive, using Eqs.~\eqref{wtransform}
and \eqref{wtransform2}, to write this gauge transformation in the form 
\begin{equation}
 \hat\delta_\xi w^{t} = \hat\delta_\xi t
 + \frac{\partial^2\tilde{\cal W}}{\partial J_{\alpha} \partial J_{t}}
 \, \hat\delta_\xi J_{\alpha}
 + \frac{\partial^2\tilde{\cal W}}{\partial x^\mu \partial J_{t}}
 \, \hat\delta_\xi x^\mu \,,
\label{eq:hdxwt}
\end{equation}
with similar expressions for the three other angle variables.

Now, we will establish that the oscillating part of 
$\omega^\alpha = \dot w^\alpha$ can be freely changed 
by the gauge transformation.
Let us assume that we have already used gauge degrees of freedom
to specify $J_{\alpha}$ (or equivalently $P_\mu$),
so that $\hat\delta_\xi J_{\alpha}$ is given. 
Then, from Eq.~\eqref{eq:hdxwt} 
the gauge transformation of the angles $w^\alpha$
is completely specified by $\xi^\mu = \hat\delta_\xi x^\mu$,
without differentiation. 
However, we recall that the value of $\Xi = u_\mu\xi^\mu$,
which is a projection of $\xi^\mu$, 
is already specified along the orbit,
up to an integration constant.
Thus, once $\hat\delta_\xi J_\alpha$ has been given, 
only three out of the four components of 
$\xi^\mu(\tau)$ are linearly independent. 
While this restriction seems to imply that 
it is not possible to freely change the oscillating part of all the frequencies
$\omega^\alpha$ by using the remaining gauge degrees of freedom,
this restriction is in fact compensated by a gauge-independent identity
obeyed by $\omega^\alpha$, as we now establish.

To do so, we notice that for a generic timelike geodesic
in the effective metric \eqref{h-R0}, the following normalization must hold: 
\begin{equation}
 -1 = u_\mu \dot x^\mu
 = u_\mu \left(\frac{\partial x^\mu}{\partial J_\alpha}\right)_{\!w} \dot J_\alpha
 + u_\mu \left(\frac{\partial x^\mu}{\partial w^\alpha}\right)_{\!J} \dot w^\alpha
 = -u_\mu \left(\frac{\partial w^\alpha}{\partial u_\mu}\right)_{\!x} \dot J_\alpha
 + u_\mu \left(\frac{\partial J_\alpha}{\partial u_\mu}\right)_{\!x} \dot w^\alpha \,.
\label{normalizationcondition}
\end{equation}
To simplify the right-hand side of this expression,
we consider the particular scaling transformation $(x^\mu,u_\mu) \to
(x^\mu, \lambda u_\mu)$ of the canonical variables, first introduced 
in Ref.~\cite{Hinderer:2008dm}.\footnote{As shown in App.~\ref{app:scaling}, 
this scaling transformation derives from the standard reparameterization invariance of 
the affinely-parameterized, four-dimensional particle action 
associated to the Hamiltonian \eqref{Hamiltonian-G0}.} 
Then, from the definitions \eqref{def-J0}, \eqref{def-WJ} and \eqref{def-q0}
of the actions $J_\alpha$, the generating function ${\cal W}$
and the angles $w^\alpha$, we find that those quantities scale as
\begin{subequations}
 \begin{align}
  J_\alpha(\lambda) &= \lambda \, J_\alpha(1) \,, \label{scaledJ} \\
  {\cal W}(\lambda) &= \lambda \, {\cal W}(1) \,, \\
  w^\alpha(\lambda) &= \lambda^0 w^\alpha(1) \,. \label{scaledw}
 \end{align}
\end{subequations}
Similarly, under that scaling transformation, we note that 
$H^{(0)}$ scales as $H^{(0)}(\lambda)= \lambda^2 H^{(0)}(1)$
and hence that $\omega_{(0)}^\alpha$ scales as
$\omega_{(0)}^\alpha(\lambda)=\lambda\,\omega_{(0)}^\alpha(1)$. 
From $\dot w^\alpha = \omega^\alpha$, this implies that 
the affine parameter $\tau$ should scale as $\tau(\lambda)=\tau(1)/\lambda$,
and thus that $\hat\mu(\lambda)=\lambda \, \hat \mu(1)$.
In particular, the scaling relations \eqref{scaledJ}
and \eqref{scaledw} imply the identities
\begin{subequations}
 \begin{align}
  u_\mu \left(\frac{\partial J_\alpha}{\partial u_\mu}\right)_{\!x}
  &= \left. \frac{\ud J_\alpha}{\ud\lambda} \right|_{\lambda=1} = J_\alpha \,, \\
  u_\mu \left(\frac{\partial w^\alpha}{\partial u_\mu}\right)_{\!x}
  &= \left. \frac{\ud w^\alpha}{\ud\lambda} \right|_{\lambda=1} = 0 \,,
 \end{align}
\end{subequations}
which are particular cases of Euler's theorem for homogeneous functions.
Substituting those identities into the normalization condition \eqref{normalizationcondition},
we obtain the following key constraint on the frequencies $\omega^\alpha$:
\begin{equation}
 -1= \dot w^\alpha J_\alpha = \omega^\alpha J_\alpha \,. 
\label{Jomeganorm}
\end{equation}
This result was previously established in the test-particle limit
in Ref.~\cite{LeTiec:2013}. Since the origin of this constraint can
be traced back to Eq.~\eqref{normalizationcondition}, 
thereafter we will refer to Eq.~\eqref{Jomeganorm} as 
the \textit{normalization condition} for the action-angle variables.

Because the relation \eqref{Jomeganorm} holds independently of the gauge choice,
it is clear that $\omega^\alpha J_\alpha$ is the gauge-invariant linear combination
of $\omega^\alpha$ for fixed $J_\alpha$, and the constraint on the gauge transformation 
mentioned above is simply $J_\alpha \, \hat\delta_\xi \omega^\alpha=0$, thus allowing
us to express one of the changes $\hat\delta_\xi \omega^\alpha$ as a function of
the three others. As a result, one can choose a convenient gauge in which all the
frequencies $\omega^\alpha$ are constant, eliminating the oscillating components.

\subsection{Gauge invariance of $\langle \omega^{\alpha} \rangle$ 
and $\langle \Hint \rangle$}
\label{subsec:gauge-inv}

Next, we establish that the averaged piece of each of the four frequencies,
$\langle \omega^\alpha \rangle$, is gauge invariant.
To do so, we will not need the explicit form of 
the gauge transformation of $\omega^{\alpha} = \dot{w}^\alpha$
itself. Instead, we only need to notice that,
as long as the gauge vector $\xi^\mu(\tau)$ 
remains small along the orbit, the gauge transformation of 
$w^{t}$ in \eqref{eq:hdxwt} remains small as well. 
This follows from the definition \eqref{def-tildeW}
of $\tilde {\cal W}$, which prevents any secular growth,
because it takes care only of the last incomplete cycle of 
the radial and polar motion.
The same argument holds as well for the other angle variables.

Still, at first glance the right-hand side of Eq.~\eqref{eq:hdxwt1}
looks divergent at the turning points of the orbit,
where $u_r(\tau)=0$ or $u_\theta(\tau)=0$, 
and thus $\hat\delta_\xi w^{\alpha}$ appears ill-defined there. 
We can show that there is no divergence at the turning points as follows. 
First, we should recall that in the expression \eqref{eq:hdxwt1}
for $\hat\delta_\xi w^{\alpha}$,
the partial derivative is taken at fixed $w^{\beta}$. 
At the turning points,
$u_r=0$ or $u_\theta=0$ holds irrespective of 
the values of $J_\alpha$ there. 
Therefore, the derivatives of $u_r$ or $u_\theta$ with respect to $J_\alpha$,
that are contained in the definition of $\Xi$, simply vanish.
Thus, there is no divergence in $(\partial\Xi/\partial J_\alpha)_w$. 

As a consequence of the above argument,
we can prove that the long-time averages of the frequencies
are gauge invariant. Indeed, the gauge transformation of the long-time average
$\langle\omega^{\alpha}\rangle$ is evaluated as 
\begin{equation}
 \hat\delta_\xi \langle \omega^{\alpha} \rangle
 = \langle \hat\delta_\xi \omega^{\alpha} \rangle 
 = \lim_{T\to\infty}\frac{1}{2T} \left[
 \hat\delta_\xi w^{\alpha}(T)-\hat\delta_\xi w^{\alpha}(-T)\right] 
 =0\,. 
\end{equation}

Finally, we discuss the transformation of the interaction Hamiltonian,
$\Hint(x,u;\gamma)$. Since the Hamiltonian $H(x,u;\gamma)$ is a scalar,
it is invariant under the transformation induced
by Eqs.~\eqref{eq:hdx_x}--\eqref{eq:hdx_u}:
\begin{equation}\label{dH}
 \hat{\delta}_\xi H(x(\tau), u(\tau); \gamma) \equiv
 \bar{H}(\bar{x}(\tau), \bar{u}(\tau); \gamma)
 - H(x(\tau), u(\tau); \gamma)
 = 0 \,.
\end{equation}
Remembering the perturbative split \eqref{Hamiltonian-eff} of the Hamiltonian,
the above condition leads to the following transformation for $\Hint$:
\begin{align}
 \hat\delta_\xi \Hint
 &\equiv
 \bar H^{(1)}(\bar{x}(\tau), \bar{u}(\tau); \gamma) - \Hint
 (x(\tau), u(\tau); \gamma)
 = - [H^{(0)}(\bar{x}(\tau), \bar{u}(\tau)) - H^{(0)}(x(\tau), u(\tau))]
 \nonumber \\
 &= - \left( \frac{\partial H^{(0)}}{\partial x^\mu }\right)_{\!u}
 \hat{\delta}_\xi x^\mu
 - \left( \frac{\partial H^{(0)}}{\partial u_\mu }\right)_{\!x}
 \hat{\delta}_\xi u_\mu
 = \dot{u}_\mu \left( \frac{\partial \Xi}{\partial u_\mu }\right)_{\!x} + \dot{x}^\mu \left( \frac{\partial \Xi}{\partial x^\mu }\right)_{\!u}
 = \dot\Xi \,,
\end{align}
where we used Hamilton's equations \eqref{H-eq0} together
with the formulas \eqref{dx-du}.
Hence, the gauge transformation of the long-time average
of the interaction Hamiltonian is obtained as
\begin{equation}
 \hat\delta_\xi  \langle \Hint \rangle = \langle \hat\delta_\xi \Hint \rangle
 = \lim_{T\to\infty}\frac{1}{2T}\left[\Xi(T)-\Xi(-T)\right]
 =0 \,,
\end{equation}
because $\Xi = u_\mu \xi^\mu$ must remain bounded.
Therefore, just like the averaged frequencies,
the averaged interaction Hamiltonian is gauge invariant.

\subsection{Meaning of $\langle\omega^\alpha\rangle$ 
and role of the conservative self-force}

We proved that the oscillating part of the actions $J_{\alpha}$ and
frequencies $\omega^{\alpha}$ can be eliminated by 
a gauge transformation, while the averaged frequencies and
the averaged rates of change of the actions are gauge
invariant. To illustrate the significance of
this last result, consider the long-time averaged part of 
Hamilton's equations \eqref{H-eq-qJ0}, namely 
\begin{equation}
 \label{H-eq-qJ-ave}
 \left \langle \omega^{\alpha} \right \rangle
 = 
 \langle \omega_{(0)}^{\alpha}(J)\rangle
 + \left\langle\left(\frac{\partial \Hint}
 {\partial J_{\alpha}}\right)_{\!w}\right\rangle , 
 \quad
\langle \dot J_{\alpha} \rangle = 0\,. 
\end{equation}
In particular, the first equation shows that the effect of 
the interaction Hamiltonian $H^{(1)}$---or the conservative self-force---in
 Hamilton's equations is to shift the values of the long-time averaged frequencies, 
for given values of the actions.

\subsubsection{Physical interpretation of the averaged frequencies 
$\langle\omega^{\alpha}\rangle$}

We proved that the long-time averaged frequencies $\langle\omega^{\alpha}\rangle$
are gauge invariant. In this subsection, we present an argument that allows us to relate
these frequencies to the observable frequencies of orbits that are encoded, for instance,
in the gravitational waveform.

Our claim is that the angle variables $w^i$ (with $i = r,\theta,\phi$)
have a definite physical meaning as the phases of the orbital motion,
even in the perturbed spacetime. Here the orbit has, of course, the usual
gauge ambiguity of $O(\eta M)$. Conversely, one can state that the orbital motion
in Boyer-Lindquist coordinates has a definite physical meaning if we neglect
the error $O(\eta M)$ in the position. Consider for instance the radial motion.
Given this gauge ambiguity, it is natural to extend
the concepts of periastron $r_{\rm min}$ and apastron $r_{\rm max}$,
which are zeros of the function $R(r,P(J))$ in Eq.~\eqref{def-R-Theta}, 
as functions of the actions $J_\alpha$,
to their neighborhoods whose size is larger than $O(\eta M)$, 
but much less than $O(M)$.
Even if the actions $J_\alpha$ fluctuate along the trajectory 
at relative $O(\eta)$ due to the gauge ambiguity, 
the neighborhoods of $r_{\rm min}$ and $r_{\rm max}$ do not vary much, 
and this fluctuation can always be absorbed into the extended periastron and apastron.
Thus, we can define the period of radial oscillations as the proper time
elapsed before returning to the neighborhood of $r_{\rm min}$,
after having left that of $r_{\rm min}$ and going through that of $r_{\rm max}$.
This period, as measured in the proper time along the orbit,
is \textit{on average} nothing but $2\pi/\omega^r$. 
The frequencies $\omega^\theta$ and $\omega^\phi$ can be discussed
along similar lines. On the other hand, it is clear that
$\langle \omega^t \rangle=\langle \ud t/ \ud \tau \rangle$ 
gives the ratio between the proper time for the asymptotic observer,
$t$, and that along the orbit, $\tau$.

Then, making use of the gauge-invariant part of the frequencies,
namely $\langle \omega^{\alpha} \rangle$,
we define the frequencies of radial, zenithal angle
and azimuthal angle oscillations for the motion in the perturbed spacetime as
\begin{equation}\label{Omega-i}
 \Omega^i \equiv 
 \frac{\langle \omega^i \rangle}{\langle \omega^{t} \rangle} \,, 
 \quad (i=r,\theta,\phi) \,. 
\end{equation}
Importantly, the definition of these orbital frequencies
is tied to the properties of the angles \eqref{def-q0},
and as such does not depend on 
one's choice of gauge for the metric perturbation.

\subsubsection{Effects of the conservative self-force}

Next, we discuss the information that is contained in
the conservative part of the self-force itself, without specifying the
gauge for the metric perturbation. In particular, we clarify
the meaning of the frequency change $(\partial H^{(1)}/\partial J_{\alpha})_w$
in Eqs.~\eqref{H-eq-qJ0} and \eqref{H-eq-qJ-ave}.

The fact that the actions $J_\alpha$ can be changed freely
by a gauge transformation [modulo the constraint \eqref{Jomeganorm}], 
as we have established in this section, might be rather disappointing. 
As we have seen, the gauge-invariant effects of the conservative self-force
in Hamilton's equations appear only in the shift of the averaged frequencies:
the first equation in Eq.~\eqref{H-eq-qJ-ave} provides the relation between
$J_\alpha$ and $\langle \omega^\alpha \rangle$. However, we proved that the actions
$J_\alpha$ do not have any gauge-invariant meaning. In fact, under a gauge transformation
the right-hand side of the first equation in \eqref{H-eq-qJ0} transforms as 
\begin{equation}
 \left(\frac{\partial H}{\partial J_{\alpha}}\right)_{\!w}
 =\omega_{(0)}^{\alpha}(J)
 + \left(\frac{\partial \Hint}{\partial J_{\alpha}}\right)_{\!w}
 \to \omega_{(0)}^{\alpha}(\bar J)
 + \left(\frac{\partial \bar H^{(1)}}{\partial \bar J_{\alpha}}\right)_{\!w} .
\end{equation}
On the other hand, we proved that the long-time average
of the left-hand side of this equation is gauge invariant.
Thus, we have 
\begin{equation}
 \langle \omega_{(0)}^{\alpha}(J)\rangle
 + \left\langle\left(\frac{\partial \Hint}
 {\partial J_{\alpha}}\right)_{\!w}\right\rangle
 = \langle\omega_{(0)}^{\alpha}(\bar J)\rangle
 + \left\langle\left(\frac{\partial \bar H^{(1)}}
 {\partial \bar J_{\alpha}}\right)_{\!w}\right\rangle , 
\end{equation}
which implies
\begin{equation}\label{pouet}
 \left\langle \left(\frac{\partial \bar H^{(1)}}
 {\partial \bar J_{\alpha}}\right)_{\!w} \right\rangle
 = \left\langle\left(\frac{\partial \Hint}
 {\partial J_{\alpha}}\right)_{\!w} \right\rangle
 - M^{\alpha \beta}(J^{(0)}) \, \langle \hat \delta_\xi J_\beta \rangle \,, 
\end{equation}
where $M^{\alpha\beta}$ is a symmetric matrix---known as the \textit{dynamical
matrix} in the theory of dynamical systems---defined by 
\begin{equation}
 M^{\alpha\beta}(J) \equiv 
 \frac{\partial \omega^{\alpha}_{(0)}}{\partial J_\beta}
 =
 \frac{\partial^2 \Hzero}{\partial J_{\alpha} \partial J_\beta} \,.
\label{defM}
\end{equation}
Equation \eqref{pouet} has four components,
but one of them should be equivalent
to the gauge transformation of the constraint \eqref{Jomeganorm}. 
As a result the shifts of frequencies
$(\partial H^{(1)}/\partial J_{\alpha})_w$ 
as functions of $J_\alpha$ will not have any physical meaning, 
even in the sense of a long-time average,
unless we specify the gauge for the metric perturbation 
so as not to allow an arbitrary shift
of the averaged actions $\langle J_\alpha \rangle$. 

\subsubsection{Gauge-invariant relationship}

Although we cannot use the action variables $J_\alpha$
to discuss a gauge-invariant relationship---except for the normalisation condition
\eqref{Jomeganorm}---still, for general orbits there is a non-trivial gauge-invariant relation. 
This is because an orbit can be specified, for instance, by the three spatial actions $J_{i}$
(with $i=r, \theta, \phi$), with $J_t$ given as a function of the
$J_i$ by Eq.~\eqref{Jomeganorm},
while there are four gauge-invariant frequencies
$\langle \omega^{\alpha} \rangle$. 
The resulting gauge-invariant relationship 
is the averaged redshift variable,\footnote{This gauge-invariant relationship was first
introduced in Ref.~\cite{Detweiler:2008ft}, in the context of circular orbits
in a Schwarzschild background. It has since then been computed for
generic (bound) orbits \cite{Barack:2011,Akcay:2015},
and for equatorial orbits in a Kerr background \cite{vandeMeent:2015lxa}.}
\begin{equation}\label{z}
 z \equiv \left\langle\frac{\ud t}{\ud \tau}\right\rangle^{-1}
 = \left\langle\frac{\ud w^{t}}{\ud \tau}\right\rangle^{-1}
 = \frac{1}{\langle \omega^t \rangle} \,, 
\end{equation}
expressed as a function of the three frequencies $\Omega^i$ 
in Eq.~\eqref{Omega-i}. 
Therefore, unless we specify gauge conditions for the metric perturbation
itself, our Hamiltonian analysis suggests that there should be no
additional (independent) gauge-invariant relationship
for generic orbits of a structureless point particle in a Kerr background.
This agrees with the conclusion reached in Ref.~\cite{Bini:2014ica}, 
in the particular case of circular orbits in a Schwarzschild background.
Further non-trivial gauge-invariant relationships would appear
only when considering extra degrees of freedom for the point particle, 
such as tidal invariants as functions
of the frequencies $\Omega^i$ \cite{Bini:2014ica,Dolan:2014pja,Bini:2015kja,Nolan:2015vpa}.
Our observation also suggests that, for given frequencies $\Omega^i$, the redshift \eqref{z}
should be related to the gauge-invariant, long-term averaged interaction
Hamiltonian $\langle H^{(1)} \rangle$. We will establish such a relation in 
Sec.~\ref{subsec:z-Hint} below.

\subsubsection{Equatorial orbits and spherical orbits}

To conclude this section, we comment on two special classes of orbits,
namely \textit{equatorial} orbits, defined by the condition $J_\theta = 0$, 
and \textit{spherical} (inclined or circular) orbits, defined by the condition $J_r = 0$.
A generic orbit is characterized by three independent frequencies $\Omega^i$,
and the metric perturbation that is sourced by 
such an orbit oscillates with those frequencies.  
For equatorial or spherical orbits, however, one of these frequencies becomes irrelevant.
For instance, an equatorial orbit does not oscillate in the $\theta$ direction, 
and the harmonics of the polar frequency $\Omega^\theta$ cannot 
be excited in the metric perturbation, 
unless the corresponding perturbations are added by hand. 
However $\Omega^\theta \neq 0$ in general, even for an equatorial orbit.

Interestingly, when discussing equatorial orbits, the reflection symmetry across the equatorial
plane implies that the condition $J_\theta=0$ 
has a gauge-invariant meaning. Similarly, the condition $J_r=0$ becomes gauge-invariant  
for spherical orbits, as shown in Sec.~\ref{subsec:Jr} below.
For these special orbits, we can establish as many additional
gauge-invariant relations as we have symmetries of the orbit.
However, since $J_\theta=0$ and $J_r=0$ are boundaries 
of the phase space, $J_\theta$ and $J_r$ become singular as phase-space coordinates
in the equatorial and spherical orbit limits. In these limits, the action of gauge transformations 
is subtle and requires a separate analysis. This issue will be addressed in details in Sec.~\ref{sec:gauge-reg}.

\section{An effective Hamiltonian in a class of canonical gauges}
\label{sec:special}

So far we have assumed that the source orbit $\gamma$ 
in the Hamiltonian \eqref{Hamiltonian-G0} is held \textit{fixed}, 
and we take the physical orbit and the source orbit to coincide 
only \textit{after} solving for Hamilton's equations. 
While such a scheme can describe 
the conservative self-force dynamics in a Kerr background, 
strictly speaking, it is not a genuine Hamiltonian formulation 
in the textbook sense, because of the explicit source-orbit dependence 
of the Hamiltonian.

In this section, we show that by making use of the time-symmetric
Green's function \eqref{symmetricG} and the gauge freedom discussed in
Sec.~\ref{sec:GT}, the dynamics encoded in the (source-dependent)
Hamiltonian \eqref{Hamiltonian-G0} is equivalent to that
deriving from the \textit{effective Hamiltonian}
\begin{equation}\label{eq:cal_H}
 {\cal H}(J)\equiv\Hzero(J)+\frac{1}{2} \, \HI(J) \, ,
\end{equation} 
if and only if we impose the canonical gauge 
conditions \eqref{gaugeconditions0} and \eqref{gaugecondition} below. 
Here, $\HI(J)$ is the gauge-invariant interaction Hamiltonian defined 
in Eq.~\eqref{HIdef} below, in which the degrees of freedom of 
the source orbit are identified with those of the physical orbit.
Crucially, this identification is done at the level of the Hamiltonian,
\textit{before} solving Hamilton's equations.

Using the effective Hamiltonian \eqref{eq:cal_H}, Hamilton's canonical
equations of motion then take the remarkably
simple form
\begin{equation}
\label{H-eq-qJ}
 \dot w^{\alpha} = \omega^{\alpha}
 = \omega^{\alpha}_{(0)}(J)
 + \frac{1}{2} \frac{\partial \HI(J)}{\partial J_{\alpha}} \,,
 \quad \dot J_{\alpha} =0 \,,
\end{equation}
irrespective of one's choice for the source orbit $\gamma$. 
By construction, the effective Hamiltonian \eqref{eq:cal_H} reproduces the functional
relationship between the frequencies $\omega^\alpha$ and the actions $J_\alpha$, for
all values of $J_\alpha$ for physical orbits obeying the constraint \eqref{Jomeganorm},
in the class of canonical gauges specified by the conditions
\eqref{gaugeconditions0} and \eqref{gaugecondition}. In that sense, the Hamiltonian
dynamics generated by the effective Hamiltonian \eqref{eq:cal_H} is equivalent to that
deriving from the original Hamiltonian \eqref{Hamiltonian-G0}.

Importantly, we note that the original Hamiltonian \eqref{Hamiltonian-eff} cannot be
reduced to the effective Hamiltonian \eqref{eq:cal_H} by means of a (infinitesimal)
canonical transformation induced by a gauge transformation. Indeed, the normalization
$H = - \tfrac{1}{2}$ of the original Hamiltonian \eqref{Hamiltonian-G0} implies
$H^{(0)} + H_\text{int} = - \tfrac{1}{2}$ on-shell, such that for physical orbits we have
\begin{equation}\label{calH-onshell}
{\cal H} = - \frac{1}{2} - \frac{H_\text{int}}{2} \, .
\end{equation}
Because a (proper-time independent) canonical transformation leaves the on-shell value
of a Hamiltonian unchanged, we conclude that the effective Hamiltonian \eqref{eq:cal_H} cannot be
related to the original Hamiltonian \eqref{Hamiltonian-G0} through such a canonical transformation.

We also note that our effective Hamiltonian \eqref{eq:cal_H} differs from that
defined in the theory of canonical perturbations, which does not assume a
functional dependence on the source orbit as in our original
Hamiltonian \eqref{Hamiltonian-G0}, and cannot alter the on-shell value of that Hamiltonian.

The objective of the remainder of this section is to prove that the gauge
conditions \eqref{gaugeconditions0} and \eqref{gaugecondition} can indeed be imposed
in a consistent manner, yielding the effective Hamiltonian \eqref{eq:cal_H},
for generic orbits in the effective spacetime, 
except in the particular cases of spherical and equatorial orbits;
the gauge transformations in these special limits are delicate, 
and we relegate their discussion to Sec.~\ref{sec:gauge-reg}. 
Appendix \ref{app:gauge} explains why it is not as obvious as it might naively appear.

\subsection{Canonical gauge conditions}

In the previous section, we have shown that it is always possible
to eliminate the oscillating part of $\omega^{\alpha} = {\dot w}^\alpha$
and $J_\alpha$ by a gauge choice, such that
\begin{equation}
 \dot \omega^{\alpha} = 0 \,, \quad \dot J_{\alpha} = 0 \,.
 \label{gaugeconditions0}
\end{equation}
From now on, we adopt Eq.~\eqref{gaugeconditions0} as a part of
our gauge conditions, and remove the symbol for the long-time average in 
$\langle \omega^\alpha \rangle$ and $\langle J_\alpha \rangle$.

Moreover, we have shown that we can arbitrarily change 
the averaged part of each $J_\alpha$ by a residual gauge transformation,
as long as the normalization condition \eqref{Jomeganorm} is satisfied.
Thereafter, we shall impose an additional gauge condition on the constant actions $J_\alpha$,
which will involve the gauge-invariant \textit{interaction Hamiltonian} 
\begin{equation}
 \HI(J) 
 \equiv 
 \hat\mu \, \langle \Hint \rangle
 = - \frac{\hat\mu}{2} \, \lim_{T\to\infty}\frac{1}{2T}\int_{-T}^{T} \ud\tau
 \int \ud\tau' \, G\left(x^{(0)}(\tau),u^{(0)}(\tau);
 x^{(0)}(\tau'), u^{(0)}(\tau')\right) ,
\label{HIdef}
\end{equation}
constructed from the gauge-invariant, long-time averaged 
part of the perturbed Hamiltonian, $\langle \Hint \rangle$. 
Here, the degrees of freedoms $(x^{(0)}(\tau),u^{(0)}(\tau))$ of the physical orbit 
are identified with those of the source orbit, $(x^{(0)}(\tau'), u^{(0)}(\tau'))$, 
so that the expression \eqref{HIdef} is symmetric under the exchange of $\tau$ and $\tau'$. 
The canonical variable $\hat\mu = \sqrt{ -2P_0}$, such that $\hat\mu \simeq 1$ 
for physical orbits [recall Eq.~\eqref{def-P}], is inserted 
in \eqref{HIdef} to promote $\langle \Hint \rangle$, 
which is constrained by \eqref{Jomeganorm},  
to a genuine (interaction) Hamiltonian that 
scales like $\lambda^4$ under the scaling transformation introduced 
in Sec.~\ref{transformationofw}. We will clarify the naturalness of 
this choice of scaling in Sec.~\ref{subsec:Jr}.

As a convenient gauge choice, we impose the following additional condition 
for the averaged part of $J_\alpha$: 
\begin{equation}
 \Delta \langle \omega^{\alpha}\rangle 
 \equiv \frac{1}{2}\frac{\partial  \HI}{\partial J_{\alpha}}
 - \left\langle \left(\frac{\partial \Hint}
 {\partial J_\alpha}\right)_{\!w} \right\rangle =0 \,.
\label{gaugecondition}
\end{equation}
Since the expression \eqref{HIdef} for $\HI$
is symmetric under the exchange of the source orbit and the field-point orbit,
the partial derivative $\partial H_\text{int} / \partial J_\alpha$
is multiplied by a factor $1/2$ to account for the fact
that the derivative acts not only on the field point,
but also on the source orbit.
Note that we could remove the long-time average in
Eq.~\eqref{gaugecondition}, 
because the gauge conditions \eqref{gaugeconditions0} imply that
$({\partial \Hint}/{\partial J_\alpha})_w = \omega^\alpha - \omega_{(0)}^\alpha$ is constant.

Since both $\langle\omega^{\alpha}\rangle$ and $\HI$ are gauge invariant,
we can immediately argue that the gauge transformation of
$\Delta \langle\omega^{\alpha}\rangle$ is given by 
\begin{equation}
 \hat\delta_\xi \Delta\langle\omega^{\alpha} \rangle 
 = - \hat\delta_\xi \left\langle\left(\frac{\partial \Hint}
 {\partial J_\alpha}\right)_{\!w} \right\rangle
 = M^{\alpha \beta}\, \hat \delta_\xi J_\beta \,, 
\label{gaugeDelta}
\end{equation}
where we used Eq.~\eqref{pouet}.

Despite the fact that
the constant shifts $\hat \delta_\xi J_\beta$
of the actions can be chosen rather freely, 
the gauge transformation \eqref{gaugeDelta} does not allow
for arbitrary shifts of $\Delta\langle\omega^{\alpha}\rangle$ 
because of the normalization condition \eqref{Jomeganorm}. 
In particular, it is not obvious that the gauge condition 
\eqref{gaugecondition} can be enforced. 
There are two potential issues that need to be addressed:
first, the constant part of the gauge transformation is restricted
to satisfy $\hat\delta_\xi \langle P_0 \rangle = 0$, 
which is equivalent to Eq.~\eqref{Jomeganorm}
under the conditions \eqref{gaugeconditions0},
and second, when $\det M=0$, which will turn out to occur
over some region in the parameter space of $J_\alpha$,  
$\Delta\langle\omega^{\alpha}\rangle$ cannot be varied
by a gauge transformation, even if we can change $\hat \delta_\xi J_\beta$ freely.
We now address these two issues in turn.

\subsection{Constraint on $\Delta\langle\omega^\alpha\rangle$}

First, we discuss the restriction on the gauge condition 
~\eqref{gaugecondition} due to the normalization condition~\eqref{Jomeganorm}.
More explicitly, since $\omega^{\alpha}$ is constant
by our gauge choice and $\langle \omega^{\alpha} \rangle$ is gauge invariant,
$\hat\delta_\xi \omega^{\alpha} 
= \hat\delta_\xi \langle \omega^{\alpha} \rangle = 0$,
Eq.~\eqref{Jomeganorm} implies the constraint
\begin{equation}
 \omega^{\alpha} \,\hat\delta_\xi J_\alpha = 0 \,.
\label{gaugerestriction0}
\end{equation}
On the other hand, thanks to the scaling relations established
in Sec.~\ref{transformationofw}, one can easily show that
\begin{equation}
 \omega^\alpha_{(0)}
 =\left.\frac{\ud\omega_{(0)}^\alpha}{\ud\lambda}\right\vert_{\lambda=1}
 =\left(\frac{\partial \omega_{(0)}^\alpha}{\partial J_\beta} \right)_{\!w} J_\beta
 = M^{\alpha\beta}J_\beta \,,
\label{scalingwJ}
\end{equation}
where we used the definition~\eqref{defM} in the last equality. 
Combining this result with Eq.~\eqref{gaugeDelta},
the restriction \eqref{gaugerestriction0} on the gauge transformation
can be rewritten as
\begin{equation}
 J_\alpha \,\hat\delta_\xi \Delta \langle\omega^{\alpha}\rangle = 0 \,.\label{gaugerestriction}
\end{equation}

We now establish that the gauge condition \eqref{gaugecondition} 
can be imposed, despite the fact that the gauge transformation of
$\Delta \langle\omega^{\alpha}\rangle$ is restricted
by Eq.~\eqref{gaugerestriction}.
To do so, we use the first of Hamilton's equations \eqref{H-eq-qJ0}
to rewrite $J_\alpha \Delta\langle \omega^{\alpha}\rangle $ 
with the definition \eqref{gaugecondition} as
\begin{equation}
 J_\alpha \, \Delta\langle \omega^{\alpha}\rangle 
 = J_\alpha \, \frac{\partial}{\partial J_\alpha} 
 \biggl( \Hzero+ \frac{1}{2}\HI \biggr)
 - J_\alpha \, \omega^\alpha \,,
\label{gaugecondition2}
\end{equation}
where we used the first part of the gauge conditions \eqref{gaugeconditions0}.
As for the first term in the right-hand side of \eqref{gaugecondition2},
since $\Hzero$ and $\HI$ scale as $\Hzero(\lambda)= \lambda^2 \Hzero (1)$
and $\HI(\lambda)= \lambda^4 H_{\rm int}(1)$
under the scaling transformation considered in Sec.~\ref{transformationofw},
we find
\begin{equation}\label{plop}
 J_\alpha \, \frac{\partial}{\partial J_\alpha} 
 \biggl( \Hzero+ \frac{1}{2}\HI \biggr)
 = \left. \frac{\ud}{\ud\lambda} 
 \biggl( \Hzero+ \frac{1}{2}\HI \biggr)
 \right|_{\lambda=1}
 = 2\Hzero + 2\HI = -1
\end{equation}
when evaluated on-shell. 
On the other hand, the normalization condition \eqref{Jomeganorm} implies that 
$J_\alpha \, \omega^\alpha=-1$. Therefore, we have established that
\begin{equation}
 J_\alpha \, \Delta\langle\omega^{\alpha}\rangle=0 \,.
\label{JDeltaw}
\end{equation}
To conclude, although the projection of $\Delta\langle\omega^{\alpha}\rangle$ 
in the direction of $J_\alpha$ cannot be shifted by the gauge transformation, 
as implied by Eq.~\eqref{gaugerestriction}, one can 
nevertheless consistently impose the gauge condition \eqref{gaugecondition}
because the projection of $\Delta\langle\omega^{\alpha}\rangle$ along $J_\alpha$
vanishes to start with. 

\subsection{Degeneracy of $M^{\alpha\beta}$}

Next, we discuss the possibility that the matrix \eqref{defM} be degenerate,
i.e., that $\det M = 0$. 
In that case, there should exist a vector $\kappa^{\alpha}$ such that
\begin{equation}
 \kappa_{\alpha} M^{\alpha\beta} = 0 \,. 
\label{degeneracycondition}
\end{equation}
Then, from Eq.~\eqref{gaugeDelta} the gauge transformation of
$\Delta \langle\omega^{\alpha}\rangle$ is further restricted to satisfy
\begin{equation}\label{plouf}
 \kappa_{\alpha} \, \hat\delta_\xi \Delta\langle\omega^{\alpha}\rangle = 0 \,. 
\end{equation}

The degeneracy condition $\det M=0$ is closely related to 
the existence of \textit{iso-frequency pairs}.
In our formalism, an iso-frequency pair stands for a pair of
physically distinct orbits, i.e., two orbits with different
$P_\mu$ (different $J_\alpha$), that nonetheless
share the same frequencies $\omega^\alpha$ in phase space,
without assuming the normalization condition \eqref{Jomeganorm}.
Once an iso-frequency pair has been found, it extends
until one of its components reaches the boundary of the parameter space of 
allowed $P_\mu$, or until the ``distance'' between those orbits vanishes.
We refer to the three-dimensional surface over which the distance
between the components of the pair vanishes as the \textit{boundary}
of the iso-frequency pair. 
Then, one can easily show that the boundary of the iso-frequency pair
corresponds to the surface over which $\det M=0$. 
Indeed, the condition \eqref{degeneracycondition} can be rewritten as 
\begin{equation}
 \kappa_\alpha M^{\alpha\beta}
 = \kappa_\alpha \, \frac{\partial\omega^\beta}{\partial J_\alpha}
 =\lim_{\varepsilon\to 0} \, \frac{1}{2\varepsilon}
 \left[ \omega^\beta (J+\varepsilon\kappa)
 -\omega^\beta (J-\varepsilon\kappa) \right] ,
\label{degenerayconditionB}
\end{equation}
which implies that $\kappa_\alpha M^{\alpha\beta} = 0$ 
if and only if $J_\alpha + \varepsilon \kappa_\alpha$ and
$J_\alpha - \varepsilon \kappa_\alpha$ coincide with 
the actions that characterize an iso-frequency pair.

If we impose the constraint \eqref{Jomeganorm} along the trajectories, 
our notion of iso-frequency pair agrees with that discussed in Ref.~\cite{Barack:2011ed} 
for a Schwarzschild background, and later generalized to the Kerr case \cite{Warburton:2013yj}.
In this work, the authors found that there are infinitely many pairs of 
``iso-frequency'' Kerr geodesics, i.e., pairs of orbits that have different values of the
constants of motion $(\hat E, \hat L_z, \hat Q)$, 
but that share the same fundamental frequencies $(\Omega^r,\Omega^\theta,\Omega^\phi)$.

In the current problem concerning the matrix $M$, 
we need to consider iso-frequency pairs for Kerr geodesics, 
including the redshift variable $z=1/\omega^t$ in addition to the frequencies $\Omega^i$. 
We argue that 
if the normalization condition \eqref{Jomeganorm} is relaxed,
the existence of an iso-frequency pair
in the sense that the three frequencies $\Omega^i$ are identical 
implies the existence of a pair
in the sense that the four frequencies $\omega^\alpha$ are identical.
To prove this claim, we use the rescaled variables
$\hat P_a=P_a/\hat \mu$, with $a=1,2,3$.
By making use of the scaling transformation 
$\omega^\alpha(\lambda) = \lambda\,\omega^\alpha(1)$ 
introduced in Sec.~\ref{transformationofw},
the frequencies $\omega^{\alpha}$ scale as 
\begin{equation}\label{omega-mu}
 \omega^{\alpha}(\hat P_a,\hat\mu) = \hat \mu \, \omega^{\alpha}(\hat P_a) \, .
\end{equation}
Hence the $t$-frequencies $\Omega^i = \omega^i / \omega^t$
are specified by $\hat P_a$ only, namely $\Omega^i = \Omega^i(\hat P_a)$.
This means that even if we freely change $\hat\mu$, the iso-frequency pair
with respect to $\Omega^i$ is maintained once the values of $\hat P_a$ are
chosen appropriately. Now, from Eq.~\eqref{omega-mu} it is obvious
that one can always tune the value of $\hat\mu$ so as to make the redshift 
$z=1/\omega^t$ identical within that pair.
Here we note that rescaling the value of $\hat \mu = \sqrt{- 2 P_0}$ 
is equivalent to a rescaling of the affine parameter along the trajectory, 
because $P_0(\lambda) = \lambda^2 P_0(1)$ under such a transformation;
see App.~\ref{app:scaling}. In this sense, our iso-frequency pairs defined
in terms of the four frequencies \eqref{omega-mu} are parameterized by 
physically distinct affine parameters. This is an important difference with
respect to the iso-frequency pairs discussed in Refs.~\cite{Barack:2011ed,Warburton:2013yj},
for which both orbits are parameterized by 
the same Boyer-Lindquist coordinate time $t$.

We now prove that the constraint \eqref{plouf} is not problematic,
because the relation
\begin{equation}
 \kappa_\alpha \,\Delta\langle\omega^{\alpha}\rangle = 0
\label{appendixeq}
\end{equation}
is satisfied identically.
To do so, we recall that the orbital motion is periodic in the angles $w^i$.
Thus, performing a three-dimensional Fourier series decomposition in space,
and a Fourier transform in time, the Fourier decomposition of the interaction 
Hamiltonian \eqref{Hamiltonian-int} in terms of the Green's function is schematically given by
\begin{equation}
 \Hint(w,J;\gamma)
 = 
 {[\omega^t_{(0)}(\tau)]}^{-1} \sum_{\bn,\bn'}
 \cG_{\bn,\bn'}(J,J^{(0)},\varpi_{\bn'}(\omega_{(0)}(\tau)))
 \, e^{-\ui\varpi_{\bn'}(\omega_{(0)}(\tau)) w^{t}+2\pi \ui\,{\bn}\cdot{\bw}} \,,
\label{eq:startHi2}
\end{equation}
where $\bn=(n_r,\,n_\theta,\,n_\phi)$ is a collection of integers, $\bw=(w^r,w^\theta,w^\phi)$,
and $\varpi_{\bn}(\omega) \equiv 2\pi \,\bn\cdot{\boldsymbol\omega/\omega^{t}}$,
with $\boldsymbol\omega = (\omega^r,\omega^\theta,\omega^\phi)$,
and we have used the fact that $\omega^\alpha_{(0)}$ and $J_\alpha^{(0)}$
are constant along timelike Kerr geodesics. 
The details of this calculation are provided in App~\ref{app:Fourier1}.
Then, the long-time average of Eq.~\eqref{eq:startHi2} 
over the specified trajectory is given by  
\begin{equation}
 \HI(J) = \hat\mu \, \langle \Hint\rangle = \frac{\hat\mu(J)}{\omega^t_{(0)}(J)}
 \sum_{\bn} \cG_{\bn,\bn}(J,J,\varpi_{\bn}(\omega_{(0)}(J))) \,. 
\label{eq:Hiapp}
\end{equation} 
Finally, inserting the expressions \eqref{eq:startHi2} and \eqref{eq:Hiapp}
for $\Hint$ and $\HI$ into Eq.~\eqref{gaugecondition},
we find (after setting $\hat\mu \simeq 1$ on-shell), 
\begin{equation}
 \Delta\langle \omega^{\alpha}\rangle
 = \left(
 \omega^\alpha - M^{\alpha\beta}\frac{\partial}{\partial \omega^\beta}
 \right)
 \frac{1}{\omega^t} 
 \sum_{\bn} \cG_{\bn,\bn}(J^{(0)},J^{(0)},\varpi_{\bn}(\omega))
 \bigg|_{\omega=\omega_{(0)}(\tau)}
 \equiv \A\,\omega^\alpha_{(0)}(J^{(0)}) - M^{\alpha\beta}\,\B_\beta \,,  
\label{defAB}
\end{equation}
where $\A$ and $\B_\beta$ are defined by the last equality. 
Recalling Eq.~\eqref{JDeltaw} and using the relation 
$J_\alpha M^{\alpha\beta} J_\beta \simeq - 1$, 
which is a consequence of the normalization condition \eqref{Jomeganorm},
we note that $\A$ can be expressed in terms of $\B_\beta$ as
$\A = -J_\alpha M^{\alpha\beta} \B_\beta$. Using Eq.~\eqref{scalingwJ},
the formula \eqref{defAB} for $\Delta \langle \omega^{\alpha}\rangle$ can be rewritten as 
\begin{equation}
\label{eq:DomHn}
 \Delta \langle \omega^{\alpha}\rangle
 = M^{\alpha\beta} \bigl( \A J_\beta^{(0)} - \B_\beta \bigr) \,.
\end{equation}
Combined with Eq.~\eqref{degeneracycondition},
the above relation proves Eq.~\eqref{appendixeq}.
Thus, whenever $\det M = 0$, the restriction \eqref{plouf}
on the gauge transformation is not problematic
because the projection of $\Delta \langle\omega^{\alpha}\rangle$
along the direction of $\kappa_\alpha$ vanishes to start with.

In summary, we conclude that the gauge conditions \eqref{gaugeconditions0} 
and \eqref{gaugecondition}, such that $\omega^{\alpha}$ and $J_{\alpha}$ are constant,
can be imposed for generic orbits in the effective spacetime.
In the remainder of this paper, we will adopt these \textit{canonical gauge 
conditions}. Hamilton's equations of motion \eqref{H-eq-qJ0} then take the simple
form \eqref{H-eq-qJ}, which derives from the effective Hamiltonian \eqref{eq:cal_H}.

\subsection{Uniqueness of the effective Hamiltonian}

A natural question is whether the effective Hamiltonian \eqref{eq:cal_H}
is uniquely defined or not.
Indeed, since we have shown in Sec.~\ref{subsec:dJ}
that the action variables $J_\alpha$ can be freely changed
by a gauge transformation 
[modulo the constraint \eqref{Jomeganorm}],
in particular their constant components,
one could in principle devise alternative effective Hamiltonians by
imposing the gauge conditions \eqref{gaugeconditions0} and \eqref{gaugecondition} for different
values of the constant actions, say $J'_\alpha = J_\alpha + \Delta J_\alpha(J)$.
Thus, one might consider alternative effective Hamiltonians of the form
\begin{equation}\label{calH'}
 {\cal H}'(J) = {\cal H}(J) + \Delta {\cal H}(J) \,,
\end{equation}
as long as those preserve the four frequencies
$\omega^\alpha = \partial {\cal H} / \partial J_\alpha$
and the normalization condition \eqref{Jomeganorm}. 
These two conditions are expressed as
\begin{subequations}\label{conditions}
\begin{align}
 \omega'{}^\alpha(J') &= \omega^\alpha(J) \,, \\
 \omega'^\alpha(J') J'_\alpha &= \omega^\alpha(J) J_\alpha \,,
\end{align}
\end{subequations}
with $\omega'^\alpha(J') \equiv \partial {\cal H}'(J') / \partial J'_\alpha$. 
Combining these expressions with Eq.~\eqref{calH'} immediately gives
\begin{subequations}
\begin{align}
 &M^{\alpha\beta} \Delta J_\beta + \frac{\partial \Delta{\cal H}}{\partial J_\alpha} = 0 \,, \label{condition2a} \\
 &\omega^\alpha(J) \, \Delta J_\alpha = 0 \,, \label{condition2b}
\end{align}
\end{subequations}
where we recall that $M^{\alpha\beta} \equiv
\partial \omega^\alpha/\partial J_\beta
= \partial^2 {\cal H} / \partial J_\alpha \partial J_\beta$ is symmetric.
Introducing the inverse matrix $M_{\alpha\beta}$,
the relation \eqref{condition2a} is equivalent to
$\Delta J_\alpha = - M_{\alpha\beta} \, \partial (\Delta{\cal H}) / \partial J_\beta$. 
Substituting this expression into the condition \eqref{condition2b},
we finally obtain 
\begin{equation}
 J_{\beta} \, \frac{\partial \Delta{\cal H}}{\partial J_\beta} = 0 \,, \label{intcondition}
\end{equation}
where we used the fact that $J_\beta=M_{\alpha\beta} \, \omega^\alpha+O(\eta)$.
This means that the correction term $\Delta{\cal H}$ is merely constrained
to be constant along the integral curves of $J_\alpha$. (Note that the notation
for covariant and contravariant indices
is opposite to that usually adopted in differential geometry.)
Thus, the effective Hamiltonian is not unique, unless some
additional property of the correction term $\Delta {\cal H}$ gets specified.

However, since we require that 
the gauge-invariant interaction Hamiltonian 
in Eq.~\eqref{HIdef} scales like $\lambda^4$,
the correction term $\Delta{\cal H}$ in Eq.~\eqref{calH'}
should also scale like $\Delta{\cal H}(\lambda) = \lambda^4 \, \Delta{\cal H}(1)$. 
Hence, the condition \eqref{intcondition} can be rewritten as
\begin{equation}
 4 \Delta {\cal H} = \frac{\ud \Delta{\cal H}}{\ud \lambda} \biggr|_{\lambda=1}
 = J_{\alpha} \, \frac{\partial\Delta{\cal H}}{\partial J_\alpha} = 0 \,.
\end{equation}
This proves that the form of the effective Hamiltonian 
and the associated constant actions $J_\alpha$ are, 
in fact, uniquely determined
under the adopted scaling ansatz for $H_\text{int}(J)$.

Before closing this section, we wish to comment on the main differences
between the effective Hamiltonian \eqref{eq:cal_H} and that 
proposed by Vines and Flanagan (henceforth VF) \cite{Vines:2015}, 
who recently devised another Hamiltonian formulation
of the conservative self-force dynamics in a Kerr background. 
Our approach differs from that of VF over (at least) three points.
First, VF's Hamiltonian formulation holds 
for a \textit{fixed} effective metric $g_{\mu\nu}(x; \gamma)$ generated
by a \textit{fixed} source orbit $\gamma$.
Since their Hamiltonian is implicitly a functional of the source orbit $\gamma$
that generates the conservative self-force,
different source orbits do define different Hamiltonians. 
Second, VF rely on non-canonical transformations 
to eliminate the source-orbit dependence in their Hamiltonian,
and define their effective Hamiltonian as 
the \textit{background Hamiltonian} $H^{(0)}(J)$.
As a result, there is no $O(\eta)$ contribution---no interaction
Hamiltonian---in VF's effective Hamiltonian. Third, VF's action-angle
variables do not satisfy the constraint \eqref{Jomeganorm},
and hence the condition that restricts 
to physical orbits is not given explicitly.

By contrast, we define a \textit{unique} effective Hamiltonian \eqref{eq:cal_H}
that is a function of the actions $J_\alpha$ only and scales like $\lambda^4$, while
our action-angle variables satisfy 
the condition \eqref{Jomeganorm} 
and define the gauge-invariant orbital frequencies $\Omega^i$,
given in Eq.~\eqref{Omega-i}, in the standard manner.
Moreover, our effective Hamiltonian \eqref{eq:cal_H} is obtained 
by making use of the gauge freedom inherent to 
the self-forced motion to impose the canonical gauge conditions
\eqref{gaugeconditions0} and \eqref{gaugecondition}.

\section{Regularity of the gauge transformation in special limits} 
\label{sec:gauge-reg}

We are left to show that the gauge conditions \eqref{gaugeconditions0} and 
\eqref{gaugecondition} can be imposed consistently in the limit $J_r \to 0$
(or $J_\theta \to 0$) of a spherical (or equatorial) orbit.
In Sec.~\ref{subsec:Jr}, we explore in details the effects of a gauge transformation 
on the radial action $J_r$, and show that the canonical gauge conditions can indeed be imposed 
in the limit $J_r \to 0$, where the $\lambda^4$-scaling of the interaction Hamiltonian
\eqref{HIdef} plays a crucial role. (The equatorial orbit limit
can be discussed similarly.) However a subtlety arises in that the dynamical matrix \eqref{defM} 
diverges at the separatrix of the Kerr spherical geodesics, 
i.e., at the location of any innermost stable spherical orbit (ISSO) 
of the Kerr geometry. In Sec.~\ref{subsec:M-ISSO}, 
we discuss the regularity of the matrix $M^{\alpha \beta}$ 
and show that the canonical gauge conditions \eqref{gaugeconditions0} and 
\eqref{gaugecondition} can be imposed, even in the ISSO limit.

\subsection{Gauge transformations in the spherical or equatorial orbit limit}
\label{subsec:Jr}

Just like the usual spherical coordinates of Euclidean geometry
become singular on the polar axis, the action-angle variables become 
singular as phase-space coordinates when one of the actions vanishes. 
Indeed, in the spherical orbit limit $J_r \to 0$,
the radial angle variable $w^{r}$ becomes degenerate as a coordinate.
(Since the equatorial orbit limit $J_\theta \to 0$ can be discussed similarly,
in what follows we restrict our analysis to the spherical orbit limit.)
Consequently, $\hat\delta_\xi J_r = - (\partial \Xi / \partial w^r)_J$ vanishes
in this limit [recall Eq.~\eqref{eq:dJ}]. 
This can also be understood from the second equality \eqref{eq:dJ},
because the amplitude of $w^{r}$-dependent oscillations
of $x^{\mu}(w,J)$ and $u_{\mu}(w,J)$ is at most $O(\sqrt{J_r})$
in the spherical orbit limit. To be more explicit, 
both $x^{\mu}(w,J)$ and $u_{\mu}(w,J)$
can be written as Fourier series expansions with respect to $w^r$, namely 
\begin{align}
\label{wrexpansion}
 x^{\mu}(w,J) = \sum_n x_{(n)}^{\mu}(w^A,J_\alpha)\, e^{\ui n w^r} \,,  \quad
 u_{\mu}(w,J) = \sum_n u_{(n)\mu}(w^A,J_\alpha)\, e^{\ui n w^r} \,, 
\end{align}
where $A=t,\theta,\phi$. 
\vspace{-0.05cm}
As shown in App.~\ref{app:Fourier2}, the coefficients $x_{(n)}^{\mu}$
and $u_{(n)\mu}$ are of $O(J_r^{|n|/2})$ when $J_r \to 0$, 
which yields $w^r$-dependant oscillations of $x^{\mu}(w,J)$ and $u_{\mu}(w,J)$ of $O(\sqrt{J_r})$.
As a result, both $(\partial x^\mu / \partial w^r)_J$
and $(\partial u_\mu / \partial w^r)_J$ are also restricted 
to be $O(\sqrt{J_r})$, and \eqref{eq:dJ} implies that $\hat\delta_\xi J_r = O(\sqrt{J_r})$.
This magnitude is suppressed in the spherical orbit limit, but still large enough to gauge-transform the perturbations of $J_r$ induced by the metric perturbation $h_{\mu\nu}$ (i.e., the self-force). Indeed, assuming that the metric perturbation is also an oscillatory function of $w^r$, the rate of change of $J_r$ driven by the metric perturbation,
\begin{equation}
 \dot J_r(\tau)
 = \left(\frac{\partial J_r}{\partial x^{\mu}}\right)_{\!u} \dot x^{\mu}(\tau)
 +\left(\frac{\partial J_r}{\partial u_{\mu}}\right)_{\!x} \dot u_{\mu}(\tau) = -\left(\frac{\partial u_{\mu}}{\partial w^{r}}\right)_{\!J} \dot x^{\mu}(\tau) +\left(\frac{\partial x^{\mu}}{\partial w^r}\right)_{\!J} \dot u_{\mu}(\tau) = O(\sqrt{J_r}) \,,
\end{equation}
is also suppressed in the spherical orbit limit, for the exact same reason. 

Meanwhile, the gauge transformation of $w^{r}$ is singular 
in the limit where $J_r\to 0$, because the factors
$({\partial x^{\mu}}/{\partial J_r})_w |_{(w,J) = (w(\tau),J(\tau))}$ 
and $({\partial u_{\mu}}/{\partial J_r})_w |_{(w,J) = (w(\tau),J(\tau))}$ 
in the second line of \eqref{eq:hdxwt1} are $O(1/\sqrt{J_r})$,
as is seen from the expressions in Eqs.~\eqref{wrexpansion}. 
This enhanced magnitude is exactly that required to gauge-transform
the perturbations of $w^r$ induced by the metric perturbation, 
because the rate of change of $w^r$ driven by the metric perturbation, 
\begin{equation}
 \dot w^r(\tau)
 = \left(\frac{\partial w^r}{\partial x^{\mu}}\right)_{\!u} \dot x^{\mu}(\tau)
 +\left(\frac{\partial w^r}{\partial u_{\mu}}\right)_{\!x} \dot u_{\mu}(\tau) = \left(\frac{\partial u_{\mu}}{\partial J_{r}}\right)_{\!w} \dot x^{\mu}(\tau)
 -\left(\frac{\partial x^{\mu}}{\partial J_r}\right)_{\!w} \dot u_{\mu}(\tau) = O(1/\sqrt{J_r}) \,,
\end{equation}
is also enhanced in the spherical orbit limit. 

To establish that we have enough degrees of freedom to gauge-transform
the action-angle variables $(w^{\alpha}, J_{\alpha})$,
we write down the inverse relations of Eqs.~\eqref{eq:dJ}
and \eqref{eq:hdxwt1} as 
\begin{subequations}
 \begin{align}
  \hat\delta_\xi x^{\mu}(\tau) 
  &= \hat\delta_\xi J_{\alpha}(\tau)
  \left(\frac{\partial x^{\mu}}{\partial J_{\alpha}} \right)_{\!w}
  + \hat\delta_\xi w^{\alpha}(\tau) 
  \left(\frac{\partial x^{\mu}}{\partial w^{\alpha}}\right)_{\!J} \,, \\
  \hat\delta_\xi u_{\mu}(\tau) 
  &= \hat\delta_\xi J_{\alpha}(\tau)
  \left(\frac{\partial u_{\mu}}{\partial J_{\alpha}} \right)_{\!w}
  + \hat\delta_\xi w^{\alpha}(\tau) 
  \left(\frac{\partial u_{\mu}}{\partial w^{\alpha}}\right)_{\!J} \,. 
 \end{align}
\end{subequations}
The coefficients that appear in the above expressions have already been discussed.
In those equations, the coefficients multiplying $\hat\delta_\xi J_{r}(\tau)$ 
are $O(1/\sqrt{J_r})$ in the spherical orbit limit, while those multiplying 
$\hat\delta_\xi w^{r}(\tau)$ are $O(\sqrt{J_r})$ in that limit.
The other coefficients are $O(J_r^0)$.
Hence, we find that the gauge transformations
of $x^\mu(\tau)$ and $u_\mu(\tau)$ are finite if, and only if,
\begin{equation}
\label{circularlimitJw}
 \hat\delta_\xi J_{r}(\tau) = O(\sqrt{J_r}) \,, \quad
 \hat\delta_\xi w^{r}(\tau) = O(1/\sqrt{J_r}) \,,
\end{equation}
while respecting the restriction coming from 
the normalization condition \eqref{Jomeganorm}, namely 
\begin{equation}
\label{gconstraint-qJ}
 0 =  \omega^{\alpha}(\tau)\,\hat\delta_{\xi} J_{\alpha}(\tau)
 + J_{\alpha}(\tau)\, \hat \delta_{\xi} \omega^{\alpha} (\tau) \,,
\end{equation}
where $\hat\delta_\xi \omega^{\alpha} (\tau)
\equiv \hat\delta_\xi {\dot w}^{\alpha} (\tau)$.
Hence, as long as the transformation respects the normalization condition \eqref{gconstraint-qJ}, 
we can freely gauge-transform the action-angle variables $(J_r,w^r)$ 
under the conditions \eqref{circularlimitJw}, 
as well as the other constant components of $O(J_r^0)$.
From these considerations, we find that the condition $J_r=0$ acquires a
\textit{gauge-invariant} meaning in the spherical orbit limit. 
Similarly, the condition $J_\theta=0$ becomes
gauge-invariant in the equatorial orbit limit.
Still, the gauge transformation of $w^{r}$ (or $w^{\theta}$) can become singular
in the spherical (or equatorial) orbit limit.
Therefore, the gauge must be chosen appropriately \textit{before} 
taking one of these limits, to ensure the regularity of the gauge transformation 
in these limits.

Our next objective is to establish the consistency of 
the canonical gauge condition \eqref{gaugecondition} under 
the regularity conditions \eqref{circularlimitJw}. 
To do so, we first show that the gauge transformation
of $\langle J_r\rangle$ scales as  $\hat\delta_\xi \langle J_r\rangle = O(J_r)$, 
which is different from the scaling that one would naively expect 
from the long-time average of Eq.~\eqref{circularlimitJw}.

When considering radial oscillations with a small amplitude, 
it is convenient to expand all equations around spherical motion,
using the eccentricity $e\equiv(r_\text{max}-r_\text{min})/(r_\text{max}+r_\text{min})$
as the small parameter, where $r_\text{max}$ and $r_\text{min}$ denote
the coordinate radii at the apastron and the periastron.
Then, from the definition of the radial action, 
it can be shown that $J_r = O(e^2)$ along the orbit,
as proven in App.~\ref{app:Fourier2}. 
More precisely, defining $\sqrt{{\cal R}(r,P)}\equiv\sqrt{R(r,P)}/\Delta(r)$,
we expand ${\cal R}(r,P)$ around the spherical orbit radius $r_0(P)$ as 
\begin{equation}\label{calR}
 {\cal R}(r,P) = A(P) - B(P)^2\left(r-r_0(P)\right)^2
 + O\left(\left(r-r_0(P)\right)^3\right) ,
\end{equation}
where the spherical orbit radius is defined by the condition
\begin{equation}\label{def-r0}
 \left.\frac{\partial{\cal R}(r,P)}{\partial r}\right\vert_{r=r_0(P)} = 0 \,.
\end{equation}
The latter relation determines $r_0$ as a function of $P_\mu$.
In the following discussion, the higher order terms in $e$
will always be neglected. 
Substituting the expression \eqref{calR} into the definition \eqref{def-J0}
of the radial action, we find $J_r = A(P) / (2B(P))$, from which we deduce
\begin{equation}
 J_r = \frac{1}{2B(P)} \, u_r^2 + \frac{B(P)}{2} \, (r-r_0(P))^2 \,.
\label{eq:Jr}
\end{equation}
This expression is quadratic in $u_r$ and $r-r_0$. Thus,
when we consider the linear perturbation of $J_r$ in $\eta$,
i.e. $J_r^{(1)}$, a factor of $u_r$ or $r-r_0$ on the right-hand side
will have to be evaluated on the background Kerr geodesic. 
These background factors oscillate with an amplitude of $O(e)$,
with an oscillation frequency that is a multiple integer of the radial frequency, $n_r \omega^r$ 
[recall the expansions \eqref{wrexpansion}]. 
Since the long-time average of the radial action 
$\langle J_r\rangle$ is non-vanishing for non-spherical orbits, 
the other factor of $u_r$ or $r-r_0$ 
should be the perturbation at $O(\eta)$ 
with the same oscillation frequency. 
From the second of Hamilton's equation \eqref{H-eq-qJ0} 
and the Fourier decomposition \eqref{eq:startHi2} of the interaction Hamiltonian $H^{(1)}$, 
such a perturbation must be caused by the radial oscillation of 
the source term in $H^{(1)}$, which is of $O(e)$. 
Therefore, we argue that $\langle J^{(1)}_r \rangle = O(e^2)$.
Now, when we consider the gauge transformation of $\langle J_r\rangle$, 
it is natural to restrict to the class of gauge transformations 
for which the components of $u_r$ and $r-r_0$ 
that oscillate at the frequency $n_r\omega^r$, 
with a non-vanishing integer $n_r$, remain of $O(e)$,
because the metric perturbation itself is an oscillatory function 
with oscillation frequency $n_r\omega^r$ and amplitude of $O(e)$.
Under this restriction, we find that 
the gauge transformation of $\langle J_r\rangle$ 
scales as $\hat\delta_\xi \langle J_r\rangle = O(e^2) = O(J_r)$.

Finally, we show that the restriction $\hat\delta_\xi \langle J_r\rangle = O(J_r)$ is 
consistent with the gauge transformation \eqref{gaugeDelta} of the gauge condition \eqref{gaugecondition},
because the corresponding shift of $\Delta\langle\omega^{\alpha}\rangle$ vanishes 
automatically as $J_r\to 0$.
Going back to Eq.~\eqref{defAB}, 
we find that ${\B}_r$ is proportional to $n_r$ and 
the contribution from the terms with 
$n_r \neq 0$ comes from the radial oscillation of $O(e)$. 
This yields ${\B}_r=O(e^2)=O(J_r)$---recall that the Fourier coefficients
$\cG_{\bn,\bn}(J^{(0)},J^{(0)},\varpi_{\bn}(\omega_{(0)}(J)))$ 
are constructed from the product of integrals over the source orbit
and the field-point orbit---, and hence the term with $\beta=r$
in \eqref{eq:DomHn} is suppressed to be $O(J_r)$. 
From the above considerations,
we conclude that one can, indeed, consistently impose
the gauge condition \eqref{gaugecondition}, 
even in the limit where $J_r\to 0$. 
A similar discussion shows that the same conclusion holds
in the equatorial orbit limit $J_\theta \to 0$.

This fact makes our special gauge choice \eqref{gaugecondition} quite privileged.
Indeed, if we only needed to satisfy the consistency condition
\eqref{JDeltaw} together with the gauge conditions \eqref{gaugeconditions0},
we could impose---for instance---a gauge condition of the type 
\begin{equation}\label{pilou}
\Delta \langle \omega^\alpha \rangle
\equiv
\frac{2}{k+2} \frac{\partial (\hat\mu^{k-2} H_{\rm int})}{\partial J_\alpha}
-
\left\langle\left(\frac{\partial H^{(1)}}{\partial J_\alpha}\right)_{\!w}\right\rangle
= 0 \, ,
\end{equation}
for any integer $k \neq -2$, instead of the condition \eqref{gaugecondition}. 
The effective Hamiltonian describing the dynamics 
in this gauge would then be given by 
$H^{(0)}(J) + 2 \hat\mu^{k-2}H_{\rm int}(J) / (k+2)$. 
However, in such a gauge, the cancellation mechanism 
in the spherical orbit limit mentioned above does not work. 
Indeed, recall that it was essential in deriving Eq.~\eqref{defAB} that the numerical 
factor in front of $H_{\rm int}(J)$ be $1/2$. 
Therefore, it is not appropriate to use a gauge condition of the type \eqref{pilou} to discuss the spherical orbit limit, except for the case $k = 2$ that we have imposed in Eq.~\eqref{gaugecondition}. The same is true for the equatorial orbit limit.

\subsection{Regularity of $M^{\alpha\beta}$ and ISSO limit}
\label{subsec:M-ISSO}

A remaining concern about spherical orbits is whether or not the matrix
$M^{\alpha\beta}$ is regular when $J_r\to 0$. 
In this section, we prove that $M^{\alpha\beta}$ is not singular in this limit, 
except in the ISSO limit where $\omega^r \to 0$. Below,
we restrict our discussion of the orbital dynamics
to the test-particle limit, focusing on the spherical orbit limit 
$J_r\to 0$. Unless stated otherwise, all formulae in this subsection implicitly
assume this limit. 

Since the expressions for $J_\alpha$ with $\alpha \neq r$ 
as functions of $P_\mu$ are regular, even in the limit $J_r\to 0$,
we have $\partial J_{\alpha}/\partial P_{\mu} = O(J_r^0)$ for those
components. For $\alpha=r$, we have 
\begin{align}
 \frac{\partial J_{r}}{\partial P_{\mu}} 
 &= \frac{1}{2\pi} \frac{\partial}{\partial P_{\mu}}
 \left(2\,\int_{r_{\rm min}}^{r_{\rm max}} \sqrt{{\cal R}(r',P)} \, \ud r'\right)
 = \frac{1}{2\pi} \int_{r_{\rm min}}^{r_{\rm max}}
 \frac{\partial {\cal R}(r',P)}{\partial P_{\mu}} \, \frac{\ud r'}
 {\sqrt{{\cal R}(r',P)}}
 \cr
 &= \frac{1}{2\pi} \int_{r_{\rm min}}^{r_{\rm max}}
 \frac{\partial {R}(r',P)}{\partial P_{\mu}} \, \frac{\ud r'}{\sqrt{{R}(r',P)}}
 = O(1/\omega^r) \,.
\label{eq:pJpP}
\end{align}
To establish the last equality in the limit $\omega^r \to 0$, 
it is convenient to replace the integration variable $r'$ by
the Carter-Mino time $\lambda'$, such that $\int \rmd \lambda' 
= \int \ud r'/\sqrt{{R}(r',P)}$ \cite{Mino:2003yg}.
Since the function ${\partial {R}(r',P)}/{\partial P_{\mu}}$ and 
the relation between the Carter-Mino time
and the proper time are not singular, 
one finds that the integration range behaves as $O( 1 / \omega^r)$.
Note that there is no contribution from 
$\partial r_{\rm min}/\partial P_{\mu}$ 
and $\partial r_{\rm max}/\partial P_{\mu}$ in Eq.~\eqref{eq:pJpP}
because ${\cal R}(r,P)=0$ at $r=r_{\rm min}(P)$ and $r = r_{\rm max}(P)$.
Therefore, the components of the inverse matrix of
$\partial J_{\alpha}/\partial P_{\mu}$ behave as 
\begin{equation}\label{dPdJ-ISSO}
  \frac{\partial P_{\mu}}{\partial J_{\alpha}}
 =\left\{ \begin{array}{c}
 O(1)~~~~{\rm for}~~\alpha \neq r \,,\cr
 O(\omega^r)~~{\rm for}~~\alpha= r \,.
 \end{array} \right.
\end{equation}

For spherical orbits ($J_r=0$), we have 
${\cal R}(r_0(P),P)=0$ in addition to Eq.~\eqref{def-r0}, which 
specifies $P_{\mu}$ for given $J_A$, with $A=t,\theta,\phi$.
Moreover, the derivative of ${\cal R}'(r_0(P),P)=0$
with respect to $J_A$, where a prime denotes a derivative
with respect to the radius, 
yields
\begin{equation}
 \frac{\partial {\cal R}'}{\partial J_A}
 = \left. \frac{\partial {\cal R}'}{\partial P_\mu}\right|_{r_0}
 \frac{\partial P_\mu}{\partial J_A}
 + {\cal R}''\frac{\partial r_0}{\partial P_\mu}
 \frac{\partial P_\mu}{\partial J_A} = 0 \,.
\end{equation}
Thus, we obtain
\begin{equation}
 \frac{\partial r_0}{\partial J_A}
 = \frac{\partial r_0}{\partial P_\mu}\frac{\partial P_\mu}{\partial J_A}
 = - \frac{1}{{\cal R}''} \left. \frac{\partial {\cal R}'}
 {\partial P_\mu}\right|_{r_0}
 \frac{\partial P_\mu}{\partial J_A} = O(1/(\omega^r)^2) \,,
\label{dr0dJA}
\end{equation}
because ${\cal R}''(r_0(P),P)= O((\omega^r)^2)$,\footnote{Strictly
speaking, ${\cal R}''(r_0(P),P) = -2 (\Upsilon^r)^2$
(see Eq.~\eqref{eq:r_sol} below), where the frequencies $\Upsilon^\alpha$
are measured in Carter-Mino time $\lambda$,
and are different from the frequencies $\omega^\alpha$ measured 
in proper time $\tau$. However, we simply have
$\Upsilon^\alpha = \Sigma(\tau) \, \omega^\alpha$, 
where $\Sigma(\tau) \equiv \Sigma(r(\tau),\theta(\tau))$ 
is defined by Eq.~\eqref{Kerr-hojo}, 
and the regularity of $\Sigma(r,\theta)$ insures that 
$\Sigma(\tau) = O(1)$, which implies $\Upsilon^r = O(\omega^r)$.}
while the other factors are $O((\omega^r)^0)$.
Similarly, we obtain 
\begin{equation}
 \frac{\partial r_0}{\partial J_r}
 = \frac{\partial r_0}{\partial P_\mu}\frac{\partial P_\mu}{\partial J_r}
 = - \frac{1}{{\cal R}''} \left. \frac{\partial {\cal R}'}{\partial P_\mu}\right|_{r_0}
 \frac{\partial P_\mu}{\partial J_r} = O(1/\omega^r) \,.
\label{dr0dJr}
\end{equation}

Using Eq.~\eqref{dr0dJA}, we now estimate the components $M^{Ar} = M^{rA}$
of the matrix $M^{\alpha\beta}$ as
\begin{equation}
 \frac{\partial \omega^A}{\partial J_r}
 = \frac{\partial \omega^r}{\partial J_A}
 = \frac{1}{2\omega^r}\frac{\partial (\omega^r)^2}{\partial J_A}
 =-\frac{{\cal A}^2 }{2\omega^r}{\cal R}''' \frac{\partial r_0}{\partial J_A}
 + O(1/\omega^r)
 = O(1/(\omega^r)^3) \,, 
\label{domegardJA}
\end{equation}
where the first equality holds thanks to Eq.~\eqref{defM} and 
we have used ${\cal R}''(r_0(P(J_A)),P(J_A))= -(\omega^r)^2 / {\cal A}^2$, 
with ${\cal A}\equiv \langle \ud\lambda/\ud\tau\rangle$, in the third equality. 
Now, the expressions for the long-time averages of $\ud w^A/\ud\lambda$
and $\ud\lambda/\ud\tau$ are regular as long as we express these quantities
using $r(\lambda)$, $\theta(\lambda)$ and $P_\mu$.
However, we proved in Eqs.~\eqref{dr0dJA} and \eqref{dr0dJr}
that the derivatives of $r_0$ with respect to the actions $J_\alpha$ are, 
in fact, singular in the ISSO limit $\omega^r \to 0$.
Since $r(\lambda)$ is almost equal to $r_0$ 
in the spherical orbit limit $J_r\to 0$, 
the singular behavior seen in Eq.~\eqref{domegardJA} in the ISSO
limit $\omega^r\to 0$ arises only through $r(\lambda)$.

To establish the consistency of the estimate \eqref{domegardJA}, 
we also compute directly the derivative of $\omega^A$ with respect 
to $J_r$, and show  that $\partial \omega^A / \partial J_r = O(1/(\omega^r)^3)$
without relying on the first equality in \eqref{domegardJA}.
To do so, we need to expand the expression for $r(\lambda)$ up to $O(e^2)$. 
From the radial equation of motion for timelike geodesics in Kerr spacetime, 
i.e. $({\ud r}/{\ud \lambda})^2 = R(r,\,P)$, we obtain the Taylor expansion
\begin{align}
 \frac{\ud^2r}{\ud\lambda^2}
 &= R''(r_0(P),P)(r-r_0(P))+\frac{1}{2}R'''(r_0(P),P)(r-r_0(P))^2
 \cr
 & +\frac{1}{6}R^{(4)}(r_0(P),P)(r-r_0(P))^3 + \cdots \,.
\end{align}
Solving this equation iteratively, we find
\begin{align}
 r(\lambda) &= r_0(P) + e \, r_0(P) \cos \left(\sqrt{-R''(r_0(P),P)}\,\lambda\right) 
 \cr
 &- \frac{e^2 \,r_0(P)^2\, R'''(r_0(P),P)}{4R''(r_0(P),P)}
 \left(1-\frac{1}{12} \cos \left(2\sqrt{-R''(r_0(P),P)}\,\lambda\right) \right) 
 + \cdots \,,
\label{eq:r_sol}
\end{align}
where the ellipsis stands for all the terms $o(e^2)$.
After taking the long-time average, it turns out that the dominant contribution
in the ISSO limit $\omega^r\to 0$ originates from the third term
in the right-hand side of Eq.~\eqref{eq:r_sol},
which behaves as $O(e^2/(\omega^r)^2)$.
This term is related to the correction of the center of oscillation
due to non-linearities.
It will prove convenient to use this corrected center,
whose coordinate radius is
\begin{equation}\label{hat-r}
 \hat r_0=r_0(P)
 -\frac{e^2\, r_0(P)^2\, R'''(r_0(P),P)}{4R''(r_0(P),P)} \,.
\end{equation}
At $O(e^2)$, we also need to take into account
the frequency renormalization. From a standard singular perturbation analysis, 
we obtain the $O(e^2)$ correction to the frequency (squared)
of radial oscillations as 
\begin{equation}\label{omegar2}
 (\omega^r)^2=(\omega^r)^2|_{J_r=0}
 +\frac{e^2\,r_0(P)^2 {\cal A}^2 (R'''(r_0(P),P))^2}{4R''(r_0(P),P)} \,.
\end{equation}
We could also solve for $\theta(\lambda)$,
but there is no singular behavior in $\theta(\lambda)$
in the limit $\omega^r\to 0$. Therefore,
we can express it as a regular function of $P_\mu$. 

At this stage, it is reasonable to assume that the frequencies
$\omega^A$ can all be expressed as $\omega^A=f^A(\hat r_0(P),e(P),P)$
by means of smooth functions $f^A$,  
even in the ISSO limit $\omega^r\to 0$.
From the definition of the radial action,
one can easily show that $J_r=O(e^2 \omega^r)$ 
along the orbit.
Therefore, if we expand the expression \eqref{hat-r} for $\hat r_0$
up to linear order in $J_r$, it contains a term of $O(J_r/(\omega^r)^3)$, i.e.,
\begin{equation}
 \frac{\partial \hat r_0}{\partial J_r} = O(1/(\omega^r)^3) \,.
\label{dhatr0dJr}
\end{equation}
Thus, we also find
\begin{equation}
 \frac{\partial \omega^A}{\partial J_r} 
 = 
 \left(
 \frac{\partial f^A}{\partial \hat r_0}
 \right)_{\!e,P} \,
 \frac{\partial \hat r_0}{\partial J_r} + O(1/\omega^r)
 = O(1/(\omega^r)^3) \,,
\end{equation}
which turns out to be consistent with the estimate \eqref{domegardJA}.
Equations \eqref{dr0dJr} and \eqref{hat-r}--\eqref{dhatr0dJr}
are also useful to evaluate the component $M^{rr}$
of the matrix $M^{\alpha\beta}$, and we immediately obtain
\begin{equation}\label{domegardJr}
 \frac{\partial \omega^r}{\partial J_r}
 = -\frac{{\cal A}^2}{2\omega^r} \, R''' \, \frac{\partial \hat r_0}{\partial J_r}
 + O(1/(\omega^r)^2) = O(1/(\omega^r)^4) \,.
\end{equation}
The remaining portion of $M^{\alpha\beta}$ is obtained as
\begin{align}\label{domegaAdJA}
 \frac{\partial \omega^A}{\partial J_B}
 &= 
 \left(
 \frac{\partial f^A}{\partial \hat r_0}
 \right)_{\!e,P} \,	
 \frac{\partial {\hat r}_0}{\partial J_B} + O((\omega^r)^0) \cr
 &= -\frac{{\cal A}^2}{2\omega^r} \, R''' 
 \left(\frac{\partial {\hat r}_0}{\partial J_r}\right)^{-1}
 \frac{\partial {\hat r}_0}{\partial J_A}
 \frac{\partial {\hat r}_0}{\partial J_B}
 + O((\omega^r)^0)
 = O(1/(\omega^r)^2)\,,
\end{align}
by making use of 
$({\partial f^A}/{\partial \hat r_0})_{e,P} = 
({\partial \omega^{r}}/{\partial J_{A}})/
({\partial {\hat r}_0}/{\partial J_{r}})$, 
the third equality in \eqref{domegardJA}, 
and the fact that 
${\partial {r}_0}/{\partial J_A} = {\partial {\hat r}_0}/{\partial J_A}$, 
because we can set $J_{r} = 0$ before taking the partial derivatives 
with respect to $J_A$.

\newpage

In summary, collecting Eqs.~\eqref{domegardJA}, \eqref{domegardJr} and \eqref{domegaAdJA},
we have the following concise expression:
\begin{equation}
 M^{\alpha\beta} = \frac{\partial \omega^{\alpha}}{\partial J_{\beta}}
 = - \frac{{\cal A}^2}{2\omega^r} \, R''' 
 \left(\frac{\partial {\hat r}_0}{\partial J_r}\right)^{-1}
 \frac{\partial \hat r_0}{\partial J_{\alpha}}
 \frac{\partial \hat r_0}{\partial J_{\beta}} + \tilde{M}^{\alpha\beta} \,,
\label{Mab}
\end{equation}
where $\tilde{M}^{\alpha\beta}$ stands for the contributions
that are $O((\omega^r)^2)$ relatively to the leading terms.

In order to discuss the impact of the singular behavior \eqref{Mab}
on the gauge transformation \eqref{gaugeDelta} of $\Delta \langle \omega^{\alpha}\rangle$,
it is convenient to use basis vectors
for $\hat\delta_\xi \Delta\langle\omega^{A}\rangle$ and $\hat\delta_\xi J_A$ 
that reflect well the degree of singularity of $M^{\alpha\beta}$
in the ISSO limit $\omega^r \to 0$. 
As is clear from Eqs.~\eqref{dr0dJA} and \eqref{Mab},
one of the basis vectors should be chosen to be the singular direction 
\begin{equation}
 e^{[1]}_A 
 \equiv \left.\frac{\partial {\hat r}_0}{\partial J_A}\right/
 \left| \frac{\partial {\hat r}_0}{\partial J_A} \right|
 =
 \left.\frac{\partial r_0}{\partial J_A}\right/
 \left| \frac{\partial r_0}{\partial J_A} \right| ,
\end{equation}
while the other unit vectors, say $e^{[2]}_A$ and $e^{[3]}_A$,
are chosen such all three vectors are orthogonal to each other. 
In terms of the components in this basis, 
the gauge transformation \eqref{gaugeDelta} of 
$\Delta \langle \omega^{\alpha}\rangle$ 
is rewritten as
\begin{equation}\label{matrix}
 \begin{pmatrix}
 \hat\delta_\xi \Delta \langle(\omega^{r})^2\rangle \\
 \hat\delta_\xi \Delta \langle\omega^{[1]}\rangle \\
 \hat\delta_\xi \Delta \langle\omega^{[2]}\rangle \\
 \hat\delta_\xi \Delta \langle\omega^{[3]}\rangle
 \end{pmatrix}
 =
 \begin{pmatrix}
 O((\omega^r)^{-3}) & O((\omega^r)^{-2}) & O((\omega^r)^0) & O((\omega^r)^0) \\
 O((\omega^r)^{-3}) & O((\omega^r)^{-2}) & O((\omega^r)^0) & O((\omega^r)^0) \\
 O((\omega^r)^{-1}) & O((\omega^r)^0) & O((\omega^r)^0) & O((\omega^r)^0) \\
 O((\omega^r)^{-1}) & O((\omega^r)^0) & O((\omega^r)^0) & O((\omega^r)^0) 
 \end{pmatrix}
 \begin{pmatrix}
 \hat\delta_\xi J_r \\
 \hat\delta_\xi J_{[1]} \\
 \hat\delta_\xi J_{[2]} \\
 \hat\delta_\xi J_{[3]}
 \end{pmatrix},
\end{equation}
where the gauge condition \eqref{gaugeconditions0} has already been imposed 
to eliminate the long-time average. As shown in Sec.~\ref{subsec:Jr} above, 
$\hat\delta_\xi J_r$ vanishes for spherical orbits. 
This gives a gauge-invariant relationship among the
$\hat\delta_\xi \Delta \langle\omega^{\alpha}\rangle$ in general.
Moreover, in the ISSO limit $\omega^r \to 0$, Eq.~\eqref{matrix} shows
that an additional gauge-invariant relation appears,
because $\hat\delta_\xi \Delta \langle\omega^{[1]}\rangle$ diverges
if $\hat\delta_\xi J_{[1]}$ is not constrained to vanish. 
Since $\Delta \langle\omega^\alpha\rangle$ is defined in Eq.~\eqref{gaugecondition}
as the difference between two finite quantities, it should not diverge.
In order to avoid that divergence, the condition
\begin{equation}\label{newgaugeinv}
 0 = \hat\delta_\xi J_{[1]} = e^{[1]}_A \, \hat\delta_\xi J_A
 \sim (\omega^r)^2 \frac{\partial r_0}{\partial J_A} \, \hat\delta_\xi J_A
\end{equation}
must hold in the ISSO limit. This provides an additional gauge-invariant relationship among 
the actions $J_A$ in that limit. 

\hspace{-0.08cm}The meaning of the gauge constraint \eqref{newgaugeinv} 
can easily be clarified by considering the gauge transformation 
of the spherical orbit radius $r_0(J_A)$, namely
\begin{equation}\label{gauge-rmxr}
{{r}_{0}}({J}_{A}) 
\to 
\bar{r}_0(\bar{J}_A) \equiv 
{r}_{0}(J_{A} + \hat \delta_{\xi} J_{A}) 
= 
{r}_{0}(J_{A}) + 
(\omega^r)^{-2}\,\hat\delta_{\xi} J_{[1]} 
+ 
O(\eta^2)\,.
\end{equation}
Clearly, $\hat\delta_\xi J_{[1]}$ must vanish in the ISSO limit $\omega^r \to 0$
to avoid an infinite shift of the coordinate radius of the spherical orbit. 
Such a singular gauge transformation is not allowed because we
restrict ourselves to gauge transformations whose generator $\xi^\mu$
remains small everywhere along the orbit. Therefore, the additional
constraint \eqref{newgaugeinv} does not imply that we lose one degree of freedom to impose 
the gauge condition \eqref{gaugecondition} in the ISSO limit; rather,
it should be seen as a consistency condition for the gauge transformation 
in that limit.

\section{Frequency shift of innermost stable spherical orbits}
\label{sec:ISSO-shift}

In this section we shall focus on spherical (inclined or circular) orbits, 
which are characterized by a vanishing radial action, $J_r=0$, and for which
the radial oscillations become irrelevant. As an application of
the effective Hamiltonian \eqref{eq:cal_H}, we derive a simple
formula providing the location of the ISSOs 
due to the conservative part of the self-force,  
in terms of the redshift variable \eqref{z}, or equivalently 
in terms of the renormalized redshift variable $\tilde z$ 
defined in Eq.~\eqref{def-moz} below.

An ISSO is defined by the gauge-invariant condition $\omega^r = 0$,
or equivalently by a vanishing restoring radial force when considering a small deviation from 
a particular spherical orbit. We will show that this ISSO condition is in fact equivalent to
\begin{equation}
\label{ISSOcond-W}
 \det\left. \frac{\partial^2 {\tilde z}}{ \partial {\Omega}^a\partial\Omega^b}
 \right|\atISSO = 0 \,,
\end{equation}
where the indices $(a,b)$ run through $\theta$ and $\phi$. 
This important result extends to non-equatorial orbits the innermost stable circular orbit condition derived 
in Ref.~\cite{Isoyama:2014mja} for equatorial orbits. 
In particular, as will be shown below, the renormalized redshift
${\tilde z}$ is related to the gauge-invariant interaction Hamiltonian 
\begin{equation}\label{eq61}
 \HI(J(\Omega^{\theta}, \Omega^{\phi})) \,.
\end{equation}
The condition \eqref{ISSOcond-W} is especially convenient for practical calculations,
because $\tilde z$ and $\HI$ in the form \eqref{eq61} depend only 
on the frequencies $(\Omega^\theta,\Omega^\phi)$ 
that  parameterize the two-parameter family of spherical orbits; 
partial differentiations with respect to $\Omega^{\theta}$
and $\Omega^{\phi}$ will correspond to variations
along this two-parameter family of spherical orbits.

Thereafter, we will also use the notation $\Omega^{A}$,
with the capital Latin index running through $t$, $\theta$ and $\phi$,
recalling that the $t$-component is trivial (by definition)
since $\Omega^{t} = 1$. Interestingly,
to evaluate the ISSO condition \eqref{ISSOcond-W}
that we shall derive below,
it is not necessary to compute variations with respect to $J_\alpha$
in the directions perpendicular to the two-surface of spherical orbits.

\subsection{Redshift variable and interaction Hamiltonian}
\label{subsec:z-Hint}

Parameterizing a given orbit in terms of the three orbital frequencies $\Omega^i$,
with $i=r,\theta,\phi$,\footnote{Recall that, for physical orbits, one of the four
components of $J_\alpha$ is constrained to satisfy Eq.~\eqref{Jomeganorm}.} we start by showing that the redshift variable $z(\Omega)$, as defined in Eq.~\eqref{z} above,
is related in a simple manner to the on-shell value $\HI(\Omega)$ of the interaction Hamiltonian \eqref{HIdef}.
Since the on-shell value of the original Hamiltonian \eqref{Hamiltonian-eff} must satisfy the normalization condition \eqref{norm-G}, we have
\begin{equation}
\label{Hnorm}
 \Hzero(\Omega)+\HI(\Omega) = - \frac{1}{2} \,,
\end{equation}
for any values of the frequencies $\Omega^i$.
Then, recalling the definition \eqref{def-P} of $P_0$, 
expanding Eq.~\eqref{Hnorm} up to $O(\eta)$ yields
the following expressions of the interaction Hamiltonian:
\begin{equation}
 \HI(\Omega) = - P_0^{(1)}(\Omega)
 = - \omega_{(0)}^{\alpha}(J^{(0)}(\Omega)) \, {J}_{\alpha}^{(1)}(\Omega) \,.
\end{equation}
On the other hand, from Eq.~\eqref{Jomeganorm}
the redshift function \eqref{z} is immediately given by 
\begin{equation}\label{zveHint}
 z(\Omega) = -\Omega^{\alpha} J_{\alpha}(\Omega) \,,
\end{equation}
with $\Omega^{\,t}=1$, and hence $z_{(1)}(\Omega)
= - \Omega^{\alpha} J^{(1)}_{\alpha}(\Omega) 
= z_{(0)}(\Omega) \, \HI(\Omega)$.
This allows us to express the redshift $z$ in terms of
the interaction Hamiltonian $\HI$ as
\begin{equation}\label{def-z1}
 z(\Omega) = z_{(0)}(\Omega) + z_{(1)}(\Omega) 
 = z_{(0)}(\Omega) \left(1+\HI(\Omega) \right) .
\end{equation}
This result shows that the on-shell interaction Hamiltonian
is nothing but the relative change in the redshift induced
by the conservative self-force, at fixed frequencies $\Omega^i$.

We note that, for physical orbits, 
$\HI(\Omega)$ coincides with 
the gauge-invariant, long-time averaged perturbed Hamiltonian,
$\langle H^{(1)} \rangle(\Omega)$. 
Equation \eqref{def-z1} can thus be viewed as 
the gauge-invariant relationship between 
the redshift $z$ and $\langle H^{(1)} \rangle$, 
irrespective of our choice \eqref{HIdef} of 
interaction Hamiltonian, and irrespective of the canonical gauge conditions
\eqref{gaugeconditions0} and \eqref{gaugecondition}.
Importantly, we emphasize that all of the results in this subsection, and in
particular the key relationship \eqref{def-z1}, are valid for
\textit{generic} orbits.

\subsection{Canonical transformation}
\label{subsec:hojo-canonical}

To identify the ISSO condition,
we need to transform back to a pair of canonical variables analogous to
the original $(r, u_r)$, instead of $(w^r, J_r)$ for the radial motion, 
because $J_r$ becomes singular as a phase-space coordinate for spherical orbits. 
This is achieved through a canonical transformation 
from $(w^\alpha, J_\alpha)$ to $(\bar r, \bar w^A, \bar u_r, \bar J_A)$, whose generating function reads
\begin{equation}
 \overline{\cal W}(\bar r,\bar w^A,J)
 =- \bar w^t J_t
 - \int^{\bar r} \frac{\sqrt{R (r',J)}}{\Delta(r')} \, \ud r'  
 - \bar w^\theta J_\theta - \bar w^\phi J_\phi \,. 
\end{equation}
The relationships between the new and old variables are given by 
\begin{equation}
 w^{\alpha} = - \left( \frac{\partial \overline{\cal W}}{\partial J_\alpha} \right)_{\!\bar{r},\bar{w}} , 
 \quad
 \bar u_r = - \left( \frac{\partial \overline{\cal W}}{\partial \bar r} \right)_{\!\bar{w},J}
 = \frac{\sqrt{R (\bar r,J)}}{\Delta(\bar r)} \,, 
 \quad
 \bar J_A = - \left( \frac{\partial \overline{\cal W}}{\partial \bar w^A} \right)_{\!\bar{r},J} = J_A \,. 
\label{coordinateT}
\end{equation}
Repeating the argument used to derive Eq.~\eqref{eq:Jr}, we obtain
\begin{equation}
 J_r = \frac{1}{2B(J)} \, \bar u_r^2 + \frac{B(J)}{2} \, (\bar r-r_0(J_A))^2 \,.
\label{eq:Jr2}
\end{equation}
To show that the coefficient $B$ is related to
the frequency of the radial oscillation,
we expand the effective Hamiltonian ${\cal H}(J)$
around $J_r=0$ as 
\begin{equation}\label{Heff-sph}
 {\cal H} = {\cal H}|_{J_r=0} + \omega^r|_{J_r=0} \, J_r + O(J_r^2) \,,
\end{equation}
where we used $\omega^r = \partial {\cal H} / \partial J_r$.
Inserting Eq.~\eqref{eq:Jr2} in the above equation yields
the standard Hamiltonian for a harmonic oscillator with position $\bar r$,
momentum $\bar u_r$ and frequency $\omega^r$ if, and only if,
\begin{equation}
 B(J) = \omega^r(J) \,. 
\end{equation}

We may now use Eq.~\eqref{eq:Jr2} to substitute $\bar{u}_r$
in terms of $J_r$, $\bar{r}$, $\omega^r$ and $r_0$
into the second equation of \eqref{coordinateT},
and derive the following expression for $w^A$ from the first equation:
\begin{align}
 w^A &= \bar w^A + \frac{\partial \omega^r(J)}{\partial J_A}
 \int^{\bar r} \sqrt{\frac{2J_r}{\omega^r(J)}-{(r'-r_0(J_B))}^2}
 \, \ud r'
 \cr
 &- J_r \, \frac{\partial \ln \omega^r(J)}{\partial J_A}
 \int^{\bar r} \frac{\ud r'}{\sqrt{2J_r/\omega^r(J)-(r'-r_0(J_B))^2}}
 \cr
 &+ \omega^r(J) \, \frac{\partial r_0(J_B)}{\partial J_A}
 \int^{\bar r} \frac{r'-r_0(J_B)}{\sqrt{2J_r/\omega^r(J)-(r'-r_0(J_B))^2}}
 \, \ud r' \,. 
\label{wA}
\end{align}
In the spherical orbit limit, the three integrals in \eqref{wA}
are proportional to $J_r$, $J_r^0$ and $J_r^{1/2}$, respectively.
Thus, in that limit we obtain
\begin{subequations}\label{eq610}
 \begin{align}
  &\left.\bar \omega^A\right|_{J_r=0} \equiv \left.\dot {\bar w}^A\right|_{J_r=0}
  = \left.\omega^A\right|_{J_r=0} \,, \\ 
  &\left.\frac{\partial \bar \omega^A}{\partial \bar r}\right|_{J_r=0}
  = \left.\frac{\partial \omega^A}{\partial \bar r}\right|_{J_r=0}
  + \left.(\omega^r)^2 \, \frac{\partial r_0(J_B)}{\partial J_A}\right|_{J_r=0} \,.
 \end{align}
\end{subequations}
Hence, we find that while the frequencies $\bar \omega^A$ and $\omega^A$
are identical for spherical orbits, their $\bar r$-derivatives are not. 
Similarly, for $w^r$ we obtain
\begin{align}
 w^r &= \frac{\partial \omega^r(J)}{\partial J_r}
 \int^{\bar r} \sqrt{\frac{2J_r}{\omega^r(J)}-{(r'-r_0(J_A))}^2}
 \, \ud r'
 \cr
 &+ \left(1-J_r \frac{\partial \ln \omega^r(J)}{\partial J_r}\right)
 \int^{\bar r}\!\! \frac{\ud r'}{\sqrt{2J_r/\omega^r(J)-(r'-r_0(J_A))^2}} \,. 
\end{align}
In the limit of a spherical orbit, this equation reduces to 
\begin{equation}
 \left.w^r\right|_{J_r=0} = \left.\omega^r(J) 
 \int^{\bar r}\frac{\ud r'}{\bar u_r}\right|_{J_r=0} \,.
\end{equation}

\subsection{Derivation of the ISSO condition}

In the remainder of this section,
we will use only the canonical coordinates
$(\bar r, \bar w^A, \bar u_r, \bar J_A)$.
However, since we found in Eq.~\eqref{coordinateT} that $\bar J_A = J_A$, 
we will get rid of the overbar for these variables.  
By replacing the argument $J_r$ in favor of $\bar r$, $\bar u_r$ and $J_A$
using Eq.~\eqref{eq:Jr2}, 
and setting $\bar u_r=0$
in the spherical orbit limit $J_r \to 0$, 
the effective Hamiltonian of interest is 

\begin{equation}\label{defHeff}
\Heff (\bar r,J_A) 
= \left. 
\Hzero + \frac{1}{2} \HI 
\right|_{J_r = J_r(\bar r,\bar u_r=0,J_A)} \,. 
\end{equation}
Now, the spherical orbit condition is given by 
\begin{equation}
 \dot {\bar u}_r 
 = 
 -\frac{\partial \Heff(\bar r,J_A)} {\partial \bar r} 
 =
 -\frac{\partial \Heff(J)}{\partial J_r}
 \left(\frac{\partial J_r}{\partial \bar r}\right)_{\!{\bar u}_r = 0, J_A}
 = 0 \,,
\label{eq:circ_r}
\end{equation}
where we used the condition 
${\partial J_r}/{\partial \bar r} = 0$ derived from Eq.~\eqref{eq:Jr2}. 
This specifies the circular orbit radius $\bar r$ for given $J_A$.
The condition for an ISSO is specified by the requirement that the restoring
force vanishes when considering a small deviation from a given spherical orbit.
Because the frequency $\omega^r$ of the radial oscillation vanishes
in the ISSO limit, it is sufficient to consider stationary perturbations
in the ``radial'' direction ${\bar r}$.
Indeed, from the effective Hamiltonian~\eqref{Heff-sph},
the ISSO condition is simply given by
\begin{equation}
 0 = 
 \left. (\omega^r)^2 \right|\atISSO 
 =
 \left. 
 \frac{\partial^2 \Heff(\bar r,J_A)}{\partial \bar r^2}
 \right|\atISSO \,. 
\label{ISSOcond-r}
\end{equation}
We note that the use of the effective Hamiltonian 
automatically takes care of the derivative acting on 
the degrees of the freedom for the source orbit; 
this derivative is essential because the source position should be
varied simultaneously while computing the restoring force. 
An ISSO condition similar to Eq.~\eqref{ISSOcond-r}
was proposed earlier, in the context of post-Newtonian
theory, in Ref.~\cite{Buonanno:2002ft}.

Next, we prove the equivalence between Eqs.~\eqref{ISSOcond-W} and \eqref{ISSOcond-r},
introducing the \textit{renormalized redshift variable}\footnote{This renormalized
redshift variable is identical to the modified redshift function defined in Ref.~\cite{Isoyama:2014mja} in the circular equatorial orbit limit.}
\begin{equation}\label{def-moz}
 {\tilde z} (\Omega^a) \equiv z_{(0)}(\Omega^a) 
 + \frac{1}{2} z_{(1)}(\Omega^a)
 = z_{(0)}(\Omega^a) \left(1+\frac{1}{2}\HI(\Omega^a) \right)
 = z(\Omega^a) \left(1-\frac{1}{2}\HI(\Omega^a) \right) .
\end{equation} 
One may interpret the factor $1/2$ here as the numerical factor necessary
to avoid the double counting of the effect of the self-field,
just like in the Newtonian two-body problem.
First, for the unperturbed case, we show that 
\begin{equation}
\label{ISSO-hojo1}
 \left. \frac{\ud J_A}{\ud \Omega} \right|\atISSO \equiv \left.
 k^a \frac{\partial J_A}{\partial \Omega^a} \right|\atISSO \simeq 0 \,,
\end{equation}
namely, for ISSOs the actions $J_A$ are constant
along a specific direction $k^a$ given by 
\begin{equation}\label{def-n}
 k^a =
 \left.
 \left( 
 \frac{\partial \bar \omega_{(0)}^a}{\partial \bar r} 
 - \frac{\partial \bar \omega_{(0)}^{t}}{\partial \bar r} \, \Omega^a 
 \right)
 \right|_{J_r = 0} ,
\end{equation}
where we recall the expressions \eqref{eq610} for
the frequencies $\bar{\omega}^A$. Thereafter, we adopt the rule that $J_A$ and $\bar r$ are replaced
with their values for the spherical orbit specified by $\Omega^a$,
before taking the derivatives with respect to $\Omega^a$.
We can easily show that Eq.~\eqref{ISSO-hojo1} follows from 
\begin{equation}
\label{ddH-dWdr}
 \frac{\partial} {\partial \Omega^a}
 \left( \frac{\partial \Heff}{\partial \bar r} \right)  = 0 \,,
\end{equation}
which is always satisfied for spherical orbits.
Evaluating this identity by using the chain rule for partial derivatives,\begin{equation}
 \frac{\partial }{\partial \Omega^a}
 = \frac{\partial \bar r}{\partial \Omega^a } \frac{\partial}{\partial \bar r}
 + \frac{\partial J_A}{\partial \Omega^a} \frac{\partial}{\partial J_A} \,,
\end{equation}
and applying the ISSO condition \eqref{ISSOcond-r}
to the resulting expression, 
together with $\partial \Heff / \partial J_A = \omega^A = 
\bar{\omega}^A$ along spherical orbits because of Eq.~\eqref{eq610}, we get
\begin{equation}
\label{ISSO-dJdW}
 \left. \frac{\partial {J}_A}{\partial \Omega^a}
 \frac{\partial \bar\omega^A}{ \partial \bar r } \right|\atISSO = 0 \,. 
\end{equation}
Applying that same chain rule to the derivative of Eq.~\eqref{Hnorm} 
with respect to $\Omega^a$, we obtain
\begin{equation}
\label{dHdOmega}
 0= \frac{\partial}{\partial {\Omega^a}} 
 \left( \Heff + \frac{1}{2} H_\text{int} \right)
 = \omega^A \frac{\partial {J}_A}{\partial \Omega^a}
 + \frac{1}{2} \frac{\partial \HI}{\partial {\Omega^a} } \,, 
\end{equation}
where we used Eq.~\eqref{defHeff},
as well as the spherical orbit condition~\eqref{eq:circ_r}.
We also notice that \eqref{zveHint} and \eqref{dHdOmega}
imply ${\partial J_a^{(0)}}/{\partial \Omega^b} 
= - \partial^2 z_{(0)} / (\partial \Omega^a \partial \Omega^b)$, 
and hence ${\partial J_a^{(0)}}/{\partial \Omega^b}$ is symmetric.
Then, combining the relation \eqref{dHdOmega} with \eqref{ISSO-dJdW}
and using \eqref{def-n}, we find that \eqref{ISSO-hojo1} holds.

For the proof of the equivalence between Eqs.~\eqref{ISSOcond-W} 
and~\eqref{ISSOcond-r}, we need to establish a few additional identities.
Focusing on 
\begin{equation}
 \frac{\ud}{\ud \Omega} \left( \frac{\partial \Heff}{\partial \bar r} \right)
 = k^a \frac{\partial}{\partial \Omega^a} \left( \frac{\partial \Heff}{\partial \bar{r}} \right) = 0 \,,
\end{equation}
and neglecting the terms of $O(\eta^2)$,
we obtain the following identity for ISSOs:
\begin{equation}
\label{ISSO-hojo2}
 \left. \frac{\ud \omega^A}{\ud \Omega}
 \frac{\ud {J}_A}{\ud \Omega} \right|\atISSO = 0 \,.
\end{equation}
Another required identity is deduced from 
${\partial^2 (\Heff + \tfrac{1}{2}\HI)}/
({\partial {\Omega^a} \partial {\Omega^b}}) = 0$.
Using the spherical orbit condition~\eqref{eq:circ_r},
as well as Eq.~\eqref{ISSO-hojo2}, straightforward computations yield 
\begin{equation}
 \left. k^a k^b \left(
 \omega^A \frac{\partial^2 {J}_A}{\partial {\Omega^a} \partial {\Omega^b}} 
 + \frac{1}{2}
 \frac{\partial^2 \HI}{\partial {\Omega^a} \partial {\Omega^b}} \right)\right|\atISSO
 = 0 \,. 
\label{2nd-dHintdW}
\end{equation}
Combining the relations~\eqref{dHdOmega} and \eqref{2nd-dHintdW}, 
we can easily derive the following result:
\begin{align}
\label{ISSO-hojo3}
 \left. k^a k^b \, \frac{\partial^2 {\tilde z}}
 {\partial {\Omega^a} \partial {\Omega^b}} \right|\atISSO 
 &= k^a k^b \biggl[
 - \left(1 -  \frac{1}{2}\HI \right) \!
 \left( \Omega_A \frac{\partial^2 {J}_A}{\partial {\Omega^a} \partial {\Omega^b}}
 + \frac{\partial {J}_a}{\partial \Omega^b}
 + \frac{\partial {J}_b}{\partial \Omega^a} \right)\cr 
 & \qquad\quad\;\, \left. 
 -\frac{1}{2} \left( J_a \frac{\partial \HI}{\partial {\Omega^b}}
 + J_b \frac{\partial \HI}{\partial {\Omega^a}}
 + z \frac{\partial^2 \HI} {\partial {\Omega^a} \partial {\Omega^b}} \right)
 \biggr] \right|\atISSO \cr 
 &= \left. - 2 k^a \left(\frac{\ud {J}_a}{\ud \Omega} 
 + \omega^A J_a \frac{\ud {J}_A}{\ud \Omega} \right)\right|\atISSO =0 \,,
\end{align}
where we used Eqs.~\eqref{zveHint} and~\eqref{def-moz},
and ignored $O(\eta^2)$ by using Eqs.~\eqref{ISSO-hojo1} and~\eqref{2nd-dHintdW}.
The last equality holds because we have
$\omega^A \ud {J}_A / \ud \Omega = - k^a / ({J}_b k^{b}) (\ud {J}_a / \ud \Omega)$,
owing to Eqs.~\eqref{zveHint} and \eqref{ISSO-hojo2}. 
We have also used the derivative of Eq.~\eqref{zveHint}
with respect to $\Omega$, i.e., $\ud z / \ud \Omega \simeq -{J}_b k^{b}$
in the test-mass limit. 

Our last task is to get rid of the projection vector $k^a$ in
\eqref{ISSO-hojo3}, which is defined in terms of the phase-space coordinates 
$(\bar r, \bar w^A, \bar u_r, \bar J_A)$. Consider the unit vector $\ell^a$
perpendicular to the projection vector $k^a$ in the two-dimensional parameter space 
spanned by $\Omega^a$. Then Eq.~\eqref{ISSO-hojo3} can be viewed as the projection
of the $2 \times 2$ matrix ${\partial^2 {\tilde z}}/{\partial \Omega^a \partial \Omega^b}$ 
in the $k^a$-direction, and we obtain
\begin{equation}\label{def-ISCO-det}
\det\left.
 \frac{\partial^2 {\tilde z}}{\partial {\Omega}^a \partial {\Omega}^b}
 \right|\atISSO
 = 
 \left( k^a k^b \frac{\partial^2 {\tilde z}}{\partial {\Omega}^a \partial {\Omega}^b} \right) \!
 \left( \ell^a \ell^b \frac{\partial^2 {\tilde z}}{\partial {\Omega}^a \partial {\Omega}^b} \right)
 -
 \left.
 \left( \ell^a k^b \frac{\partial^2 {\tilde z}}{\partial {\Omega}^a \partial {\Omega}^b} \right)^2
 \right|_{\rm ISSO}
 = 0 \, ,
\end{equation}
where we used the identity $k^a \partial^2 {\tilde z} / \partial {\Omega^a} \partial {\Omega^b}|\atISSO \simeq 0$
that follows from \eqref{ISSO-hojo1}. Equation \eqref{def-ISCO-det} is nothing but the desired
ISSO condition \eqref{ISSOcond-W}. We have thus proved the equivalence of the ISSO conditions  
\eqref{ISSOcond-W} and \eqref{ISSOcond-r}.

Finally, we note that the equatorial limit $J_\theta \to 0$ of the ISSO condition
\eqref{ISSOcond-W} is not trivial, and will be addressed in Sec.~\ref{subsec:ISCO} below.

\section{Hamiltonian first law of binary mechanics}
\label{sec:1stlaw}

In this section, starting from the Hamiltonian \eqref{eq:cal_H}, 
we derive a ``first law'' of mechanics for our black hole
$+$ point particle binary system, which is valid at relative $O(\eta)$ 
and thus accounts for all of the conservative effects of the self-force.
This variational relation compares two neighboring solutions
of the Hamiltonian dynamics, and simply reads (recall that $\Omega^t = 1$)
\begin{equation}
 \Omega^{\alpha} \, \delta\tilde J_{\alpha} + z \, \delta\mu = 0 \,,
\label{firstlaw}
\end{equation}
where the renormalized action variables $\tilde J_{\alpha}$ are
related to the original (specific) actions $J_{\alpha}$,
via the gauge-invariant interaction Hamiltonian $\HI$, through
\begin{equation}
 \tilde J_{\alpha} = \mu J_{\alpha} \left(1-\frac{1}{2}\HI\right) .
\label{deftildeJ}
\end{equation}
Importantly, the Hamiltonian first law \eqref{firstlaw} is expressed
in terms of the non-specific actions $\mu J_\alpha$,
rather than the specific actions $J_\alpha$ that we have used so far.
Note that the normalization condition \eqref{Jomeganorm} 
implies the following algebraic relationship
between the renormalized actions \eqref{deftildeJ}
and the renormalized redshift variable \eqref{def-moz} introduced
in the previous section:
\begin{equation}
 \Omega^{\alpha} \tilde J_{\alpha} + \mu \tilde z = 0 \,.
\label{firstlawbis}
\end{equation}
After having derived and discussed the first law \eqref{firstlaw},
we will use it to prove the equivalence
between the notions of innermost stable spherical orbit
and minimum energy spherical orbit.

\subsection{Derivation of the Hamiltonian first law}
\label{1st-derivation}

The Hamiltonian first law \eqref{firstlaw} allows
for a variation $\delta \mu$ with respect to the particle's mass $\mu$.
Since the Hamiltonian \eqref{eq:cal_H} 
depends implicitly on $\mu$ through the Green's function \eqref{symmetricG},
it will prove convenient to introduce the specific interaction Hamiltonian
$\hat H_{\rm int} \equiv \HI / \mu$, such that
\begin{equation}\label{calH}
 {\cal H}(J;\mu) = \Hzero(J) + \frac{\mu}{2}\,\hat{H}_{\rm int}(J) \,.
\end{equation}

We now consider a small variation $\delta \cal{H}$
of the effective Hamiltonian \eqref{calH} 
induced by small variations $\delta J_\alpha$ 
and $\delta \mu$ of the specific actions and of the particle's mass.
For two neighboring solutions of the Hamiltonian dynamics, we have
\begin{equation}\label{dcalH1}
 \delta {\cal H} = \left( \frac{\partial {\cal H}}{\partial J_\alpha} \right)_{\!\mu}
 \delta J_\alpha
 + \left( \frac{\partial {\cal H}}{\partial \mu} \right)_{\!J} \delta \mu = \omega^{\alpha} \, \delta J_{\alpha} + \frac{\delta \mu}{2} \hat{H}_{\rm int} \,,
\end{equation}
where $\omega^\alpha = (\partial {\cal H} / \partial J_\alpha)_\mu$
are the $\tau$-frequencies of the motion.
On the other hand because the original Hamiltonian
satisfies $\Hzero + \HI = -\frac{1}{2}$ on shell, 
$\mathcal{H} = - \frac{1}{2} - \frac{1}{2} \HI$ for physical orbits,
such that
\begin{equation}\label{dcalH2}
 \delta {\cal H} = - \frac{1}{2} \, \delta \HI
 = \frac{1}{2} \, \omega^\alpha J_\alpha \, \delta \HI
 = \omega^\alpha \, \delta \biggl( J_\alpha \, \frac{\HI}{2} \biggr) \,.
\end{equation}
Here, we used the normalization condition \eqref{Jomeganorm}
in the second equality, the test-mass first law
$\omega^\alpha \, \delta J_\alpha = O(\mu)$ of Ref.~\cite{LeTiec:2013}
in the third equality, and neglected the contribution $O(\mu^2)$.
Combining the expressions \eqref{dcalH1} and \eqref{dcalH2},
we readily obtain the variational relation
\begin{equation}\label{diplodocus}
 \omega^{\alpha} \, \delta \biggl[ J_\alpha
 \biggl( 1 - \frac{\HI}{2} \biggr) \biggr]
 + \frac{\hat{H}_{\rm int}}{2} \, \delta \mu = 0 \,.
\end{equation}
The final step is to consider the linear combination
$\omega^\alpha \, \delta \tilde{J}_\alpha$
of the renormalized actions \eqref{deftildeJ}.
Using Eq.~\eqref{diplodocus} and the normalization condition \eqref{Jomeganorm},
simple algebra yields
\begin{equation}
 \omega^{\alpha} \, \delta \tilde{J}_\alpha = - \frac{\HI}{2} \, \delta \mu
 + \omega^{\alpha} J_\alpha \, \biggl( 1 - \frac{\HI}{2} \biggr)
 \, \delta \mu = - \delta \mu \,.
\end{equation}
Dividing both sides of this equation by $\omega^t=z^{-1}$,
we obtain the Hamiltonian first law \eqref{firstlaw}.
Introducing the notation $\tilde{E} \equiv - \tilde{J}_t$
for the ``mechanical energy'' of the system,
this variational relationship can be written in the more familiar form
(with $i=r,\theta,\phi$)
\begin{equation}
 \delta \tilde{E} = \Omega^i \, \delta \tilde J_i + z \, \delta \mu \,,\label{firstlaw2}
\end{equation}
which lends itself to a suggestive comparison
to the ADM-type first laws of binary mechanics derived
in Refs.~\cite{LeTiec:2012,Blanchet:2013,LeTiec:2015,Blanchet:2017},
as discussed in Sec.~\ref{subsec:ADM} below.

Together, the relations \eqref{firstlawbis} and \eqref{firstlaw2}
imply an alternative form of the first law,
\begin{equation}
 \mu \, \delta \tilde z = - \tilde J_i \, \delta \Omega^i + \frac{z_{(1)}}{2}
 \, \delta \mu \,,
\end{equation}
which shows that the renormalized redshift variable $\tilde z(\Omega^i,\mu)$
can be understood as the ``free energy'' resulting from a Legendre 
transformation
of the ``mechanical energy'' $\tilde{E}(\tilde{J}_i,\mu)$
obeying Eq.~\eqref{firstlaw2}. This last expression implies the identities
\begin{equation}
 \tilde J_i = - \mu \, \frac{\partial \tilde z}{\partial \Omega^i} \,,
 \quad
 z_{(0)} \hat{H}_{\rm int} = 2 \frac{\partial \tilde z}{\partial \mu} \,.
\label{from_firstlaw2}
\end{equation}

\subsection{Uniqueness of the Hamiltonian first law}

A natural question is whether the Hamiltonian first law \eqref{firstlaw}---or equivalently Eq.~\eqref{firstlaw2}---is unique or not. By ``first law'' we mean a variational relationship $\Omega^{\alpha} \, \delta\tilde J'_{\alpha} + z \, \delta\mu = 0$ obeyed by some variables $\tilde{J}'_\alpha$, which may be different from the renormalized actions \eqref{deftildeJ}. In order to establish the uniqueness of the first law, we shall assume that there exist such variables, say $\tilde{J}'_\alpha = \tilde{J}_\alpha + \Delta J_\alpha(\tilde{J})$, and prove that the only allowed shift is $\Delta J_\alpha = 0$.

First, we note that if both $\tilde{J}_\alpha$ and $\tilde{J}'_\alpha$ obey Eq.~\eqref{firstlaw}, then the shift $\Delta J_\alpha$ must satisfy the constraint
\begin{equation}\label{toto}
\Omega^\alpha \, \delta (\Delta J_\alpha) = 0 \, .
\end{equation}
On the other hand, as we prove below, the variational relationship \eqref{firstlaw} implies the algebraic relationship \eqref{firstlawbis}, irrespective of the explicit expression \eqref{deftildeJ} of the renormalized actions $\tilde{J}_\alpha$. By the exact same logic, since the variables $\tilde{J}'_\alpha$ are assumed to obey the first law \eqref{firstlaw}, they must also obey the algebraic formula $\Omega^{\alpha} \tilde J'_{\alpha} + \mu \tilde z = 0$, from which we deduce the constraint
\begin{equation}\label{tada}
	\Omega^\beta \Delta J_\beta = 0 \, .
\end{equation}
Now, since the expression \eqref{toto} can be recast into the equivalent form $\delta (\Omega^\beta \Delta J_\beta) = \Delta J_\alpha \, \delta \Omega^\alpha$, we find that
\begin{equation}
	\Delta J_\alpha = \frac{\partial(\Omega^\beta \Delta J_\beta)}{\partial \Omega^\alpha} = 0 \, .
\end{equation}
Thus, to conclude that the first law is unique, we are left to show that the variational relationship \eqref{firstlaw} implies the algebraic relationship \eqref{firstlawbis}, irrespective of the explicit expression for the renormalized actions $\tilde{J}_\alpha$.

To do so, we simply consider the explicit $\mu$-dependence of the quantities ${\tilde J}_\alpha$ and $z$ that appear in the first law \eqref{firstlaw}, namely
\begin{subequations}
	\begin{align}
		{\tilde J}_\alpha &= \mu \left[ J^{(0)}_\alpha(\Omega) + \eta \, \tilde{J}^{(1)}_\alpha(\Omega) + O(\eta^2) \right] , \label{tildeJexp}\\
		z &= z_{(0)}(\Omega) + \eta \, z_{(1)}(\Omega) + O(\eta^2) \, ,
	\end{align}
\end{subequations}
where we expanded in powers of the usual mass ratio $\eta = \mu /M$. Here, $J^{(0)}_\alpha(\Omega)$ and $z_{(0)}(\Omega)$ are known from the geodesic motion in Kerr, while $\tilde{J}^{(1)}_\alpha(\Omega)$ and $z_{(1)}(\Omega)$ are the conservative self-force corrections at fixed frequencies $\Omega^i$. Computing the variation $\delta {\tilde J}_\alpha$ using \eqref{tildeJexp} and isolating the contribution proportional to $\delta \mu$ in the first law \eqref{firstlaw}, we find for the coefficients $O(\mu)$ and $O(\mu^2)$,
\begin{subequations}
	\begin{align}
		&\Omega^\alpha J^{(0)}_\alpha + z_{(0)} = 0 \, , \\
		&\Omega^\alpha J^{(1)}_\alpha + \frac{1}{2} \, z_{(1)} = 0 \, .
	\end{align}
\end{subequations}
Adding those two equations and multiplying by the mass $\mu$ shows that Eq.~\eqref{firstlawbis} does, indeed, hold irrespective of the explicit expression for ${\tilde J}_\alpha$.

\subsection{Generalization of the Hamiltonian first law}

Recalling that the effective Hamiltonian \eqref{calH} depends also on the background Kerr black hole mass $M$ and spin $S$ through the metric \eqref{h-R0}, one can easily generalize the Hamiltonian first law \eqref{firstlaw2} by allowing for non-zero variations of those two parameters. The derivation is almost identical to that given earlier in Sec.~\ref{1st-derivation}. The only modification that we need to take care of is the variational relation \eqref{diplodocus}, which now includes non-zero variations in the black hole mass and spin, and reads 
\begin{equation}\label{diplodocus2}
 \omega^{\alpha} \, \delta \biggl[ J_\alpha
 \biggl( 1 - \frac{\HI}{2} \biggr) \biggr]
 + \frac{\hat{H}_{\rm int}}{2} \, \delta \mu 
 +
 \biggl( 1 - \frac{\HI}{2} \biggr) 
 \biggl[ 
 \left( \frac{\partial {\cal H}}{\partial M} \right) \delta M
 +
 \left( \frac{\partial {\cal H}}{\partial S} \right) \delta S
  \biggr]
 = 0 \,,
\end{equation}
where we used the generalized test-particle first law 
$\omega^\alpha \, \delta J_\alpha 
= 
- (\partial_M {\cal H} \,\delta M + \partial_S {\cal H} \, \delta S) 
+ O(\mu)$ of Ref.~\cite{LeTiec:2013}.
Here, the partial derivatives are evaluated while holding the canonical variables $J_i$ and the mass $\mu$ fixed.
Up to uncontrolled terms $O(\mu^3)$, we then find
\begin{equation}\label{firstlaw3}
	\delta \tilde{M} = \Omega^i \, \delta \tilde{J}_i + z \, \delta \mu + z_\text{BH} \, \delta M + \Omega_\text{BH} \, \delta S \, ,
\end{equation}
where we introduced the total mass-energy $\tilde{M} \equiv M + \tilde{E}$, such that $\tilde{M} - (M + \mu) = - (\tilde{J}_t + \mu)$ can be interpreted as the binary's renormalized negative ``binding energy,'' as well as some effective black hole ``redshift'' $z_\text{BH}$ and ``spin precession frequency'' $\Omega_\text{BH}$, which are given by
\begin{equation}\label{firstlaw4}
z_\text{BH} \equiv 1 + {\tilde z} \, 
\frac{\partial (\mu {\cal H)}}{\partial M}\,, \quad 
\Omega_\text{BH} \equiv {\tilde z} \, 
\frac{\partial {(\mu \cal H})}{\partial S} \, .
\end{equation}
Here, the partial derivatives are to be evaluated at fixed renormalized action $\tilde{J}_i$ and particle mass $\mu$. The formula \eqref{firstlaw3} generalizes to (conservative) self-forced motion the Hamiltonian first law previously established in Ref.~\cite{LeTiec:2013} for a test mass orbiting a Kerr black hole.

Introducing the variable ${\cal M} \equiv \tilde{M} - \Omega^i \tilde{J}_i = M + \mu \tilde{z}$, the first law \eqref{firstlaw3} can be written in the equivalent form $\delta {\cal M} = - \tilde{J}_i \, \delta \Omega^i + z \, \delta \mu + z_\text{BH} \, \delta M + \Omega_\text{BH} \, \delta S$, such that the redshifts and the spin precession frequency are given by the alternative expressions
\begin{equation}\label{machin}
	z = \left( \frac{\partial {\cal M}}{\partial \mu} \right)_{\!\Omega} \, , \quad z_\text{BH} = \left( \frac{\partial {\cal M}}{\partial M} \right)_{\!\Omega} \, , \quad \Omega_\text{BH} = \left(\frac{\partial {\cal M}}{\partial S} \right)_{\!\Omega} \, .
\end{equation}
These expressions are especially convenient to perform practical calculations of the black-hole quantities $z_\text{BH}$ and $\Omega_\text{BH}$, because they involve the renormalized redshift $\tilde{z}(\Omega^i;\mu,M,S)$ defined in Eq.~\eqref{def-moz} above. Given the definition of $\tilde{z}$, the expression for $z$ is trivially satisfied.
 
Interestingly, by equating the expressions \eqref{firstlaw3}, \eqref{firstlaw4} and \eqref{machin} for $z_\text{BH}$ and $\Omega_\text{BH}$, one obtains some non-trivial relations between the partial derivatives of the effective Hamiltonian and ``mechanical energy'' at fixed (renormalized) actions, and those of the (renormalized) redshift variable at fixed frequencies, namely
\begin{subequations}
	\begin{align}
		\frac{1}{\mu \tilde{z}} \biggl( \frac{\partial {\tilde E}}{\partial M} \biggr)_{\!\tilde{J},\mu} &= \left( \frac{\partial {\cal H}}{\partial M} \right)_{\!\tilde{J},\mu} = \left( \frac{\partial \ln {\tilde z}}{\partial M} \right)_{\!\Omega,\mu} \, , \\
		\frac{1}{\mu \tilde{z}} \biggl( \frac{\partial {\tilde E}}{\partial S} \biggr)_{\!\tilde{J},\mu} &= \left( \frac{\partial {\cal H}}{\partial S} \right)_{\!\tilde{J},\mu} = \left( \frac{\partial \ln {\tilde z}}{\partial S} \right)_{\!\Omega,\mu} \, .
	\end{align}
\end{subequations}
 
\subsection{Comparison to the ADM first laws}
\label{subsec:ADM}

Interestingly, the particle Hamiltonian first law \eqref{firstlaw3} is reminiscent of the various first laws of mechanics that have been established in the context of arbitrary mass-ratio compact binaries. In particular, for binary systems of massive point particles moving along circular orbits with constant angular frequency $\Omega$, the spacetime's helical Killing symmetry can be used to derive a formula relating small changes in the global Arnowitt-Deser-Misner (ADM) mass $M_\text{ADM}$ and total angular momentum $J$ of the binary system to those of the individual masses $m_a$ (with $a=1,2$) of the particles, namely \cite{LeTiec:2012}
\begin{equation}\label{firstlawADM}
	\delta M_\text{ADM} = \Omega \, \delta J + \sum_a z_a \, \delta m_a \, .
\end{equation}
Here the coefficients $z_a \equiv \ud \tau_a / \ud t$ are the redshifts of the particles, with $\tau_a(t)$ the proper time elapsed along the worldline of particle $a$; each redshift is constant along a given circular orbit. The variational formula \eqref{firstlawADM} is itself a limiting case of a more general law, valid for systems of black holes and extended fluid balls \cite{Friedman:2002}.

The formula \eqref{firstlawADM} was later recovered and extended to binary systems of spinning point masses, by means of the canonical ADM Hamiltonian framework. Starting from the center-of-mass frame three-dimensional two-body ADM Hamiltonian $H_\text{ADM}({\bf r},{\bf p};m_a,{\bf S}_a)$ of a binary system of point masses $m_a$ with relative position ${\bf r}(t)$, relative momentum ${\bf p}(t)$ and canonical spins ${\bf S}_a(t)$, the authors of Ref.~\cite{Blanchet:2013} showed that, for circular orbits and spins aligned or anti-aligned with the orbital angular momentum, 
\begin{equation}\label{firstlawADM2}
	\delta M_{\rm ADM} = \Omega \, \delta L + \sum_a \left( z_a \, \delta m_a + \Omega_a \, \delta S_a \right) .
\end{equation}
This first law relates small changes in the ADM mass $M_{\rm ADM}$ and the orbital angular momentum $L$ 
of the binary system, such that $J = L + \sum_a S_a$, to those of the particle's individual masses $m_a$ and spin magnitudes $S_a$. The redshifts $z_a$ and the spin precession frequencies $\Omega_a$ appearing in the right-hand side of Eq.~\eqref{firstlawADM2} are related to the ADM Hamiltonian according to \cite{Blanchet:2013}
\begin{equation}\label{z-ADM}
	z_a = \frac{\partial H_{\rm ADM}}{\partial m_a} \, , \quad \Omega_a = \frac{\partial H_{\rm ADM}}{\partial S_a} \, ,
\end{equation}
where the partial derivatives are to be evaluated while holding the canonical variables fixed. The analogy with the expressions \eqref{firstlaw4} is, of course, striking. The occurrence of the particle's (renormalized) redshift ${\tilde z}$ in Eq.~\eqref{firstlaw4} can, heuristically, be understood from the fact that the four-dimensional effective Hamiltonian ${\cal H}$ is parameterized by the proper time $\tau$, and not by the ADM coordinate time $t$.

Moreover, the first law \eqref{firstlawADM} was extended to point-particle binaries moving along generic bound (eccentric) orbits. The first law for eccentric-orbit non-spinning binaries reads \cite{LeTiec:2015}
\begin{equation}\label{firstlawADM3}
	\delta M_{\rm ADM} = \Omega \, \delta L + n \, \delta R + \sum_a \langle z_a \rangle \, \delta m_a \, ,
\end{equation}
where $n \equiv 2\pi / P$ is the radial frequency of the motion and $\Omega \equiv \Phi/P$ the averaged azimuthal frequency, with $\Phi$ the accumulated azimuthal angle per radial period $P$, as measured by an asymptotic static observer. The radial action and the averaged redshifts read $R = \frac{1}{2 \pi} \oint p_r \, \ud r$ and $\langle z_a \rangle = \frac{1}{P} \int_0^P \! z_a(t) \, \ud t = T_a / P$, with $T_a$ the proper-time period of the radial motion.

Remarkably, the particle Hamiltonian first law \eqref{firstlaw3} with \eqref{firstlaw4} is \textit{formally identical} to the ADM first laws \eqref{firstlawADM2}--\eqref{firstlawADM3} through $O(\mu^2)$. An important difference, however, is that while Eqs.~\eqref{firstlawADM2} and \eqref{firstlawADM3} have been explicitly checked to hold true, for any mass ratio, up to 3PN order and to linear order in the spins $S_a$, Eq.~\eqref{firstlaw3} is valid without assuming any expansion with respect to the spin of the background black hole or the orbital velocity of the smaller body. We emphasize that the above agreement is far from obvious, because the ADM Hamiltonian first laws of Refs.~\cite{Blanchet:2013,LeTiec:2015,Blanchet:2017} were established starting from the three-dimensional two-body ADM Hamiltonian $H_{\rm ADM}({\bf r},{\bf p};m_a,{\bf S}_a)$, while the particle Hamiltonian first law \eqref{firstlaw3} was derived from the 4-dimensional effective Hamiltonian ${\cal H}(J;\mu,M,S)$. The relation between ${\cal H}$ and $H_{\rm ADM}$ is not trivial, and we shall not attempt to establish it here. 

There is, however, one key difference between the Hamiltonian first law \eqref{firstlaw3} and the ADM first laws \eqref{firstlawADM2} and \eqref{firstlawADM3}: while the former involves the total ``mechanical energy'' $\tilde{M}$, which is defined along the orbit of the smaller body, the latter involves the globally-defined mass $M_\text{ADM}$ of the binary system. This suggests that the notion of orbital energy provided by $\tilde{M}$ is related in a simple manner to the \textit{Bondi mass} of the perturbative spacetime at any retarded time, $\tilde{M} = M_\text{B}$, and similarly for the axial component of the angular momentum.\footnote{In our conservative dynamics setup, the ADM mass is not well defined beyond the test-mass limit, because the usual falloff conditions required for ADM quantities are not satisfied; actually, the Bondi mass is not well defined either. For an adiabatic inspiral satisfying the retarded boundary condition, however, both global masses are well defined. See Ref.~\cite{LeTiec:2012} for a related discussion in the post-Newtonian context.} 
This identification is supported by recent calculations in the case of a particle subject to a self-force along specific strong-field orbits in a Schwarzschild background \cite{Marta:2015,Marta:2015b,Marta:2016prep}. It is also supported by a variety of calculations comparing self-force, post-Newtonian and numerical-relativity predictions \cite{Isoyama:2014mja,LeTiec:2012b,Zimmerman:2016}. In particular, the recent results of Ref.~\cite{Zimmerman:2016} support the conjecture, put forward in Ref.~\cite{LeTiec:2012}, that the ADM first law \eqref{firstlawADM}---and by extension the Hamiltonian first law \eqref{firstlaw3}---holds even when dissipation is present, when formulated in terms of Bondi quantities. Making this argument rigorous using a two-timescale expansion \cite{Hinderer:2008dm,Pound:2015wva} is still an open problem. Alternatively, one might adopt the viewpoint that the first law \eqref{firstlaw3} \textit{defines}---within the self-force framework---some physically reasonable notions of total mass, total angular momentum and actions for black-hole-particle binary systems.

\subsection{Minimum energy spherical orbits}

We now prove that, just like for circular equatorial orbits,
the notion of innermost stable spherical orbit is equivalent
to that of minimum energy spherical orbit.
First, using Eq.~\eqref{from_firstlaw2}
the ISSO condition \eqref{ISSOcond-W} can be rewritten in the equivalent form
\begin{equation}\label{plouc}
 \det \, \frac{\partial {\tilde J}_a}{\partial\Omega^b} \biggr|\atISSO = 0 \,,
\end{equation}
where the gradients $\partial \tilde{J}_\theta / \partial \Omega^a$
and $\partial \tilde{J}_\phi / \partial \Omega^a$ of
the functions $\tilde{J}_\theta(\Omega^a)$ and $\tilde{J}_\phi(\Omega^a)$
are assumed not to vanish at the locations of ISSOs.
The condition \eqref{plouc} implies that these two vectors are colinear
along the one-parameter family of ISSOs.
Since the set of all ISSOs is a curve in $(\Omega^\theta,\Omega^\phi)$-space,
there exists a unit tangent vector to the level curves
of $\tilde{J}_\theta(\Omega^a)$ and $\tilde{J}_\phi(\Omega^a)$
at the locations of ISSOs, say $n^a$, such that
\begin{equation}\label{tic}
 n^a \frac{\partial {\tilde J}_\theta}{\partial \Omega^a} \biggr|\atISSO
 = n^a \frac{\partial {\tilde J}_\phi}{\partial \Omega^a} \biggr|\atISSO = 0 \,.
\end{equation}
On the other hand, the Hamiltonian first law \eqref{firstlaw2}
implies that along a sequence of spherical orbits,
for which $\tilde{J}_r = 0$,
\begin{equation}\label{tac}
 \frac{\partial {\tilde E}}{\partial \Omega^a}
 = \Omega^\theta \, \frac{\partial \tilde{J}_\theta}{\partial \Omega^a} + \Omega^\phi \, \frac{\partial \tilde{J}_\phi}{\partial \Omega^a} \,,
\end{equation}
where the partial derivatives are taken at fixed mass $\mu$.
Then, combining \eqref{tic} and \eqref{tac}, we readily obtain
\begin{equation}\label{MESOcond}
 n^a \frac{\partial {\tilde E}}{\partial\Omega^a} \biggr|\atISSO = 0 \,,
\end{equation}
which states that the energy $\tilde{E}(\Omega^a)$ is minimized
along the direction $n^a$ in the $(\Omega^\theta,\Omega^\phi)$-space.
In other words, Eq.~\eqref{MESOcond} is the condition
for a minimum energy spherical orbit (MESO).
Therefore, we have extended the known equivalence
between the notions of ISSO and MESO, in the test-mass limit,
to (conservative) self-forced motion in a Kerr background.
Moreover, the equivalence between the conditions \eqref{ISSOcond-W}
and \eqref{MESOcond} generalizes to inclined orbits the results of
Refs.~\cite{Buonanno:2002ft,Isoyama:2014mja},
which established the equivalence
between the notions of innermost stable circular orbit (ISCO)
and minimum energy circular orbit (MECO).

\subsection{Equatorial limit of the ISSO condition}
\label{subsec:ISCO}

We conclude this section with a discussion of the ISSO condition \eqref{ISSOcond-W} in the limit $J_\theta \to 0$ of an equatorial orbit. 
Since the unique ISSO for which $J_\theta = 0$ is a circular orbit (for a given black hole spin $S$), we refer to this orbit as the ISCO.

First, we notice that along spherical orbits ($J_r = 0$), we have $\tilde{J}_r = 0$ so the first law \eqref{firstlaw2} implies
\begin{align}\label{cuicui}
  \biggl( \frac{\partial \tilde E} {\partial \Omega^\phi} \biggr)_{\!\tilde{J}_\theta} 
&= \Omega^\phi \biggl( \frac{\partial \tilde{J}_\phi} {\partial \Omega^\phi} \biggr)_{\!\tilde{J}_\theta} \cr
&= \Omega^\phi \biggl[ \biggl( \frac{\partial \tilde{J}_\phi} {\partial \Omega^\phi} \biggr)_{\!\Omega^\theta} + \biggl( \frac{\partial \tilde{J}_\phi} {\partial \Omega^\theta} \biggr)_{\!\Omega^\phi} \biggl( \frac{\partial \Omega^\theta} {\partial \Omega^\phi} \biggr)_{\!\tilde{J}_\theta} \biggr]
= \Omega^\phi \biggl( \frac{\partial \tilde{J}_\theta} {\partial \Omega^\theta} \biggr)^{-1}_{\!\Omega^\phi} \det \, \frac{\partial {\tilde J}_a}{\partial\Omega^b} \, .
\end{align}
Then, assuming that the gradients 
$(\partial \tilde{J}_\theta / \partial \Omega^\theta)_{\! \Omega^\phi}$ 
do not vanish in the ISSO limit, 
the condition~\eqref{plouc} implies   
$({\partial \tilde E} /{\partial \Omega^\phi} )_{\! \tilde{J}_\theta } 
|_{\rm ISSO} = 0$ along any level curve of $\tilde{J}_\theta(\Omega^a)$.
Since the equatorial orbit condition $J_\theta = 0$ is equivalent to $\tilde{J}_\theta = 0$, this implies that, for any ISSO in the equatorial limit 
${\tilde J}_\theta \to 0$,
\begin{equation}\label{eq:MECO}
\biggl( \frac{\partial \tilde E}
{\partial \Omega^\phi} \biggr)_{\!J_\theta=0} \, \Biggr|_{\rm ISSO}
= \biggl( \frac{\partial \tilde J_\phi}
{\partial \Omega^\phi} \biggr)_{\!J_\theta=0} \, \Biggr|_{\rm ISSO} = 0 \,.
\end{equation}
This condition is nothing but the definition of the MECO,
the only equatorial MESO.
The condition \eqref{eq:MECO} is equivalent to
\begin{equation}\label{condISCO}
 \biggl( \frac{\partial^2 \tilde z}
 {\partial \Omega^\phi{}^2} \biggr)_{\!J_\theta=0} \, 
 \Biggr|_{\rm ISSO} = 0 \,,
\end{equation}
making use of the identity 
$\partial \tilde{z} / \partial \Omega^\phi = - \tilde{J}_\phi / \mu$, 
obtained from Eq.~\eqref{from_firstlaw2}.
Hence the ISCO is the circular orbit for which the renormalized redshift 
variable $\tilde z (\Omega^\phi)$ has an inflection point. 

Reference \cite{Isoyama:2014mja} gave an alternative derivation of the
condition \eqref{condISCO} and used it to evaluate 
the Kerr ISCO frequency shift induced by the conservative self-force.
Those results have recently been confirmed and extended to larger black hole spins, by means of a direct
stability analysis of the equations of motion for slightly eccentric orbits
\cite{vandeMeent:2016hel}.

\section{Summary and outlook}
\label{sec:conclusion}

\subsection{Summary}

In this paper we devised a Hamiltonian formulation of a point particle's
motion in Kerr spacetime for generic (eccentric, inclined) bound orbits, subject to
the conservative part of the gravitational self-force, at leading order beyond the
test-mass approximation. Our formulation relies on a
description of the particle's motion as geodesic in the effective spacetime
metric \eqref{Hamiltonian-G0}, in terms of action-angle variables. In particular,
we established that the action variables, which are conserved in the test-mass limit,
do not evolve secularly under the effect of the conservative self-force for generic
(non-resonant) orbits.

We then investigated how the gauge freedom of linearized gravity affects
our Hamiltonian description of the orbital motion. We showed that the gauge transformations
of the action-angle variables remain small, even after a long-term integration, 
as long as the metric perturbation itself remains small along the orbit. This implies
that the actions can be changed rather arbitrarily, as long as the gauge-invariant
constraint \eqref{Jomeganorm} is satisfied, while the averaged frequencies associated with the
angles are gauge invariant. Moreover, our Hamiltonian analysis shows that, for a
structureless point particle, the averaged orbital frequencies \eqref{Omega-i} and the average redshift variable
\eqref{z} provide a complete gauge-invariant characterization of the dynamics
for generic orbits; however, additional gauge-invariant relationships exist in
particular cases, such as spherical or equatorial orbits.

Equipped with the above understanding of the effects of gauge transformations on
the Hamiltonian dynamics, we then introduced an effective Hamiltonian,
defined in Eq.~\eqref{eq:cal_H} in terms of the gauge-invariant interaction Hamiltonian \eqref{HIdef}, 
in which the degrees of freedom of the physical orbit were identified with those
of the source orbit that generates the metric perturbation and self-force, before
deriving Hamilton's equations. The effective Hamiltonian \eqref{eq:cal_H}, which is a
function of the action variables only, reproduces the dynamics encoded in the
original source-dependent Hamiltonian \eqref{Hamiltonian-G0}. 
We showed that such an effective Hamiltonian consistently exists if 
certain gauge conditions, given by Eqs.~\eqref{gaugeconditions0} 
and \eqref{gaugecondition}, are imposed. Importantly, our effective Hamiltonian
\eqref{eq:cal_H} is unique under the assumption of a natural $\lambda^4$-scaling.
Other choices of scaling would fail to recover the conservative
self-forced dynamics in the spherical and equatorial orbit limit. 

As an application of this effective Hamiltonian formulation, we generalized the definition
of an innermost stable spherical orbit (ISSO) in order to account for the effect of the linear conservative self-force,
and derived a simple gauge-invariant condition allowing one to compute the self-force-induced shift in the frequencies of
Kerr ISSOs. The resulting condition is concisely expressed by the formula \eqref{ISSOcond-W},
given in terms of the renormalized redshift variable \eqref{def-moz} as
a function of the frequencies $\Omega^\theta$ and $\Omega^\phi$ of spherical orbits, and is quite convenient for numerical implementations. By taking the equatorial limit of this gauge-invariant condition, 
we showed that our result recovers the condition \eqref{condISCO} used in Refs.~\cite{LeTiec:2012b,Isoyama:2014mja,vandeMeent:2016hel} to compute the Kerr (and Schwarzschild) ISSO frequency shift for equatorial orbits.

Using the effective Hamiltonian \eqref{eq:cal_H}, we also derived a new ``first law of mechanics''
for our black-hole-particle binary systems moving along generic bound orbits,
Eq.~\eqref{firstlaw}. This variational relationship holds for the renormalized action
variables $\tilde J_\alpha$ defined by Eq.~\eqref{deftildeJ}, in terms of the gauge-invariant interaction
Hamiltonian \eqref{HIdef}. Allowing for non-zero variations of the black hole mass and spin, we further established the generalized 
first law \eqref{firstlaw3}. These Hamiltonian first laws account for all of the conservative effects 
of the self-force, and recover the test-mass results derived in Ref.~\cite{LeTiec:2013}. 
Interestingly, we found that the generalized first law \eqref{firstlaw3} is formally identical
to the ADM first laws of binary mechanics \eqref{firstlawADM2} and \eqref{firstlawADM3} that were previously
established for arbitrary mass-ratio compact binaries, and known to be valid up to high post-Newtonian orders.
Finally, by using the first law \eqref{firstlaw}, we proved that the notion of ISSO is equivalent to that
of minimum energy spherical orbit (MESO), in the sense that the ISSO condition \eqref{ISSOcond-W} is equivalent
to the MESO condition \eqref{MESOcond}, expressed in terms the mechanical energy ${\tilde E} = - \tilde{J}_t$ of the orbit.

\subsection{Outlook}

Our Hamiltonian formulation of the conservative self-force dynamics in the Kerr
geometry opens a number of interesting research directions. First, it could be
applied to the problem of self-forced orbital evolution in Kerr spacetime, while
including the radiative aspects of the problem. Indeed, existing evolution schemes
for an inspiralling orbit \cite{Tanaka:2005ue,Mino:2005an,Hinderer:2008dm,Flanagan:2010cd,vandeMeent:2013sza}
can easily be formulated in terms of the action-angle variables that we used in this work.
In particular, to first order in the mass ratio, the long-term
evolution of the actions $J_\alpha$ is driven only by the radiative
part of the metric perturbation (or the dissipative self-force)
\cite{Mino:2005an}. Therefore, in the framework of a two-timescale expansion
\cite{Hinderer:2008dm}, such evolution schemes for the dissipative self-force
could easily be combined with the present Hamiltonian formulation for the
conservative self-force, yielding a complete, practical and unified framework
to describe the evolution of generic inspiralling orbits in Kerr spacetime. 

A second line of research has to do with the meaning of 
the ``canonical'' gauge conditions 
\eqref{gaugeconditions0}--\eqref{gaugecondition} that define
the effective Hamiltonian \eqref{eq:cal_H}.
By construction of the action variables \eqref{def-J0}, 
this effective Hamiltonian describes an integrable dynamical system. 
This is in agreement with previous work suggesting that the
conservative self-forced motion of a particle in Kerr spacetime
should be integrable and Hamiltonian \cite{Vines:2015}.
Our analysis shows that such an effective Hamiltonian 
can, however, only be consistently defined in properly selected gauges. 
But it is not clear why the canonical gauges are preferred 
as far as integrability and regularity are concerned.
Thus far, the canonical gauge conditions are defined only along the orbit, 
and we need yet to know these gauge conditions at the level of the 
metric perturbation. It would therefore be interesting to explore how one can perform gauge transformations
from the standard gauges of black hole perturbation theory 
to the canonical gauge, and identify the canonical-gauge 
metric perturbation. This might reveal a hidden geometrical
meaning of the canonical gauge, clarifying its relation to the integrability property.

Third, to explicitly evaluate the self-force-induced shift in the
frequencies of Kerr ISSOs, a numerical implementation of the ISSO formula
\eqref{ISSOcond-W} is called for. Crucially, this condition
does not require self-force data for slightly eccentric orbits; it merely
requires evaluating the gauge-invariant renormalized redshift variable
\eqref{def-moz} along the sequence of Kerr ISSOs. 
Such a calculation would provide an ``exact'' post-geodesic correction
to the conservative dynamics of spinning, precessing binary black holes, for strong-field orbits,
and would contribute to the ongoing effort to develop increasingly sophisticated semi-analytical models 
for arbitrary mass-ratio inspiralling compact binaries \cite{Barausse:2009xi,Barausse:2011ys,Taracchini:2013rva,Hannam:2013oca,Nagar:2015xqa}. 

Fourth, the formal analogy between the Hamiltonian first law \eqref{firstlaw3} 
and the ADM first laws \eqref{firstlawADM2}--\eqref{firstlawADM3} suggests that
these variational relations could be extended beyond their current domains of validity.
In particular, one might combine our Hamiltonian formulation to that describing the
dynamics of a spinning test particle in Kerr spacetime \cite{Barausse:2009aa,Vines:2016unv},
to derive a Hamiltonian first law for spinning particles orbiting Kerr
black holes, while including conservative self-force effects. Similarly, one could possibly
extend the ADM-type first laws, valid for arbitrary mass ratios, to precessing spinning binaries
and generic orbits.

Finally, it would be interesting to use the Hamiltonian first law \eqref{firstlaw3} 
to compare self-force predictions against the results from numerical-relativity (NR)
simulations of binary inspirals with large mass ratios. Such simulations of unequal-mass,
non-spinning binary black holes have recently confirmed that the ADM-type first law
\eqref{firstlawADM} for quasi-circular binaries holds to a remarkable degree during the
inspiral phase \cite{Zimmerman:2016}. Given the formal analogy between the Hamiltonian
first law and the ADM first laws, this suggests that NR simulations could be used to test
the conjectured link between the total mechanical energy $\tilde{M}$ and the Bondi mass
$M_\text{B}$ of the perturbative spacetime, beyond the post-Newtonian regime \cite{Zimmerman:2016prep}.
Moreover, Ref.~\cite{Zimmerman:2016} showed how a quantity analogous to the redshift of a particle can be defined for dynamical black holes in NR simulations, from a certain normalized surface gravity. These ``black hole redshifts'' could be compared to the redshifts appearing in the Hamiltonian first law \eqref{firstlaw3}, whose computation only requires self-force data for $\tilde{z}(\Omega;\mu,M,S)$, together with the formulas \eqref{machin}. Such comparisons would help refine our understanding of the relativistic dynamics of compact binary systems. These insights should, accordingly, improve our ability to extract astrophysical information from such systems, in the new era of gravitational-wave astronomy.

\acknowledgments

We thank Leor Barack, \'{E}anna \'{E}. Flanagan, 
Scott Hughes, Maarten van de Meent, Eric Poisson, Adam Pound, 
Justin Vines, Chulmoon Yoo and Aaron Zimmerman for helpful discussions.
In particular, SI is deeply grateful to Eric Poisson for endless 
encouragements, insightful exchanges and feedback. 
This research received funding 
from the European Research Council under the European Union's Seventh 
Framework Programme, Grant Agreement No.~304978. 
RF's work was funded through H2020 ERC Consolidator Grant
``Matter and strong-field gravity: New frontiers in Einstein's theory''
(MaGRaTh-646597). SI acknowledges the support of the
grant for JSPS Postdoctoral Fellowship for Research Abroad. 
ALT acknowledges support from a Marie Curie FP7 Integration Grant
(PCIG13-GA-2013-630210), and the hospitality of the Yukawa Institute
for Theoretical Physics during a visit throughout the course of this work.
This work was also supported by MEXT Grant-in-Aid for Scientific Research
on Innovative Areas, ``New developments in astrophysics
through multi-messenger observations of gravitational wave sources'',
No.~24103001(TT) and No.~24103006 (HN, TT), JSPS Grant-in-Aid for Scientific Research (C),
No.~16K05347 (HN), JSPS Grant-in-Aid for Young Scientists (B), No.~25800154 (NS),
JSPS Grant-in-Aid for Scientific Research (C), No.~16K05356 (NS),
and by MEXT Grant-in-Aid for Scientific Research (A), No.~15H02087 (TT).

\appendix

\section{Resonant orbits}
\label{app:resonant}

In this appendix, we discuss the special case of resonant orbits.
For such orbits, a number of arguments used in Sec.~\ref{sec:NoSC} do not hold.
For resonant orbits, the ratio of the frequencies 
$\omega_{(0)}^r(\tau)$ and $\omega_{(0)}^\theta(\tau)$ is a rational number.
Here, the argument $\tau$ is written to indicate that those frequencies 
are not viewed as functions of the actions $J_\alpha$.

In the resonant case, a total reflection point does not exist along the orbit, 
except if the initial values $w_{\rm I}^r$ and $w_{\rm I}^\theta$
of $w_{(0)}^r(\tau)$ and $w_{(0)}^\theta(\tau)$ are fine tuned. 
If we use the convention
\vspace{-0.15cm}
that $u^{(0)}_\theta=0$ with $\pi/2 < \theta \leqslant \pi$,
i.e., when $w_{(0)}^\theta=2\pi \tilde N^\theta$ with $\tilde N_\theta$ 
an integer,
\vspace{-0.15cm}
there is a minimum absolute value of $w_{(0)}^\theta$ modulo $2\pi$
at the points where $u^{(0)}_r=0$, with $r=r_{\rm min}$.
\vspace{-0.08cm}
This value will change if we shift the initial value 
$w_{\rm I}^r$ or $w_{\rm I}^\theta$. 
Therefore, in the resonant case, the orbit is transformed
into an inequivalent orbit under shifts of 
$w_{\rm I}^r$ or $w_{\rm I}^\theta$. 
Thus, the argument that was used in Sec.~\ref{sec:NoSC} to
prove that there is no secular change in the values of the actions does
not hold for resonant orbits.

To compute the secular changes in the ``constants'' of motion,
it is more convenient to use the canonical variables $(X^{\mu}, P_{\mu})$,
because $X^i$ with $i=1,2,3$ are all constants determined by the initial conditions 
for the background geodesics, while $X^0$ coincides with 
the proper time $\tau$ along the trajectory:
\begin{equation}\label{def-X-Kerr}
X^0(\tau) \simeq {\tau - \tau_0}\,, 
 \quad
X^1(\tau) \simeq t_{0} \,, 
 \quad
X^2(\tau) \simeq \phi_{0} \,, 
 \quad
X^3(\tau) \simeq \frac{\Delta \lambda}{2} \,,
\end{equation}
where $\tau_0$, $t_0$ and $\phi_0$ are the initial values of 
the proper time $\tau$, coordinate time $t$ and azimuthal phase $\phi$.
The parameter $\Delta \lambda$ is the Carter-Mino time difference 
in reaching the minima of the $r$ and $\theta$ oscillations, 
which we call the \textit{offset phase} \cite{Flanagan:2012kg,Isoyama:2013yor,vandeMeent:2013sza}.
In Eqs.~\eqref{def-X-Kerr}, the constant terms in $X^1$ and $X^2$ 
can be absorbed by a shift of the origin of the coordinates $t$ and $\phi$;
as such they are physically irrelevant, even for resonant orbits.
By contrast, a constant shift of $X^3$ can have a physical 
meaning as the offset phase $\Delta \lambda$ 
for resonant orbits, 
as mentioned at the end of Sec.~\ref{sec:Hamilton}.
Therefore, the averaged proper-time derivative 
of the Carter constant $P_3 \equiv \hat Q$, 
as given by the partial derivative
of the Hamiltonian with respect to $X^3$, does not vanish.   
Since $X^3_{(0)}(\tau) = {\Delta \lambda}/{2} $ is constant, 
the long-time average along the orbit can be taken 
before the differentiation with respect to $X^3$. 
As a result, we obtain the averaged rate of change of the Carter constant 
driven by the conservative self-force acting on the resonant orbit as 
\begin{equation}\label{avQdot}
 \left\langle \dot {\hat Q} \right\rangle 
 = -\left\langle \left(\frac{\partial \Hint}
 {\partial X^3} \right)_{\!P} \right\rangle
 = - \frac{\partial \HI}{\partial \Delta\lambda} \,, 
\end{equation}
where $\HI$ is the gauge-invariant interaction Hamiltonian 
defined in Eq.~\eqref{HIdef}. 
In the resonant case, the different values of $\Delta\lambda$ 
characterize physically distinct orbits, 
and hence $\HI$ also depends on $\Delta\lambda$.
The expression \eqref{avQdot} was first derived in Ref.~\cite{Isoyama:2013yor}, 
using a different strategy. 

We note that the average rate of change \eqref{avQdot} could---at least in principle---vanish, 
even for resonant orbits. So far, this possibility has not been excluded due to the current
capability of existing self-force codes in Kerr spacetime \cite{Shah:2012gu,vandeMeent:2015lxa,vandeMeent:2016pee}. 
It would be interesting to address this question in the near future.

\section{Scaling transformation}
\label{app:scaling}

In this appendix we explain how the scaling transformation introduced in
Sec.~\ref{transformationofw}, and used extensively throughout this work,
is closely related to the reparameterization invariance of the
four-dimensional particle's action associated to the Hamiltonian \eqref{Hamiltonian-G0}. 

In our formulation, the degrees of freedom of the source orbit $\gamma$ 
in the Hamiltonian \eqref{Hamiltonian-G0} are assumed to be held \textit{fixed}. 
Hence, Hamilton's equations \eqref{H-eq0} follow from extremizing the particle's action 
\begin{equation}\label{action-Heff}
S[x(\tau);\gamma] \equiv 
\mu \int \rmd \tau 
\left[ u_{\mu} {\dot x}^{\mu} - H(x,u; \gamma) 
\right] ,
\end{equation}
while holding $\gamma$ fixed.\footnote{We simply make it a rule that the
degrees of freedom of $\gamma$ are \textit{not} to be varied when varying the 
action $S$. The degrees of freedom $(x^\mu,u_\mu)$ are identified with
those of $\gamma$ only \textit{after} the variation has been performed. 
This formal action principle yields the correct equation of motion 
with $\gamma$ fixed.}
Since the action \eqref{action-Heff} admits any affine parameterization,
the Hamiltonian dynamics includes a freedom to linearly rescale the affine parameter
along the orbit. Making use of this freedom, we consider the \textit{scaling transformation}
of all variables under a linear rescaling of the proper time $\tau$, namely 
\begin{equation}\label{scale-tau}
\tau \to {\hat \tau} \equiv \tau / \lambda \, ,
\end{equation}
with $\lambda$ is a positive real number. 
Under this scaling transformation, the action \eqref{action-Heff} becomes  
\begin{equation}\label{scale-action-Heff}
S[x({\hat \tau});\gamma] = \frac{\mu}{\lambda}
\int \rmd {\hat \tau} 
\left[ {\hat u}_{\mu} \dot{\hat{x}}^\mu 
-  H({\hat x}, {\hat u} ; \gamma)
 \right] ,
\end{equation}
where the scaled canonical variables are 
$({\hat x}^{\mu},{\hat u}_{\mu}) = (x^{\mu},\lambda u_{\mu})$. 
This is nothing but the scaling transformation 
$(x^\mu,u_\mu) \to (x^\mu,\lambda u_\mu)$ 
introduced in Sec.~\ref{transformationofw} and used throughout this work.
Of course, such an overall rescaling of the action does not affect the equations of motion.

From the definition \eqref{Hamiltonian-G0}, 
we find for the scaled Hamiltonian 
\begin{equation}\label{scale-Heff} 
H({\hat x},{\hat u}; \gamma) = \lambda^2 \, H(x,u\,; \gamma)\,,
\end{equation}
whose on-shell value simply reads $H({\hat x},{\hat u};\gamma)\vert_\gamma = - \lambda^2/2$.
The scaling transformation of other canonical variables, such as the action-angle
variables, is obtained in a similar manner, i.e., first by expressing them in terms of
$(x^\mu,u_\mu)$, and then by applying the scaling $u_\mu \to \lambda u_\mu$.

\section{Non-triviality of the canonical gauge conditions}
\label{app:gauge}

In this appendix, we explain why the gauge condition \eqref{gaugecondition}
is not automatically satisfied once we impose the conditions \eqref{gaugeconditions0}. 
We start from the expression
\begin{align}
 \left\langle\left(\frac{\partial \Hint}
 {\partial J_{\alpha}}\right)_{\!w}\right\rangle
 &= \left\langle\left(\frac{\partial \Hint}{\partial J_{\alpha}}\right)_{\!X}
 - \left(\frac{\partial \Hint}{\partial w^\beta}\right)_{\!J}
 \left(\frac{\partial w^\beta}
 {\partial J_{\alpha}}\right)_{\!X}\right\rangle
 \cr
 &= \left\langle\left(\frac{\partial \Hint}{\partial J_{\alpha}}\right)_{\!X}
 + \dot J_\beta \left(\frac{\partial w^\beta}
 {\partial J_{\alpha}}\right)_{\!X}\right\rangle . 
\label{dHdJ1} 
\end{align}
Here, one may think that we can eliminate the second term
on the right-hand side by choosing the gauge in which $\dot J_\alpha=0$.
However, this manipulation cannot be justified.
Indeed, when we take the derivative with respect to $J_\alpha$ for fixed $X^\mu$,
we notice that the deviation between the neighboring geodesics with $J_\alpha$
and $J_\alpha+\Delta J_\alpha$ for the same given values of $X^\mu$ 
grows secularly proportionally to $|\tau|$. Therefore, both 
$({\partial \Hint}/{\partial J_{\alpha}})_{X}$ 
and $({\partial w^\beta}/{\partial J_{\alpha}})_{X}$ show secular growth. 
(The situation is completely different when the values of
$w^\alpha$, rather than those of $X^\mu$, are fixed;
in this case there is no such secular growth.) 
Therefore, the second term on the right-hand side of \eqref{dHdJ1} 
becomes $0 \times \infty$ when $\tau\to\infty$, and hence 
it is not clear whether or not we can eliminate this term,
even if we choose the gauge in which $\dot J_\alpha=0$. 
Even if we assume that we can neglect the contribution from the second term, 
the first term on the right-hand side of Eq.~\eqref{dHdJ1}
does not appear to be well defined either. 
Here one may propose to use the fact
that for an arbitrary function $f(X,J)$, the identity
\begin{equation}\label{patata}
 \frac{\partial}{\partial J_{\alpha}} \int \ud\tau f(X,J)|_{X=X_{(0)}(\tau)}
 = \int \ud\tau \left(\frac{\partial f(X,J)}
 {\partial J_{\alpha}} \right)_{\!X=X_{(0)}(\tau)}
\end{equation}
holds, recalling that $X_{(0)}^{i}(\tau)$ is constant for $i=1,2,3$,
and that $X_{(0)}^{0}(\tau)=(\tau-\tau_{\rm I})+\Delta \tau$,
where $\Delta\tau$ is a function of $r_{\rm I},\theta_{\rm I}$ and $P^{(0)}_\mu$. 
Using this expression, the first term on the right-hand side of Eq.~\eqref{dHdJ1}
might be further rewritten as
\begin{equation}
 \left\langle\left(\frac{\partial \Hint}
 {\partial J_{\alpha}}\right)_{\!X}\right\rangle
 = \frac{\partial  \left\langle \Hint \right\rangle}{\partial J_{\alpha}} \,,
 \label{dHdJ2}
\end{equation}
where recall that $\langle \Hint \rangle$ depends both on $J_\alpha$
and the source orbit $\gamma$. The values of the actions
for the source orbit, $J_\alpha^{(\gamma)}$, should not be identified
with $J_\alpha$ before taking the derivative with respect to $J_\alpha$.
Therefore, the right-hand side in Eq.~\eqref{dHdJ2} 
requires to evaluate the long-time average while substituting different orbits
into the two arguments of the Green's function.
Since those orbits deviate from each other at large $|\tau|$,
the resulting expression is not well defined.

\section{Fourier decompositions}
\label{app:Fourier}

In this appendix we provide details on the Fourier decompositions
(series and transform) with respect to the angles of the interaction
Hamiltonian \eqref{Hamiltonian-int} and the canonical variables.

\subsection{Interaction Hamiltonian}
\label{app:Fourier1}

First, we prove that the Fourier expansion of 
the interaction Hamiltonian $H^{(1)}(w,J;\gamma)$ 
is given by the expression \eqref{eq:startHi2}.
Our starting point is the Fourier decomposition of 
the time-symmetric Green-like function $G(w,J;w',J')$,
defined from Eq.~\eqref{symmetricG} through the
canonical transformation from $(x^\mu,u_\mu)$ to $(w^\alpha,J_\alpha)$. 

It is known that the (regularized) time-symmetric tensorial Green's function 
$G^{\mu\nu\,\rho\sigma}_{\rm (sym-S)}(x; x')$ is invariant 
under displacements along Killing directions of the 
Kerr metric. Since the Kerr metric is stationary and
axisymmetric, the Green's function is invariant under constant shifts
$t \to t + \Delta t$ and $\phi \to \phi + \Delta \phi$ of the coordinates
$t$ and $\phi$, all other components of $(x^\mu,u_\mu)$ being
held fixed. Given the relationship 
$x^{\mu} (w,J) 
= 
\delta^{\mu}_{t} w^{t} + \delta^{\mu}_{\phi} w^{\phi}
+ {\hat x}^{\mu} (w^{r},w^{\theta},J)$ 
that follows from Eq.~\eqref{wtransform}, 
this shift symmetry implies
an equivalent invariance under constant shifts ${w}^{t} \to {w}^{t} + \Delta t$
and ${w}^{\phi} \to {w}^{\phi} + \Delta \phi$ of the angle variables
$w^t$ and $w^\phi$, all other components of $(w^\alpha,J_\alpha)$ being
held fixed \cite{Hinderer:2008dm}.
This, in turn, implies the following symmetry property of the Green-like function
\eqref{symmetricG}:
\begin{equation}\label{Glike-symmetry}
G({w}^{t} + \Delta t,{\bf w},{w}^{\phi} + \Delta \phi,J\,;
{w'}^{t} + \Delta t,{\bf w'},{w'}^{\phi} + \Delta \phi,J') 
=
G(w,J;w',J')\,,
\end{equation} 
where ${\bf w} \equiv (w^{r},w^{\theta})$ and ${\bf w}' \equiv ({w'}^{r},{w'}^{\theta})$.
Therefore, $G(w,J,w',J')$ depends on $w^{t}$ and ${w'}^{t}$,
as well as on $w^{\phi}$ and ${w'}^{\phi}$, only through
the differences $w^{t} - {w'}^{t}$ and $w^{\phi} - {w'}^{\phi}$. 
Consequently, the Fourier decomposition of $G(w,J;w',J')$ 
takes the form 
\begin{equation}\label{Fourier-Glike}
G(w,J;w',J') = 
\frac{1}{2 \pi} \int^{+\infty}_{-\infty} 
\rmd \varpi\, \sum_{m = -\infty}^{+\infty} 
e^{-\ui\varpi (w^{t} - {w'}^{t}) + \ui m (w^{\phi} - {w'}^{\phi})} \,
G_{\varpi m}({\bf w},J;{\bf w'},J')\,,
\end{equation}
with the reality condition $G_{\varpi (-m)} ({\bf w},J;{\bf w'},J')
= G_{\varpi m}^{\ast} ({\bf w},J;{\bf w'},J')$. 
Since the Fourier coefficients $G_{\varpi m}({\bf w},J;{\bf w'},J')$ 
are periodic functions of ${\bf w}$ and ${\bf w'}$ 
with period $2 \pi$, we may perform their Fourier series
expansions, 
\begin{equation}\label{Fourier-Glike-rth}
G_{\varpi m}({\bf w},J;{\bf w'},J') =
\sum_{{\bf n} = -\infty}^{+\infty}\,\sum_{{\bf n}' = -\infty}^{+\infty} 
e^{-\ui ({\bf n} \cdot {\bf w} + {\bf n'} \cdot {\bf w'})} \, 
G_{\varpi m, {\bf n} {\bf n}'}(J;J') \, ,
\end{equation}
where ${\bf n} \equiv (n_{r},n_{\theta})$ and 
${\bf n}' \equiv (n_{r}',n_{\theta}')$ are doublet of integers, 
the Fourier coefficients obeying the reality condition
$G_{\varpi m, (-n_{r}, n_{\theta}) {\bf n'}}(J;J')
= G_{\varpi m, (n_{r}, n_{\theta}) {\bf n'}}^{\ast}(J;J')$.

Next, we substitute for Eqs.~\eqref{Fourier-Glike} and \eqref{Fourier-Glike-rth}
into the expression \eqref{eq:startHi2} for the interaction Hamiltonian to obtain 
\begin{align}\label{Fourier-Hint0}
H^{(1)}(w,J;\gamma)
&= 
-\frac{1}{4 \pi} \int^{+\infty}_{-\infty} 
\rmd \varpi\, \sum_{(m,{\bf n})= -\infty}^{+\infty} \,
e^{-\ui\varpi w^{t} + \ui (m w^{\phi} - {\bf n} \cdot {\bf w} ) } \cr 
&\times \left[ \sum_{\bf n' = -\infty}^{+\infty}
\int^{+\infty}_{-\infty} \rmd \tau' \,
e^{\ui \varpi w^{t}_\subgamma(\tau') 
- \ui (m w_\subgamma^{\phi}(\tau') 
+ {\bf n'} \cdot {\bf w}_\subgamma(\tau'))} \,
G_{\varpi m, {\bf n} {\bf n'}} (J;J_\subgamma) \right] ,
\end{align}
where the subscript $\subgamma$ refers to the canonical variables 
for the source orbit $\gamma$, 
and we ignore the uncontrolled term at $O(\eta^2)$. 
By substituting the solution ${w}_{(0)}^{\alpha}(\tau) 
= \omega^{\alpha}_{(0)}(J^{(0)}) \,\tau 
+ {w}_{(0)\textrm{I}}^{\alpha}$
of the Kerr geodesic equations of motion into the square brackets 
in Eq.~\eqref{Fourier-Hint0}, we find
\begin{align}\label{Fourier-discretize-gamma}
&\sum_{{\bf n'} = -\infty}^{+\infty} \int^{+\infty}_{-\infty} \rmd \tau' \,
e^{\ui [\varpi w^{t}_\subgamma(\tau') 
- m w_\subgamma^{\phi}(\tau') 
- {\bf n'} \cdot {\bf w}_\subgamma(\tau')]} \, 
G_{\varpi m, {\bf n} {\bf n'}}(J;J^\subgamma)  \cr 
&=
e^{\ui \zeta^{\subgamma}_{m {\bf n'}}(\tau'_{0})}
\sum_{{\bf n'} = -\infty}^{+\infty}
{2 \pi z_{(0)}^\subgamma}
\,\delta \bigl( 
\varpi - \varpi_{m {\bf n'}}^{\subgamma}(J^\subgamma_{(0)}) 
\bigr) \, G_{\varpi m, {\bf n} {\bf n'} }(J;J^\subgamma_{(0)})\,,
\end{align}
where $z_{(0)} = 1 / \omega^t_{(0)}$ is the Kerr redshift variable. 
Here, making use of the orbital frequencies \eqref{Omega-i}, 
we defined the discretized frequency parameter $\varpi_{m {\bf n'}}^{\subgamma}$
of $\gamma$, as well as the initial phase function $\zeta_{m {\bf n'}}^{\subgamma}$, by
\begin{subequations}
\begin{align}
\varpi_{m {\bf n'}}^{\subgamma}(J^\subgamma_{(0)}) 
&\equiv m \Omega^{\phi}_{\subgamma(0)} 
+
n_r' \Omega^{r}_{\subgamma(0)} 
+
n_{\theta}' \Omega^{\theta}_{\subgamma(0)} \, ,
\label{descrete-varpi-gamma} \\
\zeta_{m {\bf n'}}^{\subgamma}(\tau'_{0}) 
&\equiv {w}_{(0)\textrm{I}}^{\subgamma t} 
- m {w}_{(0)\textrm{I}}^{\subgamma \phi}
- n'_{r} {w}_{(0)\textrm{I}}^{\subgamma r}
- n'_{\theta} {w}_{(0)\textrm{I}}^{\subgamma \theta} \, .
\label{def-zeta-gamma}
\end{align}
\end{subequations}
Then, substituting for Eq.~\eqref{Fourier-discretize-gamma} 
back into \eqref{Fourier-Hint0}, and eliminating the delta function 
via the integration with respect to $\varpi$, 
we obtain the following Fourier series decomposition of the interaction Hamiltonian: 
\begin{align}\label{Fourier-Hint}
H^{(1)}(w,J;\gamma)
= 
-\frac{z_{(0)}^\subgamma }{2 \pi} \sum_{m,\bn,\bn'}
e^{\ui \zeta^{\subgamma}_{m {\bf n'}}(\tau'_{\,0})}\, 
G_{\varpi^{\subgamma}_{m {\bf n'}} m, {\bf n}{\bf n'}} (J;J^\subgamma_{(0)}) \,
e^{-\ui \varpi_{m {\bf n'}}^{\subgamma} w^{t} 
+ \ui (m w^{\phi} - {\bf n} \cdot {\bf w})} \, .
\end{align}
Since this expression is rather large, we conveniently rewrite it as
\begin{equation}
 \Hint(w,J;\gamma)
 = 
 z_{(0)}^\subgamma
 \sum_{{\bf l},{\bf l'}}
 \cG_{{\bf l},{\bf l'}}(J,J^{(0)},\varpi_{\bf l'}(\omega_{(0)}(\tau)))
 \, e^{-\ui\varpi_{\bf l'}(\omega_{(0)}(\tau)) w^{t}
 + 2\pi \ui\,{\bf l}\cdot{\bw}} \,,
\end{equation}
where ${\bf l} \equiv (-n_r,\,-n_\theta,\,m)$, 
$\bw=(w^r,w^\theta,w^\phi)$,
$\varpi_{\bf l}(\omega) \equiv \varpi_{m {\bf n'}}^{\subgamma}$,
and we introduced the Fourier coefficients
$
\cG_{{\bf l},{\bf l'}}(J,J^{(0)},\varpi_{\bf l'}(\omega_{(0)}(\tau)))
\equiv 
-(e^{\ui \zeta^{\subgamma}_{m {\bf n'}}(\tau'_{0})} / 2 \pi) \, 
G_{\varpi^{\subgamma}_{m {\bf n'}} m, {\bf n}{\bf n'}} (J;J^\subgamma_{(0)})
$. \vspace{-0.1cm}
With those notations, we recover Eq.~\eqref{eq:startHi2}.

We note that even though the Fourier coefficients \vspace{-0.1cm}
$G_{\varpi_{m {\bf n'}^{\subgamma}} m, {\bf n}{\bf n'}}(J;J')$ 
are symmetric under the exchange of their two arguments,
$J_{\alpha}$ cannot be equated to $J^\subgamma_{\alpha(0)} = 
J^\subgamma_{\alpha}(\tau)$ in Eq.~\eqref{Fourier-Hint}.
Indeed, while the field degrees of freedom 
$(w^{\alpha},J_{\alpha})$ are the canonical variables 
in phase space, the source degrees of freedom 
$J^\subgamma_{\alpha}(\tau)$ 
are the solutions constrained by Eq.~\eqref{Jomeganorm}.

\subsection{Canonical variables}
\label{app:Fourier2}

Next, we prove the statement below \eqref{wrexpansion},
namely that in the spherical orbit limit $J_r \to 0$,
the coefficients $x_{(n)}^\mu$ and $u_{(n)\mu}$
\vspace{-0.15cm}
of the Fourier series expansions of the canonical position
and momentum with respect to the radial angle
are of $O(J_r^{|n|/2})$. We also prove that $J_r = O(e^2)$ in that limit.
These results are established in the test-particle limit only.

Equations \eqref{def-J0} and \eqref{wtransform} show that 
the relations between the Boyer-Lindquist canonical variables 
$(x^\mu,u_\mu)$ and the action-angle variables $(w^\alpha,J_\alpha)$ 
must take the form
\begin{subequations}\label{pif}
\begin{align}
 x^\mu(w,J) &= \delta^\mu_t w^t + \delta^\mu_\phi w^\phi
 + \sum_{n,m} x_{(n,m)}^\mu(J) \, e^{\ui (n w^r + m w^\theta)} \, , \\
 u_j(w,J) &= \sum_{n,m} u_{(n,m)j}(J) \, e^{\ui (n w^r + m w^\theta)} \, ,
\end{align}
\end{subequations}
for $j = r,\theta$, as well as $u_t(w,J) = -J_t$ and $u_\phi(w,J) = J_\phi$.
Since the solutions of the equations of motion for bound timelike geodesics
in Kerr are simply given by
$w^\alpha = \omega_{(0)}^\alpha(J) \, \tau + \text{const.}$
and $J_\alpha = \text{const.}$, Eqs.~\eqref{pif} can be rewritten as
\begin{subequations}\label{sols}
\begin{align}
 &x^\mu(\tau) = \delta^\mu_t \omega^t \tau + \delta^\mu_\phi \omega^\phi \tau
 + \sum_{n,m} x_{(n,m)}^\mu \, e^{\ui (n \omega^r + m \omega^\theta) \tau} \,, \\
 &u_j(\tau) = \sum_{n,m} u_{(n,m)j} \, e^{\ui (n \omega^r + m \omega^\theta) \tau} 
 \,.
\end{align}
\end{subequations}
On the other hand, these expressions should be solutions 
of the equations of motion \eqref{H-eq0} for the canonical variables
$(x^\mu,u_\mu)$ in the test-mass limit, namely 
\begin{subequations}\label{eom}
\begin{align}
 &\dot{x}^\mu(\tau) = g^{\mu\nu}_{(0)}(r(\tau),\theta(\tau)) \, u_\nu(\tau)
 \,,  \label{xdot} \\
 &\dot{u}_\mu(\tau) = \Gamma^\rho_{(0)\mu\nu}(r(\tau),\theta(\tau))
 \, \dot{x}^\nu(\tau) u_\rho(\tau) \,.
\end{align}
\end{subequations}
Substituting for the solutions \eqref{sols}
into the equations of motion \eqref{eom},
one can solve for the coefficients $X_{(n,m)} \equiv \{x_{(n,m)}^\mu,u_{(n,m)j}\}$.
The equations for the $r$ and $\theta$ components can be solved
irrespective of those for the $t$ and $\phi$ components.
Extracting each Fourier component from these equations,
we obtain (schematically) the recursion relations
\begin{equation}\label{recursion}
 X_{(n,m)} = \mathop{\sum_{n_1,\cdots,n_N}}_{\sum_i n_i = n}
 \,\, \mathop{\sum_{m_1,\cdots,m_M}}_{\sum_j m_j = m}
 C_{n,m;n_1,\cdots,n_N,m_1,\cdots,m_M} \, X_{(n_1,m_1)} \cdots X_{(n_N,m_M)} \,.
\end{equation}

For an orbit with eccentricity 
$e\equiv(r_\text{max}-r_\text{min})/(r_\text{max}+r_\text{min})$, 
where $r_\text{max}$ and $r_\text{min}$ denote
the coordinate radii at the apastron and the periastron, respectively, 
we must have $X_{(\pm 1,m)} = O(e)$.
Then, by solving the recursion relations \eqref{recursion}
we obtain $X_{(n,m)} = O(e^{|n|})$.
Moreover, we know that $X_{(0,m\neq0)}$ cannot be $O(e^0)$
because it must vanish in the spherical orbit limit $e \to 0$.
Since the dominant contribution to $X_{(0,m\neq0)}$ comes from
products of terms like $X_{(1,m')}$ and $X_{(-1,m-m')}$,
we find $X_{(0,m\neq0)} = O(e^2)$.
For the same reason we have $u_{(0,0)r} = O(e^2)$, which implies
\begin{equation}\label{u_r}
 u_r(\tau) = O(e) \,.
\end{equation}

The Fourier coefficients for the $t$ and $\phi$ components
can be obtained by substituting the solutions obtained
for the $r$ and $\theta$ components into the remaining equations of motion.
We find that the same power counting in the eccentricity $e$ holds
for the $t$ and $\phi$ components, such that
\begin{equation}\label{xn_un}
 x^\mu_{(n)} \equiv \frac{1}{2\pi}
 \int_0^{2\pi} x^\mu(w,J) \, e^{-\ui n w^r} \, \ud w^r = O(e^{|n|}) \,,
\end{equation}
and similarly for the Fourier coefficients $u_{(n)\mu}$ of the canonical momentum.

Finally, we turn to the radial action integral. From its definition \eqref{def-J0}
and the equation of motion \eqref{xdot}, we have
\begin{equation}
 J_r 
 = \frac{1}{2\omega^r T_r} \int_{-T_r}^{T_r} u_r(\tau) \, \dot{r}(\tau) \, \ud \tau
 = \lim_{T\to\infty} \frac{1}{2\omega^r T} \int_{-T}^T
 g_{(0)}^{rr}(r(\tau),\theta(\tau)) {[u_r(\tau)]}^2 \, \ud \tau \,,
\end{equation}
where $T_r \equiv 2\pi / \omega^r$ is the $\tau$-period of the radial motion,
and we have used the fact that $J_r$ is constant
to replace the integral over one radial period by the long-term average.
Then, using Eq.~\eqref{u_r} and the fact that
$g_{(0)}^{rr} = \Delta / \Sigma = O(e^0)$,
\vspace{-0.1cm}
we find that $J_r = O(e^2)$ along the orbit. 
Combining this result with \eqref{xn_un},\vspace{-0.1cm}
we conclude that $x_{(n)}^\mu = O(J_r^{|n|/2})$,
and similarly $u_{(n)\mu} = O(J_r^{|n|/2})$ in the spherical orbit limit.

\bibliographystyle{apsrev}

\begin{thebibliography}{99}

\bibitem{Abbott:2016blz} 
  B.~P.~Abbott {\it et al.},
  Phys.\ Rev.\ Lett.\  {\bf 116}, 061102 (2016),
  arXiv:1602.03837 [gr-qc].

\bibitem{Abbott:2016bis} 
  B.~P.~Abbott {\it et al.},
  Phys.\ Rev.\ Lett.\  {\bf 116}, 241103 (2016),
  arXiv:1606.04855 [gr-qc].

\bibitem{Abbott:2016ter} 
  B.~P.~Abbott {\it et al.},
  Phys. Rev. X {\bf 6}, 041015 (2016),
  arXiv:1606.04856 [gr-qc].

\bibitem{Seoane:2013qna} 
  P.~{Amaro-Seoane} {\it et al.}
  (2013),
  arXiv:1305.5720 [astro-ph.CO].

\bibitem{Seoane:2017drz} 
  P.~{Amaro-Seoane} {\it et al.}
  (2017),
  arXiv:1702.00786 [astro-ph.IM].


\bibitem{Seto:2001qf} 
  N.~Seto, S.~Kawamura and T.~Nakamura,
  Phys.\ Rev.\ Lett. {\bf 87}, 221103 (2001),
  astro-ph/0108011.

\bibitem{BBO}
  S.~Phinney {\it et al.},
  {\it The Big Bang Observer: Direct Detection of Gravitational Waves from the Birth of
  the Universe to the Present}, NASA Mission Concept Study (2014).


\bibitem{BDECIGO} 
 S. Sato, ``Space Gravitational Wave Observatory,''
(2016), presented at JPS 2016 Autumn Meeting, University of Miyazaki. 


\bibitem{Nakamura:2016hna} 
  T.~Nakamura {\it et al.},
  PTEP {\bf 2016}, 093E01 (2016),
  arXiv:1607.00897 [astro-ph.HE].


\bibitem{Armano:2016} 
  M.~Armano {\it et al.},
  Phys.\ Rev.\ Lett.\  {\bf 116}, 231101 (2016).

\bibitem{AmaroSeoane:2007aw} 
  P.~Amaro-Seoane, J.~R.~Gair, M.~Freitag, M.~C.~Miller, I.~Mandel, C.~J.~Cutler and S.~Babak,
  Class.\ Quant.\ Grav.\  {\bf 24}, R113 (2007),
  astro-ph/0703495 [astro-ph].

\bibitem{Barausse:2014tra} 
  E.~Barausse, V.~Cardoso and P.~Pani,
  Phys.\ Rev.\ D {\bf 89}, 104059 (2014),
  arXiv:1404.7149 [gr-qc].

\bibitem{Berti:2015} 
  E.~Berti {\it et al.},
  Class.\ Quant.\ Grav.\  {\bf 32}, 243001 (2015),
  arXiv:1501.07274 [gr-qc].

\bibitem{Barack:2004} 
  L.~Barack and C.~Cutler,
  Phys.\ Rev.\ D {\bf 69}, 082005 (2004),
  gr-qc/0310125.

\bibitem{Cutler:1992tc} 
  C.~Cutler {\it et al.},
  Phys.\ Rev.\ Lett.\  {\bf 70}, 2984 (1993),
  astro-ph/9208005.

\bibitem{Lindblom:2008} 
  L.~Lindblom, B.~J.~Owen and D.~A.~Brown,
  Phys.\ Rev.\ D {\bf 78}, 124020 (2008),
  arXiv:0809.3844 [gr-qc].

\bibitem{Buonanno:2014aza} 
  A.~Buonanno and B.~S.~Sathyaprakash,
  in {\it General Relativity and Gravitation: A Centennial Perspective}, edited by A. Ashtekar, B. K. Berger, J. Isenberg and M. MacCallum (Cambridge University Press, Cambridge, 2015) p. 287,
  arXiv:1410.7832 [gr-qc].

\bibitem{Teukolsky:1973ha} 
  S.~A.~Teukolsky,
  Astrophys.\ J.\  {\bf 185}, 635 (1973).

\bibitem{Kojima:1984cj} 
  Y.~Kojima and T.~Nakamura,
  Prog.\ Theor.\ Phys.\  {\bf 71}, 79 (1984).

\bibitem{Mino:1997bx} 
  Y.~Mino, M.~Sasaki, M.~Shibata, H.~Tagoshi and T.~Tanaka,
  Prog.\ Theor.\ Phys.\ Suppl.\  {\bf 128}, 1 (1997),
  gr-qc/9712057.

\bibitem{Sasaki:2003xr} 
  M.~Sasaki and H.~Tagoshi,
  Living Rev.\ Rel.\  {\bf 6}, 6 (2003),
  gr-qc/0306120.

\bibitem{Carter:1968rr} 
  B.~Carter,
  Phys.\ Rev.\  {\bf 174}, 1559 (1968).

\bibitem{Mino:2003yg} 
  Y.~Mino,
  Phys.\ Rev.\ D {\bf 67}, 084027 (2003),
  gr-qc/0302075.

\bibitem{Sago:2005gd} 
  N.~Sago, T.~Tanaka, W.~Hikida and H.~Nakano,
  Prog.\ Theor.\ Phys.\  {\bf 114}, 509 (2005),
  gr-qc/0506092.

\bibitem{Sago:2005fn} 
  N.~Sago, T.~Tanaka, W.~Hikida, K.~Ganz and H.~Nakano,
  Prog.\ Theor.\ Phys.\  {\bf 115}, 873 (2006),
  gr-qc/0511151.

\bibitem{Ganz:2007rf} 
  K.~Ganz, W.~Hikida, H.~Nakano, N.~Sago and T.~Tanaka,
  Prog.\ Theor.\ Phys.\  {\bf 117}, 1041 (2007),
  gr-qc/0702054.

\bibitem{Fujita:2009us} 
  R.~Fujita, W.~Hikida and H.~Tagoshi,
  Prog.\ Theor.\ Phys.\  {\bf 121}, 843 (2009),
  arXiv:0904.3810 [gr-qc].

\bibitem{Sago:2015rpa} 
  N.~Sago and R.~Fujita,
  PTEP {\bf 2015}, 073E03 (2015),
  arXiv:1505.01600 [gr-qc].

\bibitem{Hughes:2005qb} 
  S.~A.~Hughes, S.~Drasco, E.~E.~Flanagan and J.~Franklin,
  Phys.\ Rev.\ Lett.\  {\bf 94}, 221101 (2005),
  gr-qc/0504015.

\bibitem{Drasco:2005kz} 
  S.~Drasco and S.~A.~Hughes,
  Phys.\ Rev.\ D {\bf 73}, 024027 (2006),
  Erratum-ibid. {\bf 88}, 109905 (2013),
  Erratum-ibid. {\bf 90}, 109905 (2014),
  gr-qc/0509101.

\bibitem{Cutler:1994pb} 
  C.~Cutler, D.~Kennefick and E.~Poisson,
  Phys.\ Rev.\ D {\bf 50}, 3816 (1994).

\bibitem{Finn:2000sy} 
  L.~S.~Finn and K.~S.~Thorne,
  Phys.\ Rev.\ D {\bf 62}, 124021 (2000),
  gr-qc/0007074.

\bibitem{Hughes:2001jr} 
  S.~A.~Hughes,
  Phys.\ Rev.\ D {\bf 64}, 064004 (2001),
  Erratum-ibid. {\bf 88}, 109902 (2013),
  gr-qc/0104041.

\bibitem{Mino:1996nk} 
  Y.~Mino, M.~Sasaki and T.~Tanaka,
  Phys.\ Rev.\ D {\bf 55}, 3457 (1997),
  gr-qc/9606018.

\bibitem{Quinn:1996am} 
  T.~C.~Quinn and R.~M.~Wald,
  Phys.\ Rev.\ D {\bf 56}, 3381 (1997),
  gr-qc/9610053.

\bibitem{Barack:1999wf} 
  L.~Barack and A.~Ori,
  Phys.\ Rev.\ D {\bf 61}, 061502 (2000),
  gr-qc/9912010.

\bibitem{Lousto:1999za} 
  C.~O.~Lousto,
  Phys.\ Rev.\ Lett.\  {\bf 84}, 5251 (2000),
  gr-qc/9912017.

\bibitem{Burko:2000xx} 
  L.~M.~Burko,
  Phys.\ Rev.\ Lett.\  {\bf 84}, 4529 (2000),
  gr-qc/0003074.

\bibitem{Barack:2001gx} 
  L.~Barack, Y.~Mino, H.~Nakano, A.~Ori and M.~Sasaki,
  Phys.\ Rev.\ Lett.\  {\bf 88}, 091101 (2002),
  gr-qc/0111001.

\bibitem{Mino:2001mq} 
  Y.~Mino, H.~Nakano and M.~Sasaki,
  Prog.\ Theor.\ Phys.\  {\bf 108}, 1039 (2003),
  gr-qc/0111074.

\bibitem{Anderson:2005gb} 
  W.~G.~Anderson and A.~G.~Wiseman,
  Class.\ Quant.\ Grav.\  {\bf 22}, S783 (2005),
  gr-qc/0506136.

\bibitem{Vega:2011wf} 
  I.~Vega, B.~Wardell and P.~Diener,
  Class.\ Quant.\ Grav.\  {\bf 28}, 134010 (2011),
  arXiv:1101.2925 [gr-qc].

\bibitem{Pound:2013faa} 
  A.~Pound, C.~Merlin and L.~Barack,
  Phys.\ Rev.\ D {\bf 89}, 024009 (2014),
  arXiv:1310.1513 [gr-qc].

\bibitem{Barack:2007tm} 
  L.~Barack and N.~Sago,
  Phys.\ Rev.\ D {\bf 75}, 064021 (2007),
  gr-qc/0701069.

\bibitem{Detweiler:2008ft} 
  S.~Detweiler,
  Phys.\ Rev.\ D {\bf 77}, 124026 (2008),
  arXiv:0804.3529 [gr-qc].

\bibitem{Sago:2008id} 
  N.~Sago, L.~Barack and S.~Detweiler,
  Phys.\ Rev.\ D {\bf 78}, 124024 (2008),
  arXiv:0810.2530 [gr-qc].

\bibitem{Barack:2010tm} 
  L.~Barack and N.~Sago,
  Phys.\ Rev.\ D {\bf 81}, 084021 (2010),
  arXiv:1002.2386 [gr-qc].

\bibitem{Wardell:2014kea} 
  B.~Wardell, C.~R.~Galley, A.~Zenginoglu, M.~Casals, S.~R.~Dolan and A.~C.~Ottewill,
  Phys.\ Rev.\ D {\bf 89}, 084021 (2014),
  arXiv:1401.1506 [gr-qc].

\bibitem{Osburn:2014hoa} 
  T.~Osburn, E.~Forseth, C.~R.~Evans and S.~Hopper,
  Phys.\ Rev.\ D {\bf 90}, 104031 (2014),
  arXiv:1409.4419 [gr-qc].

\bibitem{Merlin:2014qda} 
  C.~Merlin and A.~G.~Shah,
  Phys.\ Rev.\ D {\bf 91}, 024005 (2015), 
  arXiv:1410.2998 [gr-qc].

\bibitem{Shah:2012gu} 
  A.~G.~Shah, J.~L.~Friedman and T.~S.~Keidl,
  Phys.\ Rev.\ D {\bf 86}, 084059 (2012),
  arXiv:1207.5595 [gr-qc].

\bibitem{vandeMeent:2015lxa} 
  M.~van~de~Meent and A.~G.~Shah,
  Phys.\ Rev.\ D {\bf 92}, 064025 (2015),
  arXiv:1506.04755 [gr-qc].

\bibitem{vandeMeent:2016pee} 
  M.~van~de~Meent,
  Phys.\ Rev.\ D {\bf 94}, 044034 (2016),
  arXiv:1606.06297 [gr-qc].

\bibitem{Barack:2009ux} 
  L.~Barack,
  Class.\ Quant.\ Grav.\  {\bf 26}, 213001 (2009),
  arXiv:0908.1664 [gr-qc].

\bibitem{Poisson:2011nh} 
  E.~Poisson, A.~Pound and I.~Vega,
  Living Rev.\ Rel.\  {\bf 14}, 7 (2011),
  arXiv:1102.0529 [gr-qc].

\bibitem{Harte:2014wya} 
  A.~I.~Harte,
  Fund.\ Theor.\ Phys.\  {\bf 179}, 327 (2015),
  arXiv:1405.5077 [gr-qc].

\bibitem{Pound:2015tma} 
  A.~Pound,
  Fund.\ Theor.\ Phys.\  {\bf 179}, 399 (2015),
  arXiv:1506.06245 [gr-qc].

\bibitem{Wardell:2015kea} 
  B.~Wardell,
  Fund.\ Theor.\ Phys.\  {\bf 179}, 487 (2015),
  arXiv:1501.07322 [gr-qc].

\bibitem{Detweiler:2002mi} 
  S.~Detweiler and B.~F.~Whiting,
  Phys.\ Rev.\ D {\bf 67}, 024025 (2003),
  gr-qc/0202086.

\bibitem{Mino:2005an} 
  Y.~Mino,
  Prog.\ Theor.\ Phys.\  {\bf 113}, 733 (2005),
  gr-qc/0506003.

\bibitem{Tanaka:2005ue} 
  T.~Tanaka,
  Prog.\ Theor.\ Phys.\ Suppl.\  {\bf 163}, 120 (2006),
  gr-qc/0508114.

\bibitem{Hinderer:2008dm} 
  T.~Hinderer and {\'E}.~{\'E}.~Flanagan,
  Phys.\ Rev.\ D {\bf 78}, 064028 (2008),
  arXiv:0805.3337 [gr-qc].

\bibitem{Isoyama:2012bx} 
  S.~Isoyama, R.~Fujita, N.~Sago, H.~Tagoshi and T.~Tanaka,
  Phys.\ Rev.\ D {\bf 87}, 024010 (2013),
  arXiv:1210.2569 [gr-qc].

\bibitem{Flanagan:2010cd} 
  {\'E}.~{\'E}.~Flanagan and T.~Hinderer,
  Phys.\ Rev.\ Lett.\  {\bf 109}, 071102 (2012),
  arXiv:1009.4923 [gr-qc].

\bibitem{Flanagan:2012kg} 
  {\'E}.~{\'E}.~Flanagan, S.~A.~Hughes and U.~Ruangsri,
  Phys.\ Rev.\ D {\bf 89}, 084028 (2014),
  arXiv:1208.3906 [gr-qc].

\bibitem{Isoyama:2013yor} 
  S.~Isoyama, R.~Fujita, H.~Nakano, N.~Sago and T.~Tanaka,
  PTEP {\bf 2013}, 063E01 (2013),
  arXiv:1302.4035 [gr-qc].

\bibitem{Ruangsri:2013hra} 
  U.~Ruangsri and S.~A.~Hughes,
  Phys.\ Rev.\ D {\bf 89}, 084036 (2014),
  arXiv:1307.6483 [gr-qc].

\bibitem{Berry:2016bit} 
  C.~P.~L.~Berry, R.~H.~Cole, P.~Ca{\~n}izares and J.~R.~Gair,
  Phys.\ Rev.\ D {\bf 94}, 124042 (2016)
  arXiv:1608.08951 [gr-qc].


\bibitem{vandeMeent:2013sza} 
  M.~van~de~Meent,
  Phys.\ Rev.\ D {\bf 89}, 084033 (2014),
  arXiv:1311.4457 [gr-qc].

\bibitem{Pound:2012nt} 
  A.~Pound,
  Phys.\ Rev.\ Lett.\  {\bf 109}, 051101 (2012),
  arXiv:1201.5089 [gr-qc].

\bibitem{Pound:2012dk} 
  A.~Pound,
  Phys.\ Rev.\ D {\bf 86}, 084019 (2012),
  arXiv:1206.6538 [gr-qc].

\bibitem{Gralla:2012} 
  S.~Gralla,
  Phys.\ Rev.\ D {\bf 85}, 124011 (2012),
  arXiv:1203.3189 [gr-qc].

\bibitem{Pound:2014xva} 
  A.~Pound and J.~Miller,
  Phys.\ Rev.\ D {\bf 89}, 104020 (2014),
  arXiv:1403.1843 [gr-qc].

\bibitem{Pound:2015wva} 
  A.~Pound,
  Phys.\ Rev.\ D {\bf 92}, 104047 (2015),
  arXiv:1510.05172 [gr-qc].

\bibitem{Miller:2016hjv} 
  J.~Miller, B.~Wardell and A.~Pound,
  Phys.\ Rev.\ D {\bf 94}, 104018 (2016),
  arXiv:1608.06783 [gr-qc].

\bibitem{Shah:2010bi} 
  A.~G.~Shah, T.~S.~Keidl, J.~L.~Friedman, D.~H.~Kim and L.~R.~Price,
  Phys.\ Rev.\ D {\bf 83}, 064018 (2011)
  arXiv:1009.4876 [gr-qc].

\bibitem{Barack:2009ey} 
  L.~Barack and N.~Sago,
  Phys.\ Rev.\ Lett.\  {\bf 102}, 191101 (2009),
  arXiv:0902.0573 [gr-qc].

\bibitem{Isoyama:2014mja} 
  S.~Isoyama, L.~Barack, S.~R.~Dolan, A.~Le~Tiec, H.~Nakano, A.~G.~Shah, T.~Tanaka and N.~Warburton,
  Phys.\ Rev.\ Lett.\ {\bf 113}, 161101 (2014),
  arXiv:1404.6133 [gr-qc].

\bibitem{Barack:2010} 
  L.~Barack, T. Damour and N.~Sago,
  Phys.\ Rev.\ D {\bf 82}, 084036 (2010),
  arXiv:1008.0935 [gr-qc].

\bibitem{Barack:2011} 
  L.~Barack and N.~Sago,
  Phys.\ Rev.\ D {\bf 83}, 084023 (2011),
  arXiv:1101.3331 [gr-qc].

\bibitem{vandeMeent:2016hel} 
  M.~van de Meent,
  Phys.\ Rev.\ Lett.\  {\bf 118}, 011101 (2017)
  arXiv:1610.03497 [gr-qc].

\bibitem{Dolan:2013roa} 
  S.~R.~Dolan, N.~Warburton, A.~I.~Harte, A.~Le~Tiec, B.~Wardell and L.~Barack,
  Phys.\ Rev.\ D {\bf 89}, 064011 (2014),
  arXiv:1312.0775 [gr-qc].

\bibitem{Akcay:2016dku} 
  S.~Akcay, D.~Dempsey and S.~R.~Dolan,
  Class.\ Quant.\ Grav.\  {\bf 34}, 084001 (2017),
  arXiv:1608.04811 [gr-qc].

\bibitem{Dolan:2014pja} 
  S.~R.~Dolan, P.~Nolan, A.~C.~Ottewill, N.~Warburton and B.~Wardell,
  Phys.\ Rev.\ D {\bf 91}, 023009 (2015),
  arXiv:1406.4890 [gr-qc].

\bibitem{Nolan:2015vpa} 
  P.~Nolan, C.~Kavanagh, S.~R.~Dolan, A.~C.~Ottewill, N.~Warburton and B.~Wardell,
  Phys.\ Rev.\ D {\bf 92}, 123008 (2015),
  arXiv:1505.04447 [gr-qc].

\bibitem{Blanchet:2013haa} 
  L.~Blanchet,
  Living Rev.\ Rel.\  {\bf 17}, 2 (2014),
  arXiv:1310.1528 [gr-qc].

\bibitem{Choptuik:2015mma} 
  M.~W.~Choptuik, L.~Lehner and F.~Pretorius,
  in {\it General Relativity and Gravitation: A Centennial Perspective}, edited by A. Ashtekar, B. K. Berger, J. Isenberg and M. MacCallum (Cambridge University Press, Cambridge, 2015) p. 361,
  arXiv:1502.06853 [gr-qc].

\bibitem{Buonanno:1998gg} 
  A.~Buonanno and T.~Damour,
  Phys.\ Rev.\ D {\bf 59}, 084006 (1999),
  gr-qc/9811091.

\bibitem{Buonanno:2000ef} 
  A.~Buonanno and T.~Damour,
  Phys.\ Rev.\ D {\bf 62}, 064015 (2000),
  gr-qc/0001013.

\bibitem{Blanchet:2010} 
  L.~Blanchet, S.~Detweiler, A.~Le~Tiec and B.~F.~Whiting,
  Phys.\ Rev.\ D {\bf 81}, 064004 (2010),
  arXiv:0910.0207 [gr-qc].

\bibitem{Blanchet:2010b} 
  L.~Blanchet, S.~Detweiler, A.~Le~Tiec and B.~F.~Whiting,
  Phys.\ Rev.\ D {\bf 81}, 084033 (2010),
  arXiv:1002.0726 [gr-qc].

\bibitem{Damour:2010} 
  T.~Damour,
  Phys.\ Rev.\ D {\bf 81}, 024017 (2010),
  arXiv:0910.5533 [gr-qc].

\bibitem{LeTiec:2011} 
  A.~Le~Tiec, A.~H.~Mrou{\'e}, L.~Barack, A.~Buonanno, H.~P.~Pfeiffer, N.~Sago and A.~Taracchini,
  Phys.\ Rev.\ Lett. {\bf 107}, 141101 (2011),
  arXiv:1106.3278 [gr-qc].

\bibitem{LeTiec:2012b} 
  A.~Le~Tiec, E.~Barausse and A.~Buonanno,
  Phys.\ Rev.\ Lett. {\bf 108}, 131103 (2012),
  arXiv:1111.5609 [gr-qc].

\bibitem{Barausse:2012} 
  E.~Barausse, A.~Buonanno and A.~Le~Tiec,
  Phys.\ Rev.\ D {\bf 85}, 064010 (2012),
  arXiv:1111.5610 [gr-qc].

\bibitem{Akcay:2012} 
  S.~Akcay, L.~Barack, T.~Damour and N.~Sago,
  Phys.\ Rev.\ D {\bf 86}, 104041 (2012),
  arXiv:1209.0964 [gr-qc].

\bibitem{LeTiec:2013b} 
  A.~Le~Tiec \textit{et al.},
  Phys.\ Rev.\ D {\bf 88}, 124027 (2013),
  arXiv:1309.0541 [gr-qc].

\bibitem{LeTiec:2014lba} 
  A.~Le~Tiec,
  Int.\ J.\ Mod.\ Phys.\ D {\bf 23}, 1430022 (2014),
  arXiv:1408.5505 [gr-qc].
 
\bibitem{Akcay:2015} 
  S.~Akcay, A.~Le~Tiec, L.~Barack, N.~Sago and N.~Warburton,
  Phys.\ Rev.\ D {\bf 91}, 124014 (2015),
  arXiv:1503.01374 [gr-qc].

\bibitem{Akcay:2016} 
  S.~Akcay and M.~van~de~Meent,
  Phys.\ Rev.\ D {\bf 93}, 064063 (2016),
  arXiv:1512.03392 [gr-qc].

\bibitem{Bini:2016qtx} 
  D.~Bini, T.~Damour and A.~Geralico,
  Phys.\ Rev.\ D {\bf 93}, 104017 (2016),
  arXiv:1601.02988 [gr-qc].
  

\bibitem{Kavanagh:2016idg} 
  C.~Kavanagh, A.~C.~Ottewill and B.~Wardell,
  Phys.\ Rev.\ D {\bf 93}, 124038 (2016),
  arXiv:1601.03394 [gr-qc].

\bibitem{Bini:2016dvs} 
  D.~Bini, T.~Damour and A.~Geralico,
  Phys.\ Rev.\ D {\bf 93}, 124058 (2016)
  arXiv:1602.08282 [gr-qc].

\bibitem{Barack:2001ph} 
  L.~Barack and A.~Ori,
  Phys.\ Rev.\ D {\bf 64}, 124003 (2001),
  gr-qc/0107056.

\bibitem{Gralla:2011zr} 
  S.~E.~Gralla,
  Phys.\ Rev.\ D {\bf 84}, 084050 (2011),
  arXiv:1104.5635 [gr-qc].

\bibitem{Pound:2015fma} 
  A.~Pound,
  Phys.\ Rev.\ D {\bf 92}, 044021 (2015),
  arXiv:1506.02894 [gr-qc].

\bibitem{Schmidt:2002qk} 
  W.~Schmidt,
  Class.\ Quant.\ Grav.\  {\bf 19}, 2743 (2002),
  gr-qc/0202090.


\bibitem{Wilkins:1972rs} 
  D.~C.~Wilkins,
  Phys.\ Rev.\ D {\bf 5}, 814 (1972).


\bibitem{Hughes:1999bq} 
  S.~A.~Hughes,
  Phys.\ Rev.\ D {\bf 61}, 084004 (2000)
  Erratum-ibid. {\bf 63}, 049902 (2001), {\bf 65}, 069902 (2002),
  {\bf 67}, 089901 (2003), {\bf 78}, 109902 (2008),
  {\bf 90}, 109904 (2014), 
  gr-qc/9910091.
 
\bibitem{LeTiec:2012} 
  A.~Le~Tiec, L.~Blanchet and B.~F.~Whiting,
  Phys.\ Rev.\ D {\bf 85}, 064039 (2012),
  arXiv:1111.5378 [gr-qc].

\bibitem{Blanchet:2013} 
  L.~Blanchet, A.~Buonanno and A.~Le~Tiec,
  Phys.\ Rev.\ D {\bf 87}, 024030 (2013),
  arXiv:1211.1060 [gr-qc].

\bibitem{LeTiec:2015} 
  A.~Le~Tiec,
  Phys.\ Rev.\ D {\bf 92}, 084021 (2015),
  arXiv:1506.05648 [gr-qc].

\bibitem{Blanchet:2017} 
  L.~Blanchet and A.~Le~Tiec
  (2017),
  arXiv:1702.06839 [gr-qc].

\bibitem{Vines:2015} 
  J.~Vines and {\'E}.~{\'E}.~Flanagan,
  Phys.\ Rev.\ D {\bf 92}, 064039 (2015),
  arXiv:1503.04727 [gr-qc].

\bibitem{Pound:2007th} 
  A.~Pound and E.~Poisson,
  Phys.\ Rev.\ D {\bf 77}, 044013 (2008),
  arXiv:0708.3033 [gr-qc].

\bibitem{Glampedakis:2005cf} 
  K.~Glampedakis and S.~Babak,
  Class.\ Quant.\ Grav.\  {\bf 23}, 4167 (2006),
  gr-qc/0510057.

\bibitem{Yang:2014} 
  H.~Yang, H.~Miao and Y.~Chen,
  Phys.\ Rev.\ D {\bf 89}, 104050 (2014),
  arXiv:1211.5410 [gr-qc].

\bibitem{Pound:2009sm} 
  A.~Pound,
  Phys.\ Rev.\ D {\bf 81}, 024023 (2010),
  arXiv:0907.5197 [gr-qc].

\bibitem{Harte:2011ku} 
  A.~I.~Harte,
  Class.\ Quant.\ Grav.\  {\bf 29}, 055012 (2012),
  arXiv:1103.0543 [gr-qc].

\bibitem{Brink:2013nna} 
  J.~Brink, M.~Geyer and T.~Hinderer,
  Phys.\ Rev.\ Lett.\  {\bf 114}, 081102 (2015),
  arXiv:1304.0330 [gr-qc].

\bibitem{Brink:2015roa} 
  J.~Brink, M.~Geyer and T.~Hinderer,
  Phys.\ Rev.\ D {\bf 91}, 083001 (2015),
  arXiv:1501.07728 [gr-qc].
  
\bibitem{Gal'tsov:1982zz} 
  D.~V.~Gal'tsov,
  J.\ Phys.\ A {\bf 15}, 3737 (1982).

\bibitem{LeTiec:2013} 
  A.~Le~Tiec,
  Class.\ Quant.\ Grav.\ {\bf 31}, 097001 (2013),
  arXiv:1311.3836 [gr-qc].

\bibitem{Bini:2014ica} 
  D.~Bini and T.~Damour,
  Phys.\ Rev.\ D {\bf 90}, 024039 (2014),
  arXiv:1404.2747 [gr-qc].

\bibitem{Bini:2015kja} 
  D.~Bini and A.~Geralico,
  Phys.\ Rev.\ D {\bf 91}, 084012 (2015).
 
\bibitem{Warburton:2013yj} 
  N.~Warburton, L.~Barack and N.~Sago,
  Phys.\ Rev.\ D {\bf 87}, 084012 (2013),
  arXiv:1301.3918 [gr-qc].

\bibitem{Buonanno:2002ft} 
  A.~Buonanno, Y.~B.~Chen and M.~Vallisneri,
  Phys.\ Rev.\ D {\bf 67}, 024016 (2003),
  Erratum-ibid.\ {\bf 74}, 029903 (2006),
  gr-qc/0205122.

\bibitem{Friedman:2002}
  J.~L.~Friedman and K.~Uryu and M.~Shibata,
  Phys.\ Rev.\ D {\bf 65}, 064035 (2002),
  Erratum-ibid.\ {\bf 70}, 129904 (2004),
  gr-qc/0108070.

\bibitem{Marta:2015}
 M.~Colleoni and L.~Barack,
 Phys.\ Rev.\ D {\bf 91}, 104024 (2015),
 arXiv:1501.07330 [gr-qc].

\bibitem{Marta:2015b}
 M.~Colleoni, L.~Barack, A.~G.~Shah and M.~van~de~Meent,
 Phys.\ Rev.\ D {\bf 92}, 084044 (2015),
 arXiv:1508.04031 [gr-qc].

\bibitem{Marta:2016prep}
 M.~Colleoni, L.~Barack, T.~Damour, S.~Isoyama and N.~Sago, in preparation.

\bibitem{Zimmerman:2016}
  A.~Zimmerman, A.~G.~M.~Lewis and H.~P.~Pfeiffer,
  Phys.\ Rev.\ Lett. {\bf 117}, 191101 (2016),
  arXiv:1606.08056 [gr-qc].

\bibitem{Barausse:2009xi} 
  E.~Barausse and A.~Buonanno,
  Phys.\ Rev.\ D {\bf 81}, 084024 (2010),
  arXiv:0912.3517 [gr-qc].

\bibitem{Barausse:2011ys} 
  E.~Barausse and A.~Buonanno,
  Phys.\ Rev.\ D {\bf 84}, 104027 (2011),
  arXiv:1107.2904 [gr-qc].

\bibitem{Taracchini:2013rva} 
  A.~Taracchini {\it et al.},
  Phys.\ Rev.\ D {\bf 89}, 061502 (2014),
  arXiv:1311.2544 [gr-qc].

\bibitem{Hannam:2013oca} 
  M.~Hannam, P.~Schmidt, A.~Boh{\'e}, L.~Haegel, S.~Husa, F.~Ohme, G.~Pratten and M.~Purrer,
  Phys.\ Rev.\ Lett.\  {\bf 113}, 151101 (2014),
  arXiv:1308.3271 [gr-qc].

\bibitem{Nagar:2015xqa} 
  A.~Nagar, T.~Damour, C.~Reisswig and D.~Pollney,
  Phys.\ Rev.\ D {\bf 93}, 044046 (2016),
  arXiv:1506.08457 [gr-qc].

\bibitem{Barausse:2009aa} 
  E.~Barausse, E.~Racine and A.~Buonanno,
  Phys.\ Rev.\ D {\bf 80}, 104025 (2009),
  Erratum-ibid.\ {\bf 85}, 069904 (2012),
  arXiv:0907.4745 [gr-qc].

\bibitem{Vines:2016unv} 
  J.~Vines, D.~Kunst, J.~Steinhoff and T.~Hinderer,
  Phys.\ Rev.\ D {\bf 93}, 103008 (2016),
  arXiv:1601.07529 [gr-qc].
 
\bibitem{Zimmerman:2016prep} 
  A.~Zimmerman, private communication (2016).

\end{thebibliography}


\end{document}